\documentclass[lettersize,letterpaper,journal,10pt,nofonttune]{IEEEtran}
\pdfoutput=1
\IEEEoverridecommandlockouts
\usepackage{amsmath,amssymb,amsfonts}
\usepackage{algorithm}
\usepackage{algpseudocode}
\usepackage{amsmath}
\usepackage{array}
\usepackage{soul}
\usepackage[dvipsnames]{xcolor}
\usepackage{enumitem}
\usepackage{textcomp}
\usepackage{stfloats}
\usepackage{url}
\usepackage{verbatim}
\usepackage{graphicx}
\usepackage{caption}
\usepackage{subcaption}
\usepackage{xspace}
\usepackage{xcolor}
\usepackage[utf8]{inputenc}
\usepackage{colortbl}
\usepackage{cite}
\usepackage[export]{adjustbox}
\usepackage{makecell}
\usepackage[draft]{hyperref}
\usepackage{cleveref}
\usepackage[super]{nth}
\usepackage{tikz}
\usepackage[normalem]{ulem}
\usepackage{csquotes}
\usepackage{comment}
\usepackage{xspace}
\usepackage{balance}
\usepackage{fancyhdr}
\usepackage{pgfplots}

\newcommand{\blue}[1]{\textcolor{blue}{#1}}

\newcommand{\mycomment}[1]{}

\newcommand{\quadriga}{QuaDRiGa\xspace}

\definecolor{lightblue}{rgb}{0.8, 0.9, 1}

\usepackage{multirow}
\usepackage{listings}
\usepackage{booktabs}
\usepackage{tabularx}
\usepackage{siunitx}
\pgfplotsset{compat=newest}
\pgfplotsset{plot coordinates/math parser=false}
\newlength\fheight
\newlength\fwidth
\usetikzlibrary{plotmarks,patterns,decorations.pathreplacing,backgrounds,calc,arrows,arrows.meta,spy,matrix,scopes}
\usepgfplotslibrary{patchplots,groupplots}
\usepackage{tikzscale}

\newif\ifexttikz
\exttikzfalse

\ifexttikz
	\usetikzlibrary{external}
	\usepackage{fontspec}
\fi

\usepackage[acronyms,nonumberlist,nopostdot,nomain,nogroupskip,acronymlists={hidden}]{glossaries}
\newglossary[algh]{hidden}{acrh}{acnh}{Hidden Acronyms}
\newacronym{3gpp}{3GPP}{3rd Generation Partnership Project}
\newacronym{4g}{4G}{4th generation}
\newacronym{5g}{5G}{5th generation}
\newacronym{6g}{6G}{6th generation}
\newacronym{5gc}{5GC}{5G Core}
\newacronym{adc}{ADC}{Analog to Digital Converter}
\newacronym{aerpaw}{AERPAW}{Aerial Experimentation and Research Platform for Advanced Wireless}
\newacronym{ai}{AI}{Artificial Intelligence}
\newacronym{aimd}{AIMD}{Additive Increase Multiplicative Decrease}
\newacronym{am}{AM}{Acknowledged Mode}
\newacronym{amc}{AMC}{Adaptive Modulation and Coding}
\newacronym{amf}{AMF}{Access and Mobility Management Function}
\newacronym{aops}{AOPS}{Adaptive Order Prediction Scheduling}
\newacronym{api}{API}{Application Programming Interface}
\newacronym{xapp}{xApp}{Intelligent Application}
\newacronym{apn}{APN}{Access Point Name}
\newacronym{ap}{AP}{Application Protocol}
\newacronym{aqm}{AQM}{Active Queue Management}
\newacronym{ausf}{AUSF}{Authentication Server Function}
\newacronym{avc}{AVC}{Advanced Video Coding}
\newacronym{awgn}{AGWN}{Additive White Gaussian Noise}
\newacronym{balia}{BALIA}{Balanced Link Adaptation Algorithm}
\newacronym{bbu}{BBU}{Base Band Unit}
\newacronym{bdp}{BDP}{Bandwidth-Delay Product}
\newacronym{ber}{BER}{Bit Error Rate}
\newacronym{bf}{BF}{Beamforming}
\newacronym{bler}{BLER}{Block Error Rate}
\newacronym{brr}{BRR}{Bayesian Ridge Regressor}
\newacronym{bs}{BS}{Base Station}
\newacronym{bsr}{BSR}{Buffer Status Report}
\newacronym{bss}{BSS}{Business Support System}
\newacronym{ca}{CA}{Carrier Aggregation}
\newacronym{caas}{CaaS}{Connectivity-as-a-Service}
\newacronym{cb}{CB}{Code Block}
\newacronym{cc}{CC}{Congestion Control}
\newacronym{ccid}{CCID}{Congestion Control ID}
\newacronym{cv2x}{C-V2X}{Cellular Vehicle-to-Everything}
\newacronym{cco}{CC}{Carrier Component}
\newacronym{cdd}{CDD}{Cyclic Delay Diversity}
\newacronym{cdf}{CDF}{Cumulative Distribution Function}
\newacronym{cdn}{CDN}{Content Distribution Network}
\newacronym{cn}{CN}{Core Network}
\newacronym{codel}{CoDel}{Controlled Delay Management}
\newacronym{comac}{COMAC}{Converged Multi-Access and Core}
\newacronym{cord}{CORD}{Central Office Re-architected as a Datacenter}
\newacronym{cornet}{CORNET}{COgnitive Radio NETwork}
\newacronym{cosmos}{COSMOS}{Cloud Enhanced Open Software Defined Mobile Wireless Testbed for City-Scale Deployment}
\newacronym{cots}{COTS}{Commercial Off-the-Shelf}
\newacronym{cp}{CP}{Control Plane}
\newacronym{cpu}{CPU}{Central Processing Unit}
\newacronym{cqi}{CQI}{Channel Quality Information}
\newacronym{cr}{CR}{Cognitive Radio}
\newacronym{cran}{CRAN}{Cloud \gls{ran}}
\newacronym{crs}{CRS}{Cell Reference Signal}
\newacronym{csi}{CSI}{Channel State Information}
\newacronym{csirs}{CSI-RS}{Channel State Information - Reference Signal}
\newacronym{cu}{CU}{Central Unit}
\newacronym{d2tcp}{D$^2$TCP}{Deadline-aware Data center TCP}
\newacronym{d3}{D$^3$}{Deadline-Driven Delivery}
\newacronym{dac}{DAC}{Digital to Analog Converter}
\newacronym{dag}{DAG}{Directed Acyclic Graph}
\newacronym{das}{DAS}{Distributed Antenna System}
\newacronym{dash}{DASH}{Dynamic Adaptive Streaming over HTTP}
\newacronym{dc}{DC}{Dual Connectivity}
\newacronym{dccp}{DCCP}{Datagram Congestion Control Protocol}
\newacronym{dce}{DCE}{Direct Code Execution}
\newacronym{dci}{DCI}{Downlink Control Information}
\newacronym{dctcp}{DCTCP}{Data Center TCP}
\newacronym{dl}{DL}{Downlink}
\newacronym{dmr}{DMR}{Deadline Miss Ratio}
\newacronym{dmrs}{DMRS}{DeModulation Reference Signal}
\newacronym{drlcc}{DRL-CC}{Deep Reinforcement Learning Congestion Control}
\newacronym{drs}{DRS}{Discovery Reference Signal}
\newacronym{du}{DU}{Distributed Unit}
\newacronym{e2e}{E2E}{end-to-end}
\newacronym{ecaas}{ECaaS}{Edge-Cloud-as-a-Service}
\newacronym{ecn}{ECN}{Explicit Congestion Notification}
\newacronym{edf}{EDF}{Earliest Deadline First}
\newacronym{embb}{eMBB}{Enhanced Mobile Broadband}
\newacronym{empower}{EMPOWER}{EMpowering transatlantic PlatfOrms for advanced WirEless Research}
\newacronym{enb}{eNB}{evolved Node Base}
\newacronym{endc}{EN-DC}{E-UTRAN-\gls{nr} \gls{dc}}
\newacronym{epc}{EPC}{Evolved Packet Core}
\newacronym{eps}{EPS}{Evolved Packet System}
\newacronym{es}{ES}{Edge Server}
\newacronym{etsi}{ETSI}{European Telecommunications Standards Institute}
\newacronym[firstplural=Estimated Times of Arrival (ETAs)]{eta}{ETA}{Estimated Time of Arrival}
\newacronym{eutran}{E-UTRAN}{Evolved Universal Terrestrial Access Network}
\newacronym{faas}{FaaS}{Function-as-a-Service}
\newacronym{fapi}{FAPI}{Functional Application Platform Interface}
\newacronym{fdd}{FDD}{Frequency Division Duplexing}
\newacronym{fdm}{FDM}{Frequency Division Multiplexing}
\newacronym{fdma}{FDMA}{Frequency Division Multiple Access}
\newacronym{fed4fire}{FED4FIRE+}{Federation 4 Future Internet Research and Experimentation Plus}
\newacronym{fir}{FIR}{Finite Impulse Response}
\newacronym{cir}{CIR}{Channel Impulse Response}
\newacronym{fit}{FIT}{Future \acrlong{iot}}
\newacronym{fpga}{FPGA}{Field Programmable Gate Array}
\newacronym{fr2}{FR2}{Frequency Range 2}
\newacronym{fr1}{FR1}{Frequency Range 1}
\newacronym{fs}{FS}{Fast Switching}
\newacronym{fscc}{FSCC}{Flow Sharing Congestion Control}
\newacronym{ftp}{FTP}{File Transfer Protocol}
\newacronym{fw}{FW}{Flow Window}
\newacronym{ge}{GE}{Gaussian Elimination}
\newacronym{gnb}{gNB}{Next Generation Node Base}
\newacronym{nextg}{NextG}{Next Generation}
\newacronym{gop}{GOP}{Group of Pictures}
\newacronym{gpr}{GPR}{Gaussian Process Regressor}
\newacronym{gpu}{GPU}{Graphics Processing Unit}
\newacronym{gtp}{GTP}{GPRS Tunneling Protocol}
\newacronym{gtpc}{GTP-C}{GPRS Tunnelling Protocol Control Plane}
\newacronym{sca}{SCA}{Successive Convex Approximation}
\newacronym{gtpu}{GTP-U}{GPRS Tunnelling Protocol User Plane}
\newacronym{gtpv2c}{GTPv2-C}{\gls{gtp} v2 - Control}
\newacronym{gw}{GW}{Gateway}
\newacronym{harq}{HARQ}{Hybrid Automatic Repeat reQuest}
\newacronym{hetnet}{HetNet}{Heterogeneous Network}
\newacronym{hh}{HH}{Hard Handover}
\newacronym{hol}{HOL}{Head-of-Line}
\newacronym{hqf}{HQF}{Highest-quality-first}
\newacronym{hss}{HSS}{Home Subscription Server}
\newacronym{http}{HTTP}{HyperText Transfer Protocol}
\newacronym{ia}{IA}{Initial Access}
\newacronym{iab}{IAB}{Integrated Access and Backhaul}
\newacronym{ic}{IC}{Incident Command}
\newacronym{ietf}{IETF}{Internet Engineering Task Force}
\newacronym{imsi}{IMSI}{International Mobile Subscriber Identity}
\newacronym{imt}{IMT}{International Mobile Telecommunication}
\newacronym{iot}{IoT}{Internet of Things}
\newacronym{ip}{IP}{Internet Protocol}
\newacronym{itu}{ITU}{International Telecommunication Union}
\newacronym{kpi}{KPI}{Key Performance Indicator}
\newacronym{kpm}{KPM}{Key Performance Measurement}
\newacronym{kvm}{KVM}{Kernel-based Virtual Machine}
\newacronym{los}{LOS}{Line-of-Sight}
\newacronym{lsm}{LSM}{Link-to-System Mapping}
\newacronym{lstm}{LSTM}{Long Short Term Memory}
\newacronym{lte}{LTE}{Long Term Evolution}
\newacronym{lxc}{LXC}{Linux Container}
\newacronym{m2m}{M2M}{Machine to Machine}
\newacronym{mac}{MAC}{Medium Access Control}
\newacronym{manet}{MANET}{Mobile Ad Hoc Network}
\newacronym{mano}{MANO}{Management and Orchestration}
\newacronym{mc}{MC}{Multi-Connectivity}
\newacronym{mcc}{MCC}{Mobile Cloud Computing}
\newacronym{mchem}{MCHEM}{Massive Channel Emulator}
\newacronym{mcs}{MCS}{Modulation and Coding Scheme}
\newacronym{mec}{MEC}{Multi-access Edge Computing}
\newacronym{mec2}{MEC}{Mobile Edge Cloud}
\newacronym{mfc}{MFC}{Mobile Fog Computing}
\newacronym{mgen}{MGEN}{Multi-Generator}
\newacronym{mi}{MI}{Mutual Information}
\newacronym{mib}{MIB}{Master Information Block}
\newacronym{miesm}{MIESM}{Mutual Information Based Effective SINR}
\newacronym{mimo}{MIMO}{Multiple Input, Multiple Output}
\newacronym{ml}{ML}{Machine Learning}
\newacronym{mlr}{MLR}{Maximum-local-rate}
\newacronym[plural=\gls{mme}s,firstplural=Mobility Management Entities (MMEs)]{mme}{MME}{Mobility Management Entity}
\newacronym{mmtc}{mMTC}{Massive Machine-Type Communications}
\newacronym{mmwave}{mmWave}{millimeter wave}
\newacronym{mpdccp}{MP-DCCP}{Multipath Datagram Congestion Control Protocol}
\newacronym{mptcp}{MPTCP}{Multipath TCP}
\newacronym{mr}{MR}{Maximum Rate}
\newacronym{mrdc}{MR-DC}{Multi \gls{rat} \gls{dc}}
\newacronym{mse}{MSE}{Mean Square Error}
\newacronym{mss}{MSS}{Maximum Segment Size}
\newacronym{mt}{MT}{Mobile Termination}
\newacronym{mtd}{MTD}{Machine-Type Device}
\newacronym{mtu}{MTU}{Maximum Transmission Unit}
\newacronym{mumimo}{MU-MIMO}{Multi-user \gls{mimo}}
\newacronym{mvno}{MVNO}{Mobile Virtual Network Operator}
\newacronym{nalu}{NALU}{Network Abstraction Layer Unit}
\newacronym{nas}{NAS}{Non-Access Stratum}
\newacronym{nbiot}{NB-IoT}{Narrow Band IoT}
\newacronym{nfv}{NFV}{Network Function Virtualization}
\newacronym{nfvi}{NFVI}{Network Function Virtualization Infrastructure}
\newacronym{nic}{NIC}{Network Interface Card}
\newacronym{nlos}{NLOS}{Non-Line-of-Sight}
\newacronym{now}{NOW}{Non Overlapping Window}
\newacronym{nsm}{NSM}{Network Service Mesh}
\newacronym[type=hidden]{nr}{NR}{New Radio}
\newacronym{nrf}{NRF}{Network Repository Function}
\newacronym{nsa}{NSA}{Non Stand Alone}
\newacronym{nse}{NSE}{Network Slicing Engine}
\newacronym{nssf}{NSSF}{Network Slice Selection Function}
\newacronym{o2i}{O2I}{Outdoor to Indoor}
\newacronym{oai}{OAI}{OpenAirInterface}
\newacronym{oaicn}{OAI-CN}{\gls{oai} \acrlong{cn}}
\newacronym{oairan}{OAI-RAN}{\acrlong{oai} \acrlong{ran}}
\newacronym{oam}{OAM}{Operations, Administration and Maintenance}
\newacronym{ofdm}{OFDM}{Orthogonal Frequency Division Multiplexing}
\newacronym{olia}{OLIA}{Opportunistic Linked Increase Algorithm}
\newacronym{omec}{OMEC}{Open Mobile Evolved Core}
\newacronym{onap}{ONAP}{Open Network Automation Platform}
\newacronym{onf}{ONF}{Open Networking Foundation}
\newacronym{onos}{ONOS}{Open Networking Operating System}
\newacronym{oom}{OOM}{\gls{onap} Operations Manager}
\newacronym{opnfv}{OPNFV}{Open Platform for \gls{nfv}}
\newacronym{orbit}{ORBIT}{Open-Access Research Testbed for Next-Generation Wireless Networks}
\newacronym{os}{OS}{Operating System}
\newacronym{oss}{OSS}{Operations Support System}
\newacronym{pa}{PA}{Position-aware}
\newacronym{pase}{PASE}{Prioritization, Arbitration, and Self-adjusting Endpoints}
\newacronym{pawr}{PAWR}{Platforms for Advanced Wireless Research}
\newacronym{pbch}{PBCH}{Physical Broadcast Channel}
\newacronym{pcef}{PCEF}{Policy and Charging Enforcement Function}
\newacronym{pcfich}{PCFICH}{Physical Control Format Indicator Channel}
\newacronym{pcrf}{PCRF}{Policy and Charging Rules Function}
\newacronym{pdcch}{PDCCH}{Physical Downlink Control Channel}
\newacronym{pdcp}{PDCP}{Packet Data Convergence Protocol}
\newacronym{pdsch}{PDSCH}{Physical Downlink Shared Channel}
\newacronym{pdu}{PDU}{Packet Data Unit}
\newacronym{pf}{PF}{Proportionally Fair}
\newacronym{pgw}{PGW}{Packet Gateway}
\newacronym{phich}{PHICH}{Physical Hybrid ARQ Indicator Channel}
\newacronym{phy}{PHY}{Physical}
\newacronym{pmch}{PMCH}{Physical Multicast Channel}
\newacronym{pmi}{PMI}{Precoding Matrix Indicators}
\newacronym{powder}{POWDER}{Platform for Open Wireless Data-driven Experimental Research}
\newacronym{ppo}{PPO}{Proximal Policy Optimization}
\newacronym{ppp}{PPP}{Poisson Point Process}
\newacronym{prach}{PRACH}{Physical Random Access Channel}
\newacronym{prb}{PRB}{Physical Resource Block}
\newacronym{rbg}{RBG}{Resource Block Group}
\newacronym{psnr}{PSNR}{Peak Signal to Noise Ratio}
\newacronym{pss}{PSS}{Primary Synchronization Signal}
\newacronym{pucch}{PUCCH}{Physical Uplink Control Channel}
\newacronym{pusch}{PUSCH}{Physical Uplink Shared Channel}
\newacronym{qam}{QAM}{Quadrature Amplitude Modulation}
\newacronym{qci}{QCI}{\gls{qos} Class Identifier}
\newacronym{qoe}{QoE}{Quality of Experience}
\newacronym{qos}{QoS}{Quality of Service}
\newacronym{quic}{QUIC}{Quick UDP Internet Connections}
\newacronym{rach}{RACH}{Random Access Channel}
\newacronym{ran}{RAN}{Radio Access Network}
\newacronym[firstplural=end to endcess Technologies (RATs)]{rat}{RAT}{end to endcess Technology}
\newacronym{rcn}{RCN}{Research Coordination Network}
\newacronym{rec}{REC}{Radio Edge Cloud}
\newacronym{ra}{RA}{Resource Allocation}
\newacronym{red}{RED}{Random Early Detection}
\newacronym{renew}{RENEW}{Reconfigurable Eco-system for Next-generation End-to-end Wireless}
\newacronym{rf}{RF}{Radio Frequency}
\newacronym{rfc}{RFC}{Request for Comments}
\newacronym{rfr}{RFR}{Random Forest Regressor}
\newacronym{ric}{RIC}{\gls{ran} Intelligent Controller}
\newacronym{rlc}{RLC}{Radio Link Control}
\newacronym{rlf}{RLF}{Radio Link Failure}
\newacronym{rlnc}{RLNC}{Random Linear Network Coding}
\newacronym{rmr}{RMR}{RIC Message Router}
\newacronym{rmse}{RMSE}{Root Mean Squared Error}
\newacronym{rnis}{RNIS}{Radio Network Information Service}
\newacronym{rr}{RR}{Round Robin}
\newacronym{rrc}{RRC}{Radio Resource Control}
\newacronym{rrm}{RRM}{Radio Resource Management}
\newacronym{rru}{RRU}{Remote Radio Unit}
\newacronym{rs}{RS}{Remote Server}
\newacronym{rsrp}{RSRP}{Reference Signal Received Power}
\newacronym{rsrq}{RSRQ}{Reference Signal Received Quality}
\newacronym{rss}{RSS}{Received Signal Strength}
\newacronym{rssi}{RSSI}{Received Signal Strength Indicator}
\newacronym{rtt}{RTT}{Round Trip Time}
\newacronym{ru}{RU}{Radio Unit}
\newacronym{rw}{RW}{Receive Window}
\newacronym{rx}{RX}{Receiver}
\newacronym{s1ap}{S1AP}{S1 Application Protocol}
\newacronym{sa}{SA}{standalone}
\newacronym{sack}{SACK}{Selective Acknowledgment}
\newacronym{sap}{SAP}{Service Access Point}
\newacronym{sc2}{SC2}{Spectrum Collaboration Challenge}
\newacronym{scef}{SCEF}{Service Capability Exposure Function}
\newacronym{sch}{SCH}{Secondary Cell Handover}
\newacronym{scoot}{SCOOT}{Split Cycle Offset Optimization Technique}
\newacronym{sctp}{SCTP}{Stream Control Transmission Protocol}
\newacronym{sdap}{SDAP}{Service Data Adaptation Protocol}
\newacronym{sdk}{SDK}{Software Development Kit}
\newacronym{sdm}{SDM}{Space Division Multiplexing}
\newacronym{sdma}{SDMA}{Spatial Division Multiple Access}
\newacronym{sdn}{SDN}{Software-defined Networking}
\newacronym{sdr}{SDR}{Software-defined Radio}
\newacronym{seba}{SEBA}{SDN-Enabled Broadband Access}
\newacronym{sgsn}{SGSN}{Serving GPRS Support Node}
\newacronym{sgw}{SGW}{Service Gateway}
\newacronym{si}{SI}{Study Item}
\newacronym{sib}{SIB}{Secondary Information Block}
\newacronym{sinr}{SINR}{Signal to Interference plus Noise Ratio}
\newacronym{sip}{SIP}{Session Initiation Protocol}
\newacronym{siso}{SISO}{Single Input, Single Output}
\newacronym{sla}{SLA}{Service Level Agreement}
\newacronym{sm}{SM}{Service Model}
\newacronym{smf}{SMF}{Session Management Function}
\newacronym{smo}{SMO}{Service Management and Orchestration}
\newacronym{sms}{SMS}{Short Message Service}
\newacronym{smsgmsc}{SMS-GMSC}{\gls{sms}-Gateway}
\newacronym{snr}{SNR}{Signal-to-Noise-Ratio}
\newacronym{son}{SON}{Self-Organizing Network}
\newacronym{sptcp}{SPTCP}{Single Path TCP}
\newacronym{srb}{SRB}{Service Radio Bearer}
\newacronym{srn}{SRN}{Standard Radio Node}
\newacronym{srs}{SRS}{Sounding Reference Signal}
\newacronym{ss}{SS}{Synchronization Signal}
\newacronym{sss}{SSS}{Secondary Synchronization Signal}
\newacronym{st}{ST}{Spanning Tree}
\newacronym{svc}{SVC}{Scalable Video Coding}
\newacronym{tb}{TB}{Transport Block}
\newacronym{tcp}{TCP}{Transmission Control Protocol}
\newacronym{tdd}{TDD}{Time Division Duplexing}
\newacronym{tdm}{TDM}{Time Division Multiplexing}
\newacronym{tdma}{TDMA}{Time Division Multiple Access}
\newacronym{tfl}{TfL}{Transport for London}
\newacronym{tfrc}{TFRC}{TCP-Friendly Rate Control}
\newacronym{tft}{TFT}{Traffic Flow Template}
\newacronym{tgen}{TGEN}{Traffic Generator}
\newacronym{tip}{TIP}{Telecom Infra Project}
\newacronym{tm}{TM}{Transparent Mode}
\newacronym{to}{TO}{Telco Operator}
\newacronym{tr}{TR}{Technical Report}
\newacronym{trp}{TRP}{Transmitter Receiver Pair}
\newacronym{ts}{TS}{Technical Specification}
\newacronym{tti}{TTI}{Transmission Time Interval}
\newacronym{ttt}{TTT}{Time-to-Trigger}
\newacronym{tx}{TX}{Transmitter}
\newacronym{uas}{UAS}{Unmanned Aerial System}
\newacronym{uav}{UAV}{Unmanned Aerial Vehicle}
\newacronym{udm}{UDM}{Unified Data Management}
\newacronym{udp}{UDP}{User Datagram Protocol}
\newacronym{udr}{UDR}{Unified Data Repository}
\newacronym{ue}{UE}{User Equipment}
\newacronym{uhd}{UHD}{\gls{usrp} Hardware Driver}
\newacronym{ul}{UL}{Uplink}
\newacronym{um}{UM}{Unacknowledged Mode}
\newacronym{uml}{UML}{Unified Modeling Language}
\newacronym{upa}{UPA}{Uniform Planar Array}
\newacronym{upf}{UPF}{User Plane Function}
\newacronym{urllc}{URLLC}{Ultra Reliable and Low Latency Communications}
\newacronym{usa}{U.S.}{United States}
\newacronym{usim}{USIM}{Universal Subscriber Identity Module}
\newacronym{usrp}{USRP}{Universal Software Radio Peripheral}
\newacronym{utc}{UTC}{Urban Traffic Control}
\newacronym{vim}{VIM}{Virtualization Infrastructure Manager}
\newacronym{vm}{VM}{Virtual Machine}
\newacronym{vnf}{VNF}{Virtual Network Function}
\newacronym{volte}{VoLTE}{Voice over \gls{lte}}
\newacronym{voltha}{VOLTHA}{Virtual OLT HArdware Abstraction}
\newacronym{vr}{VR}{Virtual Reality}
\newacronym{vran}{vRAN}{Virtualized \gls{ran}}
\newacronym{vss}{VSS}{Video Streaming Server}
\newacronym{wbf}{WBF}{Wired Bias Function}
\newacronym{wf}{WF}{Waterfilling}
\newacronym{wlan}{WLAN}{Wireless Local Area Network}
\newacronym{osm}{OSM}{Open Source \gls{nfv} Management and Orchestration}
\newacronym{pnf}{PNF}{Physical Network Function}
\newacronym{drl}{DRL}{Deep Reinforcement Learning}
\newacronym{rl}{RL}{Reinforcement Learning}
\newacronym{mtc}{MTC}{Machine-type Communications}
\newacronym{osc}{OSC}{O-RAN Software Community}
\newacronym{rc}{RC}{RAN Control}
\newacronym{dqn}{DQN}{Deep Q-Network}
\newacronym{v2x}{V2X}{Vehicle-to-everything}
\newacronym{gbsm}{GBSM}{Geometry-Based Stochastic Model}
\newacronym{gbs}{GBSM}{Geometry-Based Stochastic}
\newacronym{quadriga}{QuaDRiGa}{QUAsi Deterministic RadIo channel GenerAtor}
\newacronym{relu}{ReLU}{Rectified Linear Unit} 
\newacronym{mpc}{MPC}{Multipath Component}
\newacronym{xpr}{XPR}{Cross-polarization Ratio}
\newacronym{lsp}{LSP}{Large Scale Parameter}
\newacronym{ssp}{SSP}{Small Scale Parameter}
\newacronym{fbs}{FBS}{First Bounce Scatterer}
\newacronym{lbs}{LBS}{Last Bounce Scatterer}
\newacronym{d2d}{D2D}{Device-to-Device}
\newacronym{rsu}{RSU}{Road Side Unit}
\newacronym{toa}{ToA}{Time-of-Arrival}
\newacronym{ris}{RIS}{Reconfigurable Intelligent Surface}
\newacronym{aoa}{AoA}{Angle of Arrival}
\newacronym{aod}{AoD}{Angle of Departure}
\newacronym{pl}{PL}{Path-Loss}
\newacronym{noma}{NOMA}{Non-Orthogonal Multiple Access}
\newacronym{sic}{SIC}{Successive Interference Cancellation}
\newacronym{gps}{GPS}{Global Positioning System}
\newacronym{ids}{IDS}{Independent Diffusive Scatterer-based}
\newacronym{inw}{INW}{Impedance Network-based}
\newacronym{minlp}{MINLP}{Mixed Integer Non-Linear Programming}
\newacronym{star}{STAR}{Simultaneous Transmitting And Reflecting}
\glsdisablehyper

\usepackage{adjustbox}
\def\BibTeX{{\rm B\kern-.05em{\sc i\kern-.025em b}\kern-.08em
    T\kern-.1667em\lower.7ex\hbox{E}\kern-.125emX}}
    
\renewcommand{\blue}[1]{\textcolor{black}{#1}}

\usepackage{tikzpagenodes,etoolbox}
\usetikzlibrary{calc}
\usepackage[contents={}]{background}
\AddEverypageHook{%
\ifnumequal{\thepage}{1}{%
 \tikz[remember picture,overlay]{%
     \node[draw,
     text width=0.95\textwidth,
     font=\footnotesize
     ]
     at ($(current page header area) - (0,5pt)$)
     {%
     This paper has been accepted for publication on \textit{IEEE Transactions on Vehicular Technology}.\\
\copyright~2025 IEEE. Personal use of this material is permitted. Permission from IEEE must be obtained for all other uses, in any current or future media, including reprinting/republishing this material for advertising or promotional purposes, creating new collective works, for resale or redistribution to servers or lists, or reuse of any copyrighted component of this work in other works..
     };
 }%
}{}
}

\begin{document}

\title{System-Level Experimental Evaluation of Reconfigurable Intelligent Surfaces for NextG
Communication Systems}
\author{{Maria Tsampazi,~\IEEEmembership{Student Member, IEEE}, and Tommaso Melodia,~\IEEEmembership{Fellow, IEEE}}

\thanks{Copyright (c) 2025 IEEE. Personal use of this material is permitted. However, permission to use this material for any other purposes must be obtained from the IEEE by sending a request to pubs-permissions@ieee.org.}
\thanks{This article is based upon work partially supported by the U.S.\ National Science Foundation under grants CNS-1925601 and CNS-2112471.}
\thanks{M. Tsampazi, and T. Melodia are with the Institute for the Wireless Internet of Things, Northeastern University, Boston, MA, U.S.A. E-mail: \{tsampazi.m, t.melodia\}@northeastern.edu.}}

\makeatletter
\patchcmd{\@maketitle}
  {\addvspace{0.5\baselineskip}\egroup}
  {\addvspace{-1.5\baselineskip}\egroup}
  {}
  {}
\makeatother

\maketitle

\glsunset{usrp}


\maketitle

\begin{abstract}

\glspl{ris} are a promising technique for enhancing the performance of \gls{nextg} wireless communication systems in terms of both spectral and energy efficiency, as well as resource utilization. However, current \gls{ris} research has primarily focused on theoretical modeling and \gls{phy} layer considerations only. Full protocol stack emulation and accurate modeling of the propagation characteristics of the wireless channel are necessary for studying the benefits introduced by \gls{ris} technology across various spectrum bands 
and use-cases.
In this paper, we propose, for the first time:
(i) accurate \gls{phy} layer \gls{ris}-enabled channel modeling through \glspl{gbsm}, leveraging the \gls{quadriga} open-source statistical ray-tracer; (ii) optimized resource allocation with \glspl{ris} by comprehensively studying energy efficiency and power control on different portions of the spectrum through a single-leader multiple-followers Stackelberg game theoretical approach; (iii) full-stack emulation and performance evaluation of \gls{ris}-assisted channels with SCOPE/srsRAN for \gls{embb} and \gls{urllc} applications in the world's largest emulator of wireless systems with hardware-in-the-loop,
namely Colosseum. Our findings indicate (i) the significant power savings in terms of energy efficiency achieved with \gls{ris}-assisted topologies, especially in the \gls{mmwave} band; and  (ii) the benefits introduced for Sub-6 GHz band \glspl{ue}, where the deployment of a relatively small \gls{ris} (e.g., in the order of $100$ \gls{ris} elements) can result in decreased levels of latency for \gls{urllc} services in resource-constrained environments. 
\end{abstract}


\section{Introduction}\label{Section I}
Undoubtedly, research in \gls{5g} wireless communication systems has been following an upward trend in recent years. \gls{5g} promises better connectivity anywhere, anytime, for everyone and everything, with increased coverage, better system capacity, extremely low latency~\cite{attar20225g}, improved performance in terms of both energy and spectral efficiency~\cite{shehab20215g,sobhi2020energy}, while also simultaneously satisfying the diverse \gls{qos} demands of its heterogeneous users~\cite{zhang2017network} across a variety of deployment scenarios and network topologies \cite{wang2018survey}. Hence, \gls{5g} and beyond technologies are expected to become an indispensable component of future smart environments~\cite{lopes20235g,liu2021promoting,sharif2019compact,marabissi2018real,jiang2019smart,zhao2021nanogenerators,sanchez2021review}.
Another prominent technology anticipated to 
accommodate the needs
of future \gls{nextg} communication networks are \glspl{ris}. \glspl{ris} have recently attracted widespread interest in the research community as a part of a smart city environment for \gls{5g}, and beyond \gls{5g} wireless communication systems~\cite{rana2023review,zivuku2022maximizing,al2023emerging,kisseleff2020reconfigurable,liu2023integrated,salah2022paving,shehab20215g,makarfi2020reconfigurable,kamruzzaman2022key,bariah2023digital,dagiuklas2023journey,li2023liquid,mishra20236g,masouros2023guest,renzo2019smart}. Indeed, \glspl{ris} technologies have been proven to extend coverage~\cite{li2023liquid,liu2023integrated,salah2022paving,makarfi2020reconfigurable}, enhance both spectral and energy efficiency~\cite{shehab20215g,hou2020reconfigurable,9079457,guan2022irs}, while at the same time enabling the vision of smart and reconfigurable propagation environments~\cite{ji2022reconfigurable,renzo2019smart,dagiuklas2023journey}, which can be controlled through~\gls{sdr} techniques~\cite{liaskos2022software,banchs2021network}. Therefore, due to both the tremendous advantages that they introduce (e.g., decreased levels of latency~\cite{basharat2022exploring,liu2021reconfigurable}) and due to their presence in a multitude of use-cases scenarios (e.g., resource allocation in multi-\gls{ue} \gls{noma} systems~\cite{mu2021capacity}, localization and positioning~\cite{elzanaty2021reconfigurable}, assisting anti-jamming  methods~\cite{yang2020intelligent}, and enhancing \gls{uav}-enabled communications~\cite{salih2023enhancing,diamanti2021energy,liu2021reconfigurable,li2020reconfigurable,basharat2021reconfigurable,ning2023intelligent,yu2022fair} among others), there is a clear need for further understanding of \gls{ris}-assisted channels through effective performance evaluation campaigns in different spectrum bands and use cases.

\blue{In this article, we aim to provide, for the first time, a holistic, system-level evaluation of \gls{ris}-assisted channels for \gls{nextg} communication systems in \gls{cv2x} environments across different portions of the spectrum, while also paving the way for their evaluation across various network slices on experimental platforms. Our key contributions are as follows:}
\blue{
\begin{enumerate}
    \item We leverage the capabilities of the fully reconfigurable, MATLAB-based~\cite{MATLAB:2021a} statistical ray-tracer \gls{quadriga}~\cite{quadrigacode}, which uses \gls{rf} data collected under various conditions to cascade the wireless links and include the effect of a \gls{ris} in the topology.
    \item We focus on a \gls{cv2x} setup comprising of a \gls{uav} serving as a flying \gls{gnb} for multiple \glspl{ue}.  
    \item Through a single-leader multiple-followers Stackelberg game we jointly optimize the overall received signal strength at the \gls{gnb}, and each \gls{ue}'s individual energy efficiency in a \gls{noma} system.
    \item Through this hierarchical game-theoretic approach, we determine both the phase shifts of the \gls{ris} reflecting elements, as well as the \gls{ul} transmission power of the \glspl{ue} for both the Sub-6 GHz and the \gls{mmwave} bands.
    \item We install the generated wireless topologies for the band that requires only a minimal number of \gls{ris} elements to achieve a significant path gain enhancement in the Colosseum Wireless Network and Channel Emulator~\cite{bonati2021colosseum}.
\item We modify the \gls{ran} functionalities in terms of scheduling profiles (i.e., \gls{rr} and \gls{wf}) and slicing (i.e., allocation of the available \glspl{prb}) and we observe the performance gain provided through the \gls{ris} technology on the \gls{embb} and \gls{urllc} network slices.
\end{enumerate}
}

\subsection{Related Work}\label{Section IA}
The current literature hosts a plethora of \gls{ris} simulators for \gls{phy} layer channel modeling across various operating frequencies. Additionally, substantial efforts have been reported in the design, creation, and real-world deployment of such surfaces.
In~\cite{basar2021reconfigurable,basar2020simris}, the authors present the \gls{mmwave} SimRIS simulator for indoor and microcell environments. The aforementioned works state that accurate \gls{phy} layer simulators that can operate in a variety of frequencies and applications, ideally using real data, are of utmost importance. The authors note that further research is needed to develop practical channel models for various propagation environments and to identify compelling use cases. Finally, the authors emphasize the need to explore the coexistence of \gls{embb} and \gls{urllc} \glspl{ue} in various frequency domains, as well as the importance of standardization and integration into existing networks. In~\cite{sihlbom2022reconfigurable} a commercial \gls{ris} simulator is leveraged to perform system-level simulation in the mid and \gls{mmwave} bands within a real-world urban environment. The authors conclude by mentioning the importance of analytical frameworks for system-level analysis by using tools leveraging stochastic geometry. Regarding the sub-6 GHz band, in~\cite{sun20213d}, a 3D \gls{gbsm} relying on the statistical properties of the wireless channels is discussed, taking into account both multipath propagation and continuous-time evolution. A \gls{gbsm} channel model incorporating a \gls{ris} is also considered in~\cite{dang2021geometry}, highlighting the importance of leveraging small panels. Other \gls{nextg} \gls{ris}-assisted channel models, which discuss path-loss propagation and take into account geometry, are also presented in~\cite{dajer2022reconfigurable,liu2021reconfigurable}. Rayleigh and Rician channels are used to model the multipath effect, and their integration in \gls{ris}-assisted communication systems is discussed in detail in \cite{zhou2023survey,kong2021channel}.  In~\cite{tian2023reconfigurable}, a frequency-independent channel model for \gls{ris} is discussed, whereas a \gls{ris}-assisted channel model for \gls{v2x} environments that incorporates the Rician channel matrix is presented in~\cite{chen2021qos}.
A \gls{ris}-assisted channel modeling approach for high frequencies is discussed in~\cite{boulogeorgos2021coverage}, while a signal processing perspective on multipath propagation with \gls{ris} is given in~\cite{bjornson2022reconfigurable}. Physics-based \gls{phy} models are presented in~\cite{yu2021smart,xu2023exploiting}, while other approaches include deploying \gls{ris} panels in the Sub-6 GHz~\cite{tang2021channel} and \gls{mmwave}~\cite{da2023varactor} bands~\cite{liu2022reconfigurable, liu2022simulation}, with a focus on energy efficiency~\cite{rossanese2022designing}.

Additionally, the contribution of \gls{ris} technology to energy efficiency through power control is a well-investigated topic. \blue{In~\cite{8741198}, a non-convex optimization framework for energy efficiency that jointly optimizes the \gls{tx} power of the \gls{bs} and the reflecting coefficients of the \gls{ris} on the \gls{dl} is considered. In~\cite{9950621}, the paper tackles energy efficiency maximization in an \gls{ul} \gls{ris}-aided \gls{mmwave} \gls{noma} network by jointly optimizing the \gls{ue} \gls{tx} power and the \gls{ris} phase shifts using iterative optimization techniques.
In~\cite{9810453}, the usage of both \glspl{ris} and \gls{uav} is discussed in terms of providing energy-efficient communications on the \gls{dl} while ensuring the \gls{qos} demands of the \glspl{ue}. The proposed method uses \gls{sca} to iteratively determine a joint optimal solution for the \gls{uav} trajectory, \gls{ris} phase shifts, and the \gls{uav} \gls{tx} power.
In~\cite{9628234}, the problem of energy efficiency maximization is modeled as a \gls{minlp} problem, with the objective of enhancing overall energy efficiency on the \gls{dl} by optimizing the \gls{bs}'s \gls{tx} power and \gls{ris} phase shifts. 
The work in~\cite{10096617} discusses the topic of energy efficiency maximization in a \gls{ris}-assisted network, where convex optimization and sequential approximation methods are employed to jointly optimize the \gls{tx} powers of the \gls{ue}s and the reflection coefficients of the \gls{ris} in the \gls{ul}. In~\cite{10107766}, a \gls{ml} approach based on the \gls{ppo} algorithm for the maximization of energy efficiency in \gls{ris}-aided networks is discussed. The optimization problem is formulated under constraints of the rate requirement of the \glspl{ue}, the power budget at the \gls{bs}, and the discrete phase shift coefficients of each reflecting element at the \gls{ris} on the \gls{dl}. In this approach, the \gls{bs}'s \gls{tx} power and the phase shift matrix of the \gls{ris} are jointly optimized. Similarly, In~\cite{9964251}, the joint energy efficiency maximization problem for a \gls{dl} \gls{noma} network is discussed, with the goal of jointly optimizing the beamforming vectors at the \gls{bs} and the coefficient matrices at the \gls{ris}, while also controlling the \gls{tx} power at the \gls{bs} using \gls{drl}. In~\cite{diamanti2021energy}, a two-tier Stackelberg game is used to jointly maximize the received signal strength at the \gls{enb} by calculating the \gls{ris} phase shifts and the \glspl{ue}' \gls{ul} transmission power in \gls{ris}-assisted, \gls{uav}-enabled networks. However, the aforementioned framework does not consider the frequency of operation, and therefore provides limited insights on energy-efficient resource allocation on different portions of the spectrum. In~\cite{diamanti2021prospect}, a Stackelberg game is also discussed for \gls{e2e} energy-efficient optimization through power control and \gls{ris} reconfiguration in \gls{ul} \gls{ris}-assisted, \gls{uav}-enabled \gls{iab} networks. However, the focus is directed towards the low Sub-6 GHz portion of the spectrum, disregarding emerging \gls{5g} frequency bands, such as the \gls{mmwave}.}

Lastly, the generation of \gls{rf} scenarios for Wireless Network Channel Emulators has been a trending topic. In~\cite{tehrani2021creating}, a framework for creating \gls{rf} scenarios for large-scale, real-time wireless emulators such as Colosseum, leveraging a commercial ray-tracer is introduced. It leverages efficient clustering techniques and channel impulse response re-sampling to scale down the large input set of \gls{rf} data to fewer parameters. 
 In~\cite{villa2022cast}, the CaST Toolchain is presented. It comprises a framework for creating \gls{rf} scenarios from ray-tracing models for \gls{fpga}-based emulation platforms and a \gls{sdn} channel sounder to characterize the emulated channels.
 In~\cite{rusca2023mobile}, a framework for creating mobile \gls{rf} scenarios, including the presence of \glspl{uav}, based on real collected data, is discussed.

The contributions discussed above clearly demonstrate the significance of \gls{ris}-assisted \gls{nextg} channel models, as well as the substantial gains achieved in terms of energy efficiency through the deployment of \glspl{ris} in various heterogeneous scenarios and use cases. At the same time, the interest in the validation and testing of channel models, leveraging the full cellular protocol stack, in high-fidelity and accurate network channel emulators is ever-growing. 

\blue{However, despite the established contributions of hierarchical game-theoretic approaches to energy efficient power control in wireless networks~\cite{lasaulce2009introducing, he2011stackelberg}, there has been limited research on the benefits of energy-efficient power control in \gls{ris}-assisted and \gls{uav}-enabled  \gls{cv2x} networks in the \gls{fr1} and \gls{fr2}. Moreover, despite the availability of frameworks for generating \gls{rf} scenarios in channel emulators, no prior work has focused on installing \gls{ris}-assisted scenarios in these emulators. Consequently, a research gap exists in cross-layer experimental evaluation of \gls{ris}-assisted channels utilizing the full protocol stack on experimental platforms. Therefore, the evaluation of \gls{ris}-assisted networks across different network slices (e.g., \gls{embb} and \gls{urllc}) is currently missing. This paper aims to fill these gaps in the literature with the following original contributions: (i) we propose accurate \gls{phy} layer \gls{ris}-enabled channel modeling through \glspl{gbsm} and \gls{quadriga}; (ii) we cascade the generated channels to include the effect of a \gls{ris}; (iii) we leverage hierarchical game theory to optimize the \gls{ris} elements' phase shifts; (iv) we study energy-efficient power control across different spectrum bands; (v) we implement the \gls{ris}-assisted channels on the Colosseum testbed; and (vi) we perform full-protocol stack emulation using srsRAN~\cite{srsran} to evaluate the performance gain in \gls{embb} and \gls{urllc} network slicing by fine-tuning the \gls{bs}'s network functions.}

The remainder of this paper is organized as follows. Section~\ref{Section II} provides an overview on \gls{gbsm}-based \gls{nextg} channel modeling with \gls{quadriga}. Section~\ref{Section III} discusses the simulation of a \gls{ris}-assisted \gls{cv2x} setup
with \gls{quadriga}, as well as its evaluation on energy efficiency in the Sub-6 GHz and the \gls{mmwave} bands. Section~\ref{Section IV} details the installation and evaluation of a \gls{ris}-assisted topology on the Colosseum wireless network and channel emulator, as well as its evaluation on the \gls{embb} and \gls{urllc} network slices. Finally, Section~\ref{Section V} concludes the paper and discusses possible future directions.

\section{\glspl{gbsm} with \quadriga}\label{Section II}

\subsection{Channel Modeling with QuaDRiGa}\label{Section IIB}

\begin{figure*}[ht]
\centering
\begin{subfigure}{0.30\textwidth}
    \includegraphics[width=\textwidth]{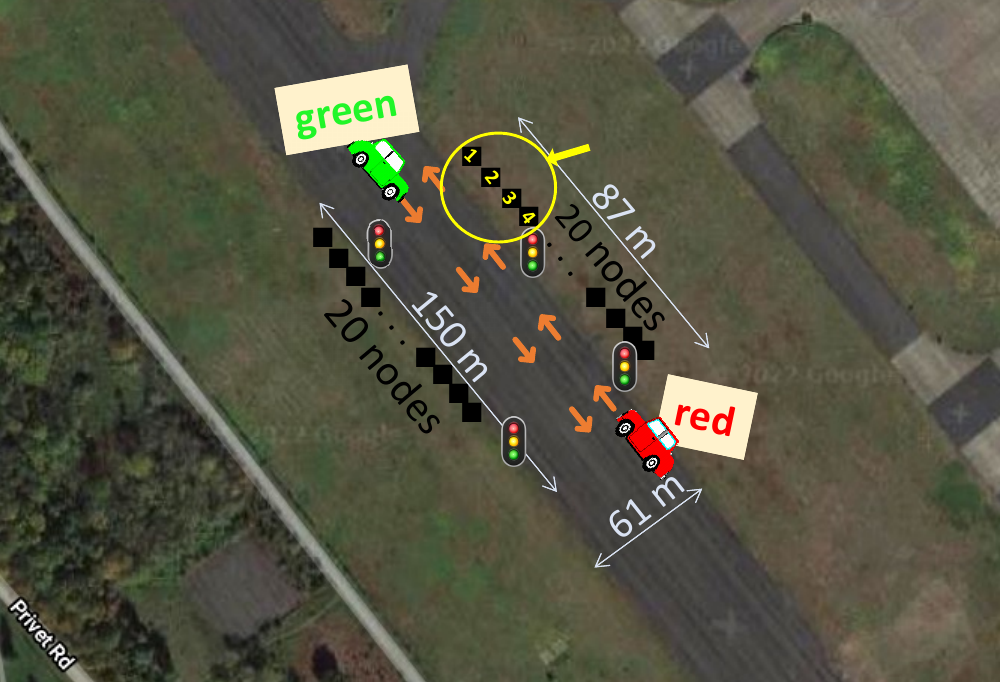}
    \caption{\centering Location from Google Maps \cite{googlemaps}.}
\label{fig:Penn_topologies}
\end{subfigure}
\begin{subfigure}{0.38\textwidth}
    \includegraphics[width=\textwidth]{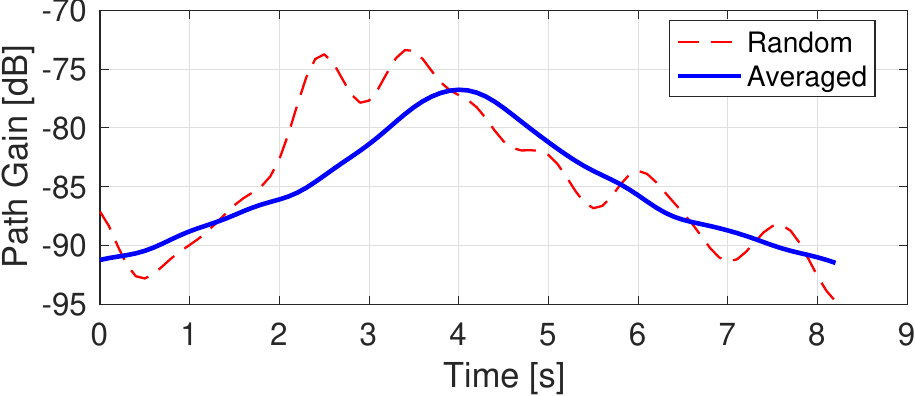}
    \caption{\centering Path Gain}
\label{fig:v1_to_v2_100sims_0.1_updateRate}
\end{subfigure}
\hfill
\begin{subfigure}{0.32\textwidth}
    \includegraphics[width=\textwidth]{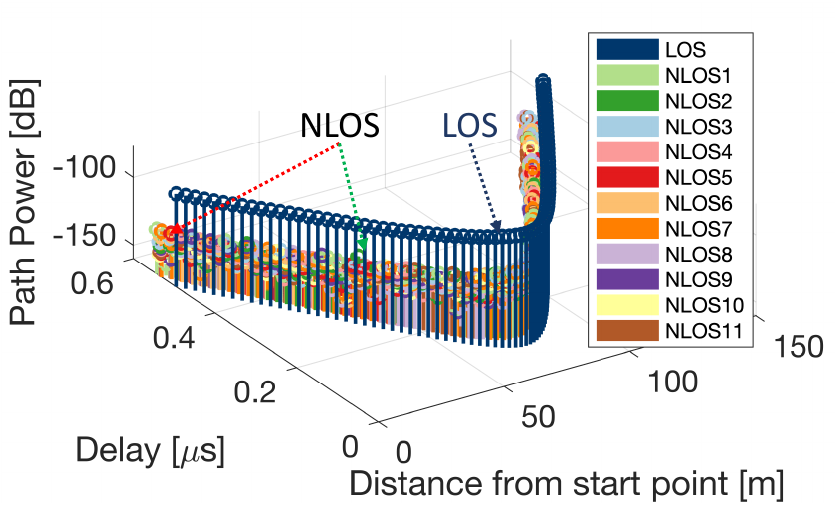}
    \caption{\centering  Path Power VS. Delay VS. Distance}
\label{fig:total}
\end{subfigure}
\begin{subfigure}{0.32\textwidth}
    \includegraphics[width=\textwidth]{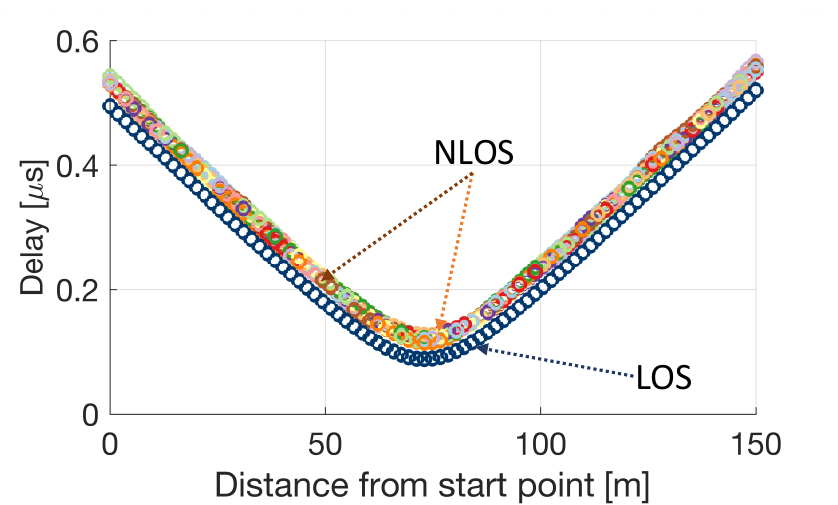}
    \caption{\centering Delay VS. Distance}
\label{fig:total2}
\end{subfigure}
\begin{subfigure}{0.32\textwidth}
    \includegraphics[width=\textwidth]{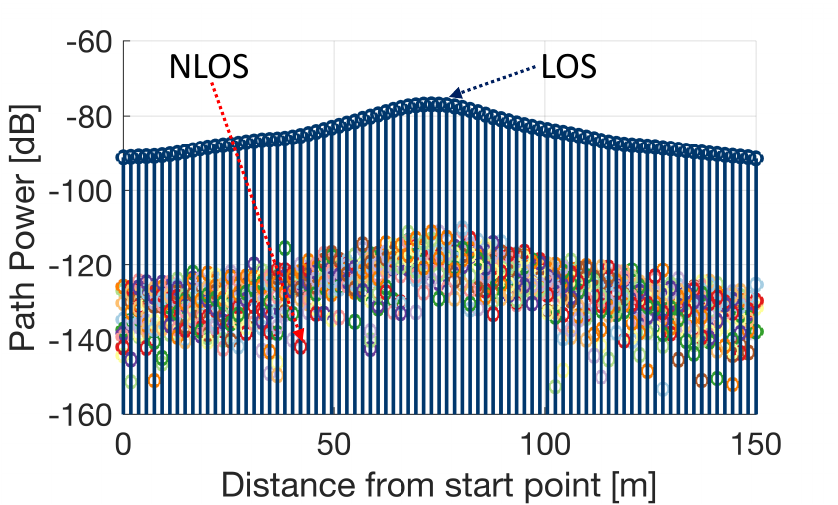}
    \caption{\centering  Path Power VS. Distance}
\label{fig:total3}
\end{subfigure}
\caption{\centering Results averaged over 100 channel realizations for the link describing green vehicle's transmission to the red in Willow Grove Naval Air Base in Horsham Township,
Pennsylvania, USA.}
\label{fig:V1toV2_pathpowers_LOS}
\vspace{-0.45cm}
\end{figure*}

The implementation of \glspl{gbsm} is fairly simple and computationally inexpensive, and
can be done through code-based simulators \blue{(e.g., NYUSIM\cite{ju2019millimeter}, WINNER II/+ \cite{kyosti20074} and 3GPP-3D\cite{ademaj20163gpp}), and  Standardized Channel Models (e.g., the Spatial Channel Model (SCM), IMT-Advanced, METIS, mmMAGIC and the 5G-ALLSTAR \cite{pang2022investigation}).
\gls{quadriga} extends the WINNER and 3GPP-3D models\cite{mondal20153d}, \blue{with the advantage that it achieves a high level of precision~\cite{jaeckel2019f}.}}
It takes as input configuration files to describe a propagation scenario, and each file contains a parameter table with statistical information about the wireless channels (i.e., delay spread, Rician K-factor, shadow fading, \gls{xpr}, angular spread values and their properties such as correlation and distribution). \blue{The aforementioned files were created through the process of channel sounding either with measurements performed in urban cities~\cite{jaeckel2014quadriga, ranjkesh2015optimized}, or by the incorporation of statistical tables such those included in \cite{etsi19deliv}.}
The modeling procedure can be divided into two parts: (i) a stochastic part that generates the \glspl{lsp}, e.g., path loss, shadowing and fading, and (ii) a geometry-based part that calculates the updated \glspl{ssp}, e.g., delay spreads~\cite{he2020investigation}. 
\blue{In contrast to the classical ray-tracing approach, \gls{quadriga} does not use an exact geometric representation of the propagation environment. 
The distribution of the scattering clusters (e.g., tree foliage and buildings) is random.
This adds simplicity into the implementation and decreases the computational cost without accuracy loss.}
%
\blue{\gls{quadriga} simulates complicated \gls{nextg} heterogeneous environments \cite{etsi5g}},
by considering effects such as dual-mobility, 3D propagation and continuous time evolution. The generated parameters are spatially correlated,
while the wide-sense stationary (WSS) properties are kept both within a segment and during mobility. Finally, the simulation steps can be summarized as follows\cite{arunachalaperumal2018enhanced, jaeckel2019f, jaeckel2014quadriga}.

\begin{enumerate}
\item \textbf{\textit{Definition of the \gls{ue} Input Variables.}}
The topology layout is defined  \blue{(i.e., \glspl{tx}/\glspl{rx})}, location (i.e., coordinates) and antenna characteristics (i.e., carrier frequency and type). \blue{Second, the nodes' trajectories and their speed profile are determined. Last, a propagation scenario is set \blue{(\gls{los}/\gls{nlos}, or a combination)}}. Examples of supported scenarios \cite{jaeckel2019f,jaeckel2014quadriga,giordani2019path} are: 

\begin{enumerate}
\item \textit{\textbf{Berlin/Dresden UMa}} 
\begin{itemize}
\item Terrestrial \textit{macrocell} parameters extracted from measurements in  Germany.
\end{itemize}

\item \textit{\textbf{WINNER model}}
\begin{itemize}
\item \textit{UMa C1 \& C2}: For \textit{macrocell} \glspl{bs} in urban \& sub-urban areas, respectively.
\item \textit{UMi B1}: For \textit{microcell} \glspl{bs} in urban areas.
\end{itemize}
\item \textit{\textbf{3GPP TR 37.885 Highway V2X }}
\begin{itemize}
    \item \textit{NLOS} due to dynamic blockages (e.g., vehicles).
\end{itemize}
\item \textit{\textbf{3GPP TR 38.901 NR model - \textit{UMa \& UMi}}}
\begin{itemize}
\item For center frequencies between 0.5 and 100 GHz.
\end{itemize}
\end{enumerate}

\item \textbf{\textit{Calculation of \glspl{lsp}.}} Out of the aforementioned parameter tables, the \glspl{lsp} are calculated. At this step, sources of scattering have been created, according to the multi-bounce  approach, comprising a \gls{fbs} and a \gls{lbs}.

\item \textbf{\textit{Generation of the \gls{xpr}.}}
Antenna characteristics and polarization effects are taken into account.

\item \textbf{\textit{Calculation of the Channel Coefficients and Path Gain Generation.}}
The results of the previous steps alongside with \glspl{ssp} (e.g., angular characteristics of the paths) are combined to calculate the channel coefficients (i.e., the \glspl{mpc}). A last fine-tuning of the shadow fading and path loss parameters, alongside with the Rician factor are applied to produce the final path gains. 

\item \textbf{\textit{Repetition of steps $1-4$ for successive channel traces and final Propagation Channels.}}
\blue{In case of mobility the process is repeated.} All individual components produced by every step of the aforementioned procedure are combined to obtain the final channels. 
\end{enumerate} 

\IEEEpubidadjcol

\subsection{Experimentation through Simulation with \quadriga}\label{Section IIC}
\blue{To test the software's effectiveness, we generate a \gls{cv2x} setup~\cite{mecklenbrauker2011vehicular} in which two vehicles approach each other under the 3GPP 37.885 Highway \gls{los} transmission conditions (Section~\ref{Section IIB}).} Both have a steady speed of $66$~km/h for a distance of $150$~m. The vehicles are equipped with omni-directional antennas mounted on their roof at $1.5$~m and operate at the center frequency of $5.9$~GHz. As it can be seen from the topology in Fig. \ref{fig:Penn_topologies}, scattering objects such as $4$ \glspl{rsu} alongside with $40$ nodes on the highway's sides, placed $5$~m apart, representing other parked cars, are also present. \blue{The simulation's location corresponds to the Willow Grove Naval Air Base in Horsham Township, Pennsylvania, USA (\gls{gps} coordinates for latitude and longitude are $[40.202951, -75.149008]$), and the conversion from \gls{gps} to Cartesian coordinates has been done with MATLAB's \cite{MATLAB:2021a} \texttt{latlon2local} function.} Given the stochastic nature of the channels, and in order to obtain a good estimate, we observe the results every $0.1$~s and average them over $100$ channel realizations. In Fig.~\ref{fig:V1toV2_pathpowers_LOS}, we present the simulation's results, when the green transmits to the red vehicle, as depicted in Fig.~\ref{fig:Penn_topologies}. In Fig.~\ref{fig:v1_to_v2_100sims_0.1_updateRate} the continuous bell-shaped curve represents the averaged path gains, while the dotted represents the path gains for a random setup. The log-distance path loss \cite{mecklenbrauker2011vehicular} model leveraged in the channel generation, explains the increase in path gains as the path loss decreases, particularly for a minimal distance between the vehicles. In the final path gain values in Fig.~\ref{fig:v1_to_v2_100sims_0.1_updateRate}, the main contribution comes from the direct \gls{los} component, while the contribution of the \gls{nlos} \glspl{mpc} is weighted to be insignificant, based on the input configuration file.
In Fig.~\ref{fig:total}, we include a 3D plot of the generated path powers versus delay and travelled distance, while in Fig.~\ref{fig:total2} and in Fig.~\ref{fig:total3}, we observe Fig.~\ref{fig:total} from two different perspectives. 12 paths are generated, one for the direct path, and 11 for the multiple copies of the signal, for each position of the mobile \gls{tx}/\gls{rx} (i.e., Fig.~\ref{fig:total3}). In Fig.~\ref{fig:total2}, the path power of the \gls{los} path attains a lower delay value compared to its \gls{nlos} counterparts for a given distance. 
As the vehicles begin to approach each other, the delay values decrease, reaching their minimum around the midpoint of the track, where the two vehicles cross. 


\subsection{\glspl{gbsm} for Wireless Network Channel Emulators }\label{Section IID}

We install the aforementioned scenario in Colosseum, a publicly available testbed comprising $128$~\glspl{srn}. It consists of pairs of Dell PowerEdge R730 servers and NI \gls{usrp} X310 \glspl{sdr} that
enable large-scale experimentation in diverse network deployments via the \gls{mchem} component. \gls{mchem} leverages \gls{fpga}-based \gls{fir} filters to replicate wireless environment conditions such as path loss, fading, attenuation, mobility, and interference pre-modeled through ray-tracing software, analytical models, or real-world measurements. The Colosseum \gls{tgen}, built on top of the \gls{mgen}
\gls{tcp}/\gls{udp} traffic generator \cite{mgen}, emulates different traffic profiles (e.g., multimedia content), demand, and distributions (e.g., Poisson, periodic), by emulating \gls{ip} traffic flows between the \glspl{srn}.

\begin{figure}[ht]
    \centering
   \begin{subfigure}{0.22\textwidth}
    \includegraphics[width=\textwidth]{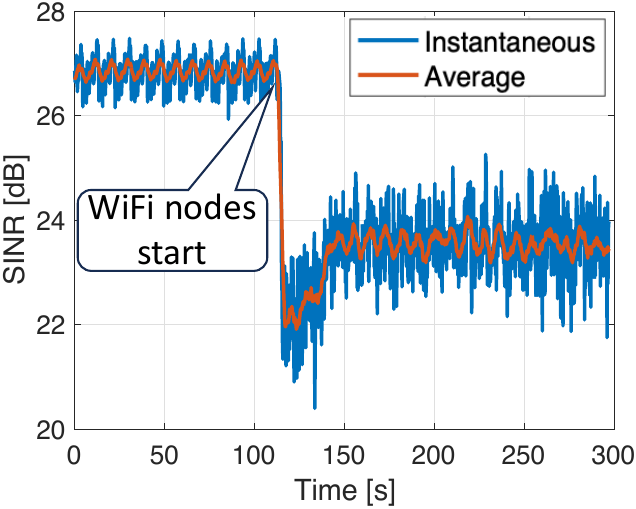}
    \caption{\centering \gls{ul} Throughput [Mbps]}
    \label{fig:throughput}
\end{subfigure}\hspace{0.02\textwidth}
\begin{subfigure}{0.22\textwidth}
    \includegraphics[width=\textwidth]{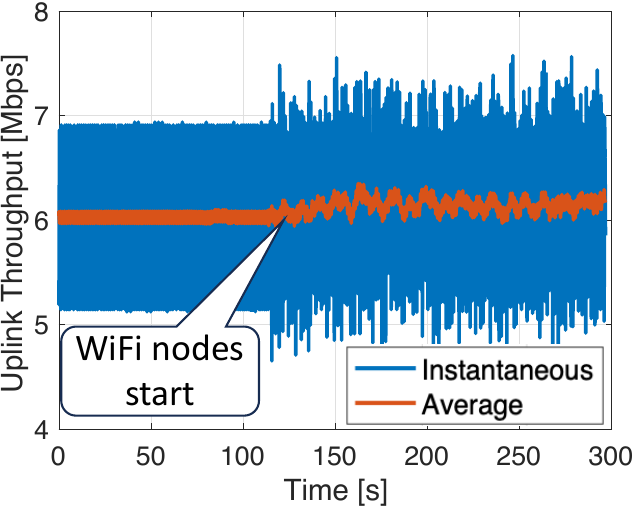}
    \caption{\centering \gls{ul} \gls{sinr}}
    \label{fig:snir}
\end{subfigure}\hspace{0.02\textwidth}
     \caption{\gls{ul} cellular metrics between \gls{bs}-node~1 and \gls{ue}-node~2.}
     \label{fig:43011cellularmetrics}
     \vspace{-10.75pt}
\end{figure}

We focus on nodes $1\rightarrow4$ as depicted in Fig.~\ref{fig:Penn_topologies}, which share a bandwidth of $10$~MHz. Nodes $1\rightarrow2$ are deployed through the srsRAN protocol stack~\cite{srsran}, while nodes $3\rightarrow4$ are deployed with the Wi-Fi Stack, which is based on the open-source GNU Radio implementation of the IEEE 802.11a/g/p standard\cite{bloessl2017performance}. Cellular node $1$ and Wi-Fi node $3$ are \glspl{bs}, while nodes $2$ and $4$ serve as the respective cellular and Wi-Fi \glspl{ue}. Finally, 
\gls{udp} traffic is generated with \texttt{iPerf}. 
In Figs.~\ref{fig:throughput} and~\ref{fig:snir}, we illustrate the \gls{ul} cellular \gls{sinr} and throughput  between the cellular nodes. 
Initially, an average \gls{sinr} of $27~\mathrm{dB}$ is observed, which drops to $22~\mathrm{dB}$, when the WiFi nodes start transmitting at $20$~dB (at approximately $t=120~\mathrm{s}$), before slightly recovering to $\sim24~\mathrm{dB}$. Similarly, the \gls{ul} throughput becomes unstable once WiFi transmissions commence. 


\blue{Conclusively, with this simple yet richly detailed use-case, we are able to verify the suitability of \gls{quadriga} and \glspl{gbsm} for generating \gls{rf} scenarios for
Colosseum, as well as the efficiency of the latter in generating and managing traffic among heterogeneous nodes.} Notably, although the aforementioned scenario comprises multiple nodes, the focus is directed towards the aforementioned to demonstrate the software's capabilities, and hence 
an extensive evaluation is omitted. Readers interested in  the creation of \gls{rf} scenarios for Colosseum can refer to~\cite{bonati2021colosseum,tehrani2021creating,villa2022cast}.

\section{Investigating Energy Efficiency with frequency-dependent \glspl{gbsm} and \glspl{ris} in \gls{cv2x}}\label{Section III}

\blue{Evidently, \quadriga is effective in providing accurate geometric-based stochastic channel modeling.} Additionally, considering that the generated channels are frequency-dependent, \gls{quadriga} poses as an ideal candidate for investigating energy efficient power allocation  across different spectrum bands and various use-cases. In this Section, we focus particularly  on a \gls{cv2x} use-case~\cite{khan2022vehicle} involving a \gls{uav} acting as a flying \gls{gnb} serving multiple \glspl{ue}, to investigate energy efficiency in the Sub-6 GHz and \gls{mmwave} bands. Without loss of generality, we assume that the \gls{bs} simulated with \gls{quadriga} is a static \gls{uav} hovering at a certain height, while the generated channels are cascaded to simulate the inclusion of a \gls{ris} in the topology. We believe that the strong \gls{los} links provided by the \gls{uav} and the cascaded channels offered by \gls{ris} technology fit into the description of a \gls{cv2x} scenario for \gls{nextg} wireless environments and smart cities as described in~\cite{diamanti2021energy}.
\subsection{System Model: A \gls{ris}-assisted \gls{cv2x} setup with \quadriga}\label{Section IIIA}

\begin{figure}[ht]
\centering
    \includegraphics[width=0.3\textwidth]{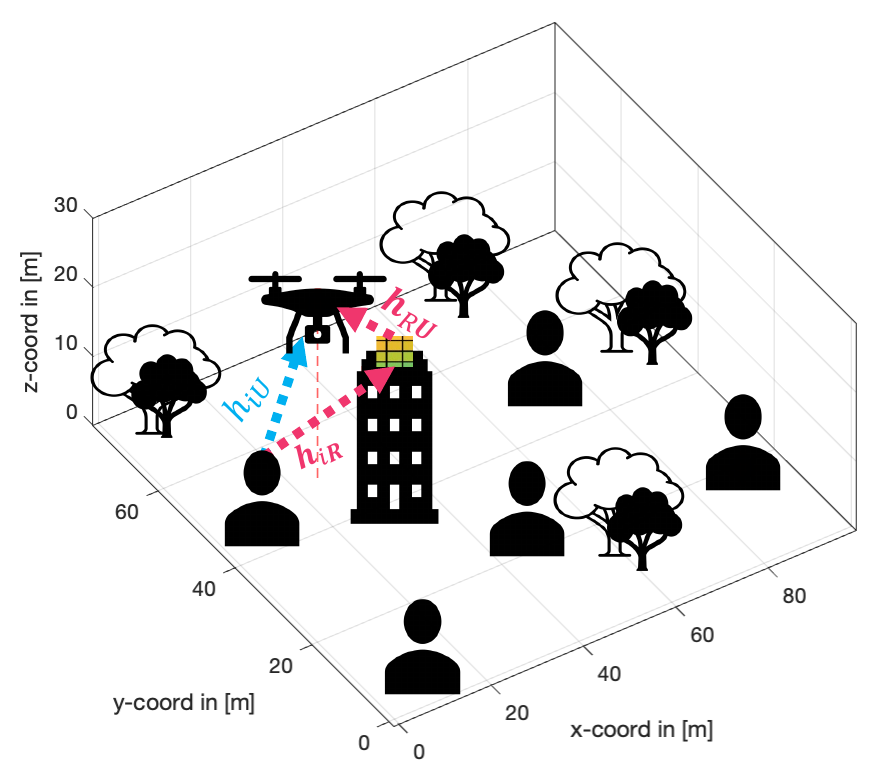}
    \caption{\gls{ris}-assisted and \gls{uav}-enabled communications system.}
    \label{fig:ris-uav}
    \vspace{-0.25cm}
\end{figure}
 As illustrated in Fig. \ref{fig:ris-uav}, a \gls{ris}-assisted and \gls{uav}-enabled wireless communications system is evaluated on the \gls{ul} direction. Specifically, the topology comprises of a \gls{uav}, a building facade hosting a \gls{ris} and a set of mobile \glspl{ue}, denoted as $I=\{1,\dots,i,\dots, |I|\}$, which communicate directly with the \gls{uav}. The RIS consists of $|M|$ reflecting elements, the set of which is defined as $M=\{1, \dots , m, \dots , |M|\}$, and each RIS element's phase shift is $\theta_m \in [0,2\pi]$, $\forall~m~\in~M$. Finally, the corresponding diagonal phase-shift matrix is defined as $\boldsymbol{\Theta}=diag( e^{j\theta_1},\dots,e^{j\theta_m},\dots,e^{j\theta_{|M|}} )$.
The incoming signals at the \gls{uav} are combined, as in \cite{diamanti2021energy}, so as each \gls{ue}'s $i$ signal is the outcome of the coherent addition of the direct and the reflected signal. The channel gain of the direct link between a user $i$ and the \gls{uav} is described as $h_{iU}$, the channel gain between a user $i$ and the \gls{ris} is defined as $\mathbf{h}_{iR}$, where $\mathbf{h}_{iR}$ can also be written as $\mathbf{h}_{iR}=[|h_{iR,1}| e^{j \omega_{1}}, \dots,|h_{iR,m}| e^{j \omega_{m}}, \dots,|h_{iR,|M|}| e^{j\omega_{|M|}}]^T$, and finally, the channel gain of the reflected link from the \gls{ris} to the \gls{uav} is denoted as $\mathbf{h}_{RU}$. Subsequently, the cascaded channel is given as $\mathbf{{h}}_{RU}^H\boldsymbol{\Theta}\mathbf{{h}}_{iR}$. Last, the steering vector for the \gls{ue} $i$ to the \gls{ris} \gls{los} link is defined as $\mathbf{h}_{iR}^{\gls{los}}=[1, e^{-j \frac{2 \pi}{\lambda} d \phi_{iR}}, \dots, e^{-j \frac{2 \pi}{\lambda}(|M|-1) d \phi_{i R}}]^{T}$, while the steering vector for the \gls{ris} to the \gls{uav} \gls{los} link is denoted as $\mathbf{h}_{RU}^{\gls{los}}=[1, e^{-j \frac{2 \pi}{\lambda} d \phi_{RU}}, \dots, e^{-j \frac{2 \pi}{\lambda}(|M|-1) d \phi_{RU}}]^T$. Moving forward, we keep using the same notation as in \cite{diamanti2021energy}, which is also reported in Table \ref{table:not-chmod}. The first \gls{ris} element, i.e., $m=1$, with its respective coordinates, i.e., $(x_R, y_R, z_R)$ [m], is used as a reference point in the following calculations, while for the \gls{uav}, its coordinates are given as follows $(x_U, y_U, z_U)$ [m]. It is noted that in \cite{diamanti2021energy}, the featured channel model is frequency-independent 
In our study, we aim at observing the impact of the frequency of operation in terms of energy efficiency on realistic \gls{ris}-assisted wireless channels on two different 5G bands, namely the Sub-6 GHz and the \gls{mmwave} band. \blue{It is highlighted that \glspl{ris} are expected to work across multiple frequency ranges, including both the \gls{fr1} and \gls{fr2} bands~\cite{ETSI_RIS001}, to ensure interoperability across diverse deployment scenarios and applications. Given that \gls{quadriga} can reliably support the simulation of channels up to $100$~GHz, as mentioned in Section~\ref{Section II}, the Sub-6~GHz and \gls{mmwave} bands are chosen.} Therefore, we experiment with two different values for the center frequency, i.e., $5.9$ GHz and $28$ GHz, respectively. Finally, both the \glspl{ue} and the \gls{uav} are equipped with single-antenna omni-directional \glspl{tx}/\glspl{rx}. It is noted that although omni-directional antennas are commonly used in the Sub-6 GHz band~\cite{paul2022omni,azim2021multi,rani2022development,sharawi2017two,de2021radio,zaidi2020wide}, and they can 
be deployed on \gls{mmwave} communication systems~\cite{mao2018planar,ranvier2008low,fan2018wideband,maccartney2015millimeter,hasan2019dual}, this is not typical~\cite{sun2015synthesizing}. Instead, high-directional antennas are most commonly used in such scenarios~\cite{abirami2017review,niu2015survey,pan2011dual}, \blue{to combat severe propagation loss and achieve high antenna gains. Within the scope of this work, we focus on the use of omni-directional antennas for the \gls{mmwave} to provide insights into the performance enhancement introduced by \gls{ris} in cases where omni-directional antennas are convenient for deployment, such as in a \gls{cv2x} environment~\cite{mao2020compact}. While it is indeed true that in a dynamic environment such as the aforementioned, a directional antenna can improve \gls{rssi}, this increased directionality comes at the cost of precise alignment with the \gls{ris} and \glspl{ue}. Therefore, continuous beam tracking and adaptation are needed, which introduce additional overhead and latency that may outweigh the benefits of directive gains. Additionally, highly accurate and precise channel modeling with directional antennas in urban \gls{cv2x} environments, where the effects of multipath propagation are strong, can prove rather challenging, mainly due to the narrow \gls{mmwave} beams, which lead to position deviations, inaccuracies, and alignment errors, resulting in beam misalignment, signal loss, and significant performance drops.}

\begin{table}[htb]
\centering
\Huge
\setlength\abovecaptionskip{-.01cm}
\caption{Channel Modeling Notation}
\begin{adjustbox}{width=1\linewidth}
\begin{tabular}{@{}l@{\hspace{1.65 mm}}l@{\hspace{1.65 mm}}l@{\hspace{1.65 mm}}l@{}}
\toprule
\textbf{Symbol} & \textbf{Description} \\
\midrule
$|I|$ & $\text{Number of mobile \glspl{ue}}$ \\
$I$ & $\text{Set of mobile \glspl{ue}}$\\
$|M|$ & $\text{Number of reflecting elements on the \gls{ris}}$ \\
$M$ &\text{Set of \gls{ris}'s reflecting elements}\\
$\theta_m$,  $\forall m \in M$ & $\text{Reflecting element's phase shift}$ \\
$\mathbf{{h}}_{RU}$ & $\text{Channel Gain for the \gls{ris} to \gls{uav} reflected link}$ \\
$\boldsymbol{\Theta}$ & $\text{Diagonal phase-shift matrix}$ \\
$\mathbf{{h}}_{iR}$ & $\text{Channel Gain for the \gls{ue} $i$ to \gls{ris} link}$ \\
$\mathbf{\mathbf{{h}}}_{RU}^H\boldsymbol{\Theta}\mathbf{{h}}_{iR}$ & $\text{Cascaded Channel Gain}$ \\
$h_{iU}$ & $\text{Channel Gain for the \gls{ue} $i$ to \gls{uav} direct link}$ \\
$d_{RU} [m]$ & $\text{Euclidean distance between the \gls{ris} and the \gls{uav}}$\\
$d_{iR} [m]$ & $\text{Distance between the \gls{ue} $i$ and the \gls{ris}}$\\
$d_{iU} [m]$ & $\text{Distance between the \gls{ue} $i$ and the \gls{uav}}$\\
$\phi_{RU}$ & $\text{Cosine of the signal's \gls{aod} for the \gls{ris} to the \gls{uav} link}$\\
$\phi_{iR}$ & $\text{Cosine of the signal's \gls{aoa} for the \gls{ue} $i$ to \gls{ris} link}$\\
$\lambda [m]$ & $\text{Carrier wavelength}$\\
$d [m]$ & $\text{Antenna separation}$\\
$\mathbf{h}_{RU}^{\gls{los}}$ & \text{Steering vector for the \gls{ris} to the \gls{uav} \gls{los} link.}\\
$\mathbf{h}_{iR}^{\gls{los}}$ & $\text{Steering vector for the \gls{ue} $i$ to the \gls{ris} \gls{los} link.}$\\
$\textit{h}^\prime_{xy}$ & $\text{\gls{mpc} Gain, where \(x \in \{i, R\}, \ y \in \{R, U\}\)}$\\
$|\textit{h}^\prime_{xy}|, \varphi^\prime_{xy} $ & $\text{Magnitude and phase of the \gls{mpc}}$\\
$|N|$ &  $\text{Number of generated \glspl{mpc}}$\\
\bottomrule
\end{tabular}
\end{adjustbox}
\label{table:not-chmod}
\vspace{-0.22cm}
\end{table}

Last, the channels in \cite{diamanti2021energy}, \blue{from which our use-case is inspired}, rely on the Rice and Rayleigh mathematical models to emulate the wireless links. However, their simple yet effective implementation may not accurately represent real-world wireless environments~\cite{dang2021geometry}. For these reasons, we proceed by creating the channels with \gls{quadriga}. The channel generation process is described in Section~\ref{Section IIB}, where for each generated link a number of \glspl{mpc} is produced based on the Channel Generator's input scenario, in our case the 3GPP-compliant scenario, 3GPP\textunderscore38.901\textunderscore UMa. Finally, we assume that the links between the \glspl{ue} and the \gls{uav} are \gls{nlos}, while those remaining, i.e., the channels between the \glspl{ue} and the \gls{ris}, and the \gls{ris} to \gls{uav}, are \gls{los}. \blue{The transmission conditions of the aforementioned links were selected to ensure that the cascaded channel created by the \gls{ris} is strong enough to redirect the \glspl{ue}' signals to the \gls{uav}, for both the Sub-6 GHz and, especially, the \gls{mmwave} band, which suffers from transmission losses.} The wireless channels are observed every $0.1$~s and for their creation, the steps listed below are followed:

\begin{itemize}
 
\item Generate scenario-based \glspl{mpc} (i.e., $h_{xy}^\prime$) for every link.

\item Perform coherent summation of the \glspl{mpc} \cite{tehrani2021creating} to obtain the link's channel gain, as reported in~\eqref{eq:mpc-def}
\begin{equation}\label{eq:mpc-def}
    \mathrm{h}^{\prime\prime}_{xy}=\sum_{j=1}^{|N|}\left|\textit{h}_{xy}^\prime\right| \cdot e^{j \varphi_{xy}^\prime}.
\end{equation}
\noindent

\item Average the channel gains (i.e., $\mathrm{h^{\prime\prime}_{xy}}$) over a $100$ different channel realizations to obtain the link's \textit{\textbf{final}} channel gain (e.g., $h_{iU}$).

\item Multiply the generated channels with the corresponding steering vectors, i.e., $\mathbf{h}_{RU}^{\gls{los}}$ and $\mathbf{h}_{iR}^{\gls{los}}$, to get the respective final channel gains (i.e., $\mathbf{h}_{RU}$, $\mathbf{h}_{iR}$).
\end{itemize}

The overall channel power gain for the \gls{ue} $i$ to \gls{uav} link is denoted as $G_{i}=|h_{iU}+\mathbf{h}_{RU}^H \boldsymbol{\Theta} \mathbf{h}_{iR}|^2$, and the \glspl{ue} are able to communicate concurrently with the \gls{uav} thanks to the \gls{noma} technique. The received channel gains (i.e., $G_{i}$) at the \gls{uav} are sorted 
and decoding starts from the \gls{ue} with the highest channel gain. Finally, the \gls{uav}'s \gls{rx} decodes the received superposed signal with the help of the \gls{sic} technique, which is implemented on its end. The interference sensed by each \gls{ue} is defined as follows in~\eqref{eq:eq1}:

\begin{equation} \label{eq:eq1}
    I_{i}=\sum_{j<i} G_{j} P_{j}+I_{0},
\end{equation}

\noindent where $P_{j}$ [W] is the \gls{ul} \gls{tx} power of the \gls{ue} $j$ and $I_0$ is the power of the zero-mean \gls{awgn}. Each \gls{ue}'s achieved \gls{sinr} is given as follows in~\eqref{eq:eq2}:
\begin{equation} \label{eq:eq2}
\gamma_{i}=\frac{P_{i} G_{i}}{I_{i}}.
\end{equation}

\subsection{Hierarchical Game-Theoretic Power Control}\label{Section IIIB}

In this Section, we aim to study how the frequency-dependent \gls{gbsm}-based system model presented in Section \ref{Section IIIA} will impact the energy efficiency in both the Sub-6 GHz and \gls{mmwave} bands. In brief, both the \glspl{ue} and the \gls{uav} indulge in a single-leader multiple-followers Stackelberg Game, where the target is to jointly maximize the \gls{uav}'s overall received signal strength and each \gls{ue}'s energy efficiency in a distributed manner. In detail, the \gls{uav} acting as a leader computes the \gls{ris} elements' effective phase shifts (and intelligently steers the signals reflected by the \gls{ris}) with a goal to enhance its received signal quality. In turn, the \glspl{ue}, acting as followers, observe the leader's action and through the formulation of a non-cooperative game aim at maximizing their energy efficiency, by determining their optimal \gls{ul} \gls{tx} power. Each \gls{ue}'s utility function is defined in~\eqref{eq:eq3}:
\begin{equation} \label{eq:eq3}
    U_{i}\left(P_{i}, \mathbf{P}_{-i}\right)=\frac{W \cdot\left(1-e^{-\alpha \gamma_{i}}\right)^{M}}{P_{i}},
\end{equation}

\noindent
where $\mathbf{P}_{-i}$ is the \gls{tx} power vector of all the \glspl{ue} except for \gls{ue} $i$, $W$ [Hz] is the system's bandwidth, and $\alpha,M \in \mathbb{R}^{+}$ are parameters which control the utility function's slope. The adopted utility function for energy efficiency represents the trade-off between the achieved \gls{qos} satisfaction, i.e., \gls{sinr}, and the invested \gls{ul} \gls{tx} power \cite{diamanti2021energy, tsiropoulou2015combined}. \blue{ In detail, the numerator of the utility function, i.e., \( (1 - e^{-\alpha \gamma_i})^M \) is a sigmoidal curve which expresses the \gls{ue} satisfaction, with respect to its \gls{qos} demands, while the denominator expresses the associated cost with reference to the transmission power levels. Because of the nature of the sigmoidal curve, small increases in \(\gamma_i\) (by increasing the \gls{tx} power) will initially improve the utility. However, increasing \(\gamma_i\) further requires significant power while providing diminishing returns in utility. As a result, the optimal strategy is to transmit at lower power levels in order to maintain higher energy efficiency.}

Finally, the joint optimization problem is formulated as follows, with the respective problem solved by the \gls{uav} written as:
\begin{subequations}
\begin{equation}
\max_{\boldsymbol{\theta}} \sum_{\forall i \in I} P_{i} |h_{iU}+\mathbf{h}_{RU}^H \boldsymbol{\Theta} \mathbf{h}_{iR}|^2
\label{eq4a}
\end{equation}
\begin{equation}
\text{s.t.\quad}0\leq\theta_m \leq 2\pi, \forall m\in M.
\label{eq4b}
\end{equation}
\end{subequations}


With regards to the optimization problem solved by the \glspl{ue}, the latter ones participate in a non-cooperative game defined as $G=[I,\{A_i\}_{\forall i \in I}, \{U_i\}_{\forall i \in I}]$. In detail, $I$ is the set of \glspl{ue} (e.g., players), $A_{i}=[P_i^{min}, P_{i}^{max}]$ is each \gls{ue}'s strategy set, and $U_{i}$ is their utility function. $P_{i}^{min}$ [W] and $P_{i}^{max}$ [W] correspond to each \gls{ue}'s minimum and maximum \gls{ul} \gls{tx} power levels respectively, where $P_i^{min}$, is determined by the \gls{sic} prerequisite at the \gls{uav}'s \gls{rx} as follows:
\begin{equation} \label{eq5}
G_i P_i - I_i \geq P_{tol}, \forall i \in \{2,\dots,i,\dots, |I|\},
\end{equation}

\noindent
such that each \gls{ue}'s signal is decoded successfully. According to~\eqref{eq5}, each \gls{ue}'s signal strength, i.e., $G_i P_i$, should be greater than or equal to the \gls{ue}'s $i$ sensed interference $I_i$,  which is broadcasted by the \gls{uav} to all the \glspl{ue}, plus a minimum acceptable power level $P_{tol} \in \mathbb{R}^+$, which represents
the \gls{sic} \gls{rx}'s tolerance/sensitivity. Finally, the maximum \gls{ul} \gls{tx} power level $P_i^{max}$ is imposed by the maximum power budget of the \gls{ue}'s device. The optimization problem is finally determined as follows:


\begin{subequations}\label{eq:7total}
\begin{equation}
\max_{P_{i} \in A_{i}} U_{i}\left(P_{i}, \mathbf{P}_{-i}\right)=\frac{W \cdot\left(1-e^{-\alpha \gamma_{i}}\right)^{M}}{P_{i}}, \forall i \in I
\label{eq7a}
\end{equation}
\begin{equation}
\text{s.t.\quad}G_i P_i - I_i \geq P_{tol}, \forall i \in \{2,\dots,i,\dots, |I|\}
\label{eq7b}
\end{equation}
\begin{equation}
P_i \leq P_i^{max}, \forall i \in I.
\label{eq7c}
\end{equation}
\end{subequations}

The non-cooperative game's, $G$, outcome is a Nash equilibrium, which determines the \glspl{ue}' optimal \gls{tx} power vector, i.e., $\mathbf{P}^{*}=[P_{1}^{*}, \dots, P_{i}^{*}, \dots, P_{|I|}^{*}]$.

 The solution of the optimization problem taking place on the \gls{uav}'s side and defined in~(\ref{eq4a})-(\ref{eq4b}) is the \gls{ris} elements' effective phase shifts' vector $\boldsymbol{\theta^{*}}$. \blue{It is noted that to maximize the overall received signal strength, the \gls{uav} only needs to calculate the effective \gls{ris} elements’ phase shifts  (i.e., \( \boldsymbol{\theta^*} \)) that maximize the overall channel power gain of the \glspl{ue}.} \blue{The aforementioned optimization problem is non-convex, making it challenging to find a globally optimal solution.} 
\blue{As described in \cite{diamanti2021energy,li2020reconfigurable}, when there is a single \gls{ue} in the system, the optimal phase shifts of the \gls{ris} elements are given by the following closed-form solution:}
\begin{equation} \label{eq8}
\theta_{m}^{*}=\angle{h_{iU}}+\omega_{m}+\frac{2 \pi}{\lambda} d(m-1) \phi_{RU}, \forall m \in \ M.
\end{equation} 
\blue{To derive the closed-form phase-shift solution in \eqref{eq8} and maximize the received signal strength at the \gls{uav}, we aim to achieve phase alignment of the received signals from different transmission paths. Notably, this can be achieved through coherent signal construction of the incoming signals at the \gls{uav}. Indeed, the \gls{ue}'s channel power gain is maximized when the direct signal and the signal reflected by the \gls{ris} are perfectly aligned and coherently combined at the \gls{uav}'s \gls{rx}~\cite{li2020reconfigurable}. This generally
holds true when the phase shifts of the direct and the cascaded
signals are equal, as follows $\angle h_{iU} = \angle \left( \mathbf{h}_{RU}^{H} \mathbf{\Theta} \mathbf{h}_{iR} \right)$. Therefore, there's an optimal \(1 \times |M|\) phase-shift vector \(\boldsymbol{\theta^*} = \angle \mathbf{v}\) for the single-\gls{ue} case. Similarly, in the multi-\gls{ue} case, there exists a distinct reflection-coefficient vector \( \mathbf{v}_i = [v_{i,1}, \ldots, v_{i,|M|}] \in \mathbb{C}^{|M| \times 1} \) for each \gls{ue} \( i \) that maximizes its channel power gain.} \blue{In the multi-\gls{ue} case, the solution to the leader's maximization problem is a linear combination of the phase shifts of the \gls{ris} elements that enhances each \gls{ue}'s signal strength, with the reflection-coefficient vectors \( \mathbf{v}_i \) being distinct for each \gls{ue}.}

On the other hand, the optimization problem solved on the \glspl{ue}' side will result in the determination of their optimal \gls{ul} \gls{tx} power vector $\mathbf{P^{*}}$, as mentioned, since given the \gls{ris} elements' effective phase shifts, each \gls{ue} determines its optimal \gls{ul} \gls{tx} power $P_{i}^{*}$ via non-cooperatively interacting with the rest of the \glspl{ue} in the action space. The Stackelberg equilibrium of this hierarchical game-theoretic approach can be given as $(\boldsymbol{\theta^{*}}, \mathbf{P^{*}})$. 

\newtheorem{Definition}{\textbf{Definition}}
\begin{Definition}\label{Definition 1}
\blue{\textbf{\textit{(Nash Equilibrium)}} The \gls{ul} \gls{tx} power vector $\mathbf{P^{*}}=[P_{1}^{*}, \dots, P_{i}^{*}, \dots, P_{I}^{*}]$ is a Nash equilibrium of the non-cooperative game $G=[I,\{A_i\}_{\forall i \in I}, \{U_i\}_{\forall i \in I}]$ if for every \gls{ue} $i\in I$, it holds that $U_i(P_i^*,\mathbf{P_{-i}})\geq U_i(P_i,\mathbf{P_{-i}}), \forall P_i\in A_i$.}
\end{Definition}

\blue{At the Nash equilibrium, the \gls{ue} has no incentive to improve its achieved utility by altering its transmission power strategy, given the strategies of the rest of the players (i.e., the other \glspl{ue}).}

\newtheorem{Theorem}{\textbf{Theorem}}
\begin{Theorem}\label{Theorem 1}
\blue{\textbf{\textit{(Existence \& Uniqueness of a Nash Equilibrium)}} 
For the non-cooperative game given as $G=[I,\{A_i\}_{\forall i \in I}, \{U_i\}_{\forall i \in I}]$, there exists a unique Nash equilibrium point, which is defined as follows:}
\begin{equation} \label{eqnash}
    \blue{P_{i}^{*}= \max\{P_{i}^{min},
    \min\{\frac{\gamma_{i}^{*} \cdot I_{i}}{W G_{i}}, P_{i}^{max}\}\}, \forall i \in I,} 
\end{equation}
\noindent
\blue{where $\gamma_{i}^{*}$ represents the unique positive solution to the equation $\frac{\partial f(\gamma_i)}{\partial \gamma_{i}} \gamma_{i}-f(\gamma_{i})=0, f(\gamma_{i})=(1-e^{-a \gamma_i})^{M}$.}
\end{Theorem}

\begin{figure}[t!]
  \centering
  \includegraphics[width=2.65in, height=2.6in]{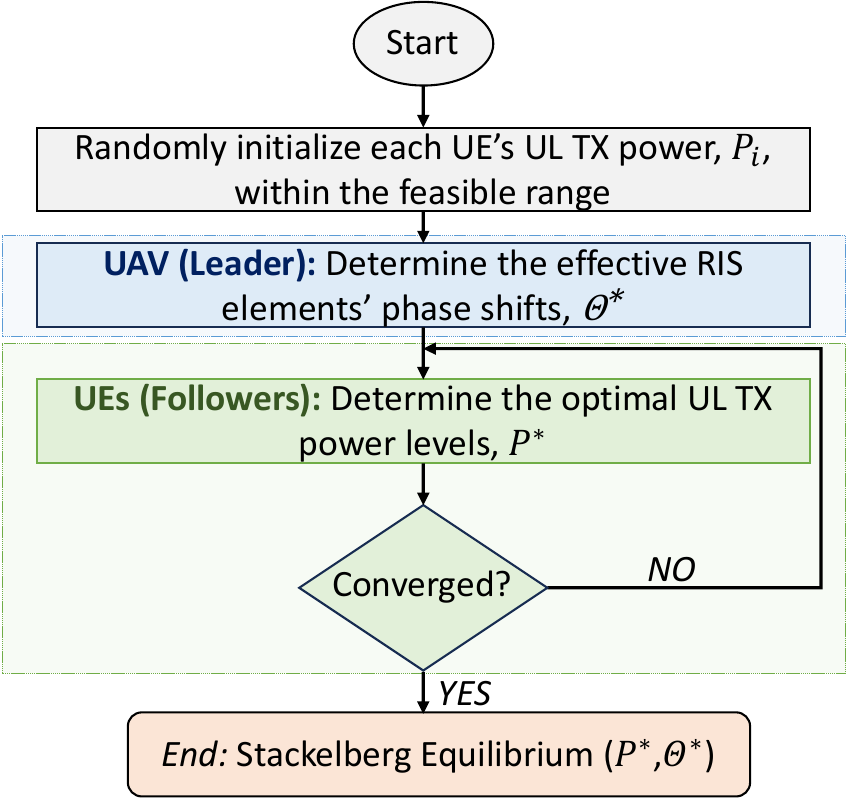} 
  \setlength\abovecaptionskip{-.002cm}
  \caption{\blue{A Hierarchical Stackelberg Game-Theoretic Approach to Energy-Efficient Power Control.}}
  \label{fig:game-flow}
  \vspace{-0.25cm}
\end{figure}

\blue{
The Theorem's proof relies on the quasi-concavity property of the utility function $U_{i}(P_{i}, \mathbf{P}_{-i})$ with respect to the \gls{ul} \gls{tx} power $P_i$. The quasi-concavity property is proven and further explained in \cite{tsiropoulou2015combined}, which also provides a reference guide with detailed steps of the proof. Therefore, given the \gls{ris} elements' effective phase shifts $\theta^{*}_{m}, \forall m \in M$, and the optimal transmission power vector $P_{i}^{*}, \forall i \in I$, the Stackelberg equilibrium is iteratively determined as presented in Fig.~\ref{fig:game-flow}. Notably, the convergence of the \glspl{ue}’ strategies to the Nash equilibrium point is achieved through the implementation of a Best Response Dynamics algorithm~\cite{tsiropoulou2015combined}, as further detailed in Algorithm~\ref{alg:stackelberg_optimization}. Note that the superscript ($j$) indicates the iterations required for the non-cooperative game played among the \glspl{ue}. Additionally, it is highlighted that the maximum \gls{tx} power levels for the \glspl{ue} (as defined in Step $3$ of Algorithm 1) are determined by 3GPP standards and are set to $23$~dBm\cite{etsi2022,etsi2022b,etsi2017}. The corresponding minimum value is set to $-20$~dBm.}

\begin{algorithm}
\caption{\blue{Stackelberg Game-Theoretic Optimization}}
\label{alg:stackelberg_optimization}
\begin{algorithmic}[1]
    \State Initialize the network topology, including the locations of \glspl{ue}, \gls{ris}, and the \gls{uav}.
    \State Generate the channels using \gls{quadriga}, as described in Section~\ref{Section IIIA}.
    \State Initialize randomly $P_i \in [10^{-5}, P_i^{\text{max}}]$.
    \State  Determine RIS elements’ phase-shift adaptation by solving \eqref{eq4a}-\eqref{eq4b} and calculate $G_i, \forall i \in I$.
    \State Sort users according to $G_i$, so decoding starts
\parbox[t]{.85\linewidth}{from the \gls{ue} with the highest channel gain.}
    \State Set $j = 0$.
    \Repeat
        \State Set $j = j + 1$.
        \For{$i \in I$}
            \State \parbox[t]{.85\linewidth}{Determine the optimal \gls{ul} \gls{tx} power $P_i^*(j)$ by  solving \eqref{eq7a}-\eqref{eq7c}.}
        \EndFor
    \Until{$|P_i^*(j) - P_i^*(j-1)| \leq \epsilon, \forall i \in I$, where $\epsilon \approx 10^{-4}$.}
\end{algorithmic}
\end{algorithm}

We consider the following setup, where a \gls{uav} hovers at the point $(x_U=25,y_U=50,z_U=25)$ [m] of the three-dimensional space serving a \gls{noma} cluster of $|I|=5$ users. Their distance from the \gls{ris} is given as $d_{i,R}=[20, 27, 37, 58, 66]$~[m] and all are randomly distributed around the \gls{ris}'s reference point, which is given as $(x_R=30,y_R=40,z_R=20)$ [m]. The system's bandwidth is equal to $W=10$~MHz, and the \gls{awgn} power is $I_0=-140$~dBm for the Sub-6 GHz and $I_0=-160$d~Bm for the \gls{mmwave} band, respectively. It is noted that \gls{awgn} noise is already added in the final path gain values through \gls{quadriga}~\cite{jaeckel2019f}. \blue{Specifically, in \gls{quadriga}, \gls{awgn} noise is used to generate spatially correlated \glspl{lsp}, which statistically characterize the propagation environment. These \glspl{lsp} indirectly shape the multipath environment and influence path gain.
Thus, the \gls{awgn} values (i.e., $I_0$), which are included in the calculation of the \gls{ue}'s interference, represent reference noise levels.}
Indeed, these two distinct values are a correction applied to the the \glspl{ue}' sensed interference, i.e., $I_{i}$, defined in~\eqref{eq:eq1}, so as the optimization framework can be evaluated for the respective spectrum bands under the same configuration of the utility's slope control parameters, defined as $a=0.3$ and $M=3$. Finally, the frequencies are centered to $5.9$ and $28$ GHz for the respective bands, the mobile \glspl{ue}' maximum power budget is set to $P_i^{max}=23$~dBm, and the sensitivity of the \gls{uav}'s \gls{rx} is set to be equal to $P_{tol}=-150$~dBm. 

\blue{It is finally noted that in the following, the performance evaluation of the \gls{ris}-assisted and \gls{uav}-enabled communication system depicted in Fig.~\ref{fig:ris-uav} will take place under $10$, $100$, and $1000$ \gls{ris} elements for \textit{both} the Sub-6 GHz and the \gls{mmwave} band. Although it generally holds true that for a given \gls{ris} dimension, a \gls{mmwave} \gls{ris} will host a higher number of \gls{ris} elements than a Sub-6 GHz \gls{ris}, given the same element spacing (e.g., $\lambda/2$), increasing the number of elements onboarded on the \gls{ris} also increases the complexity of its handling~\cite{rossanese2022designing}. Consequently, a trade-off emerges between hosting thousands of \gls{ris} elements and maintaining ease of operation, particularly when experimenting in the \gls{mmwave} band, where location inaccuracies are of utmost importance. For the aforementioned reasons, investigating the number of the elements onboarded on the \gls{ris} is deemed necessary.}

\begin{figure}[ht]
    \centering
    \begin{subfigure}[b]{0.4899\columnwidth}
        \centering
        \includegraphics[width=4.5cm, height=3cm]{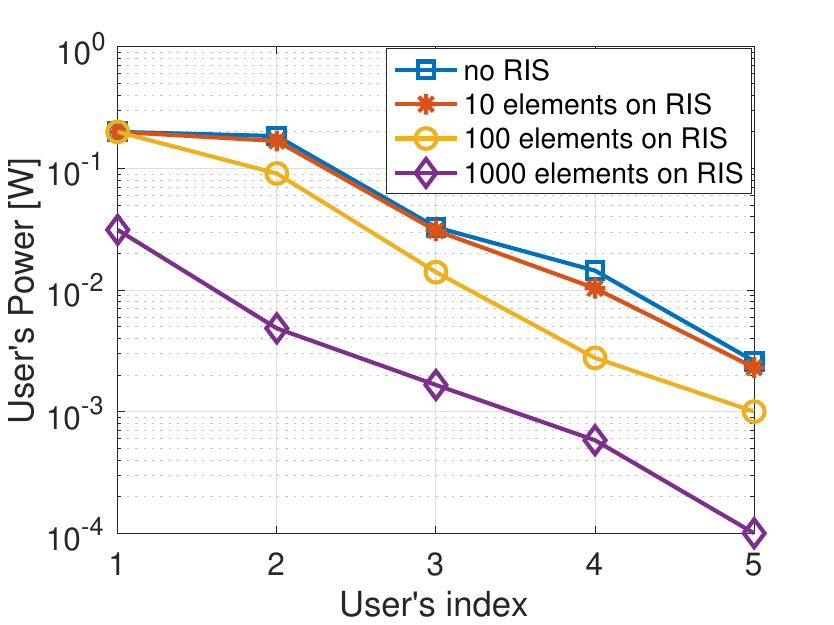} 
        \caption{Sub-6 GHz}
        \label{fig:sub6ghz_p}
    \end{subfigure}
    \hspace{0.000001\textwidth}
    \begin{subfigure}[b]{0.4899\columnwidth}
        \centering
        \includegraphics[width=4.5cm, height=3cm]{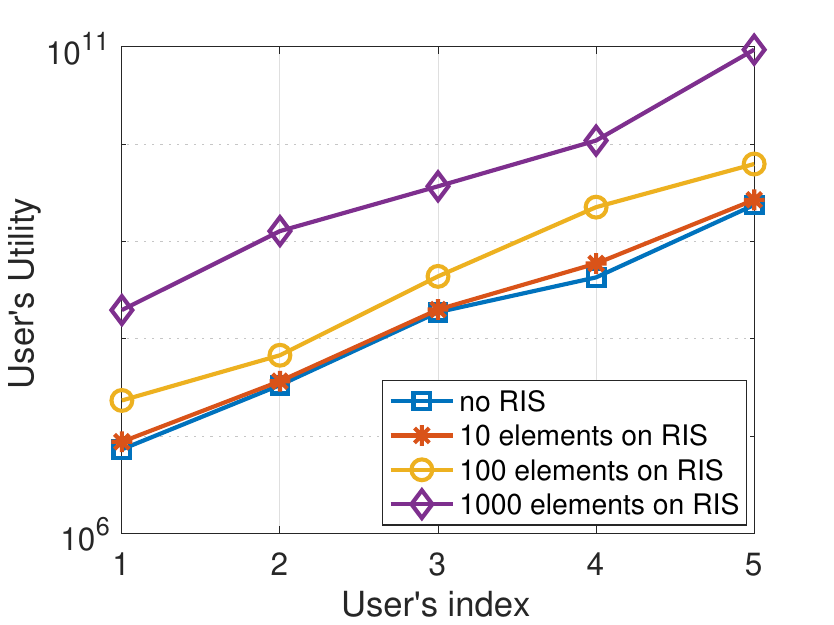} 
        \caption{Sub-6 GHz}
        \label{fig:sub6ghz_utility}
    \end{subfigure}
    \hspace{0.000001\textwidth}
    \begin{subfigure}[b]{0.4899\columnwidth}
        \centering
        \includegraphics[width=4.5cm, height=3cm]{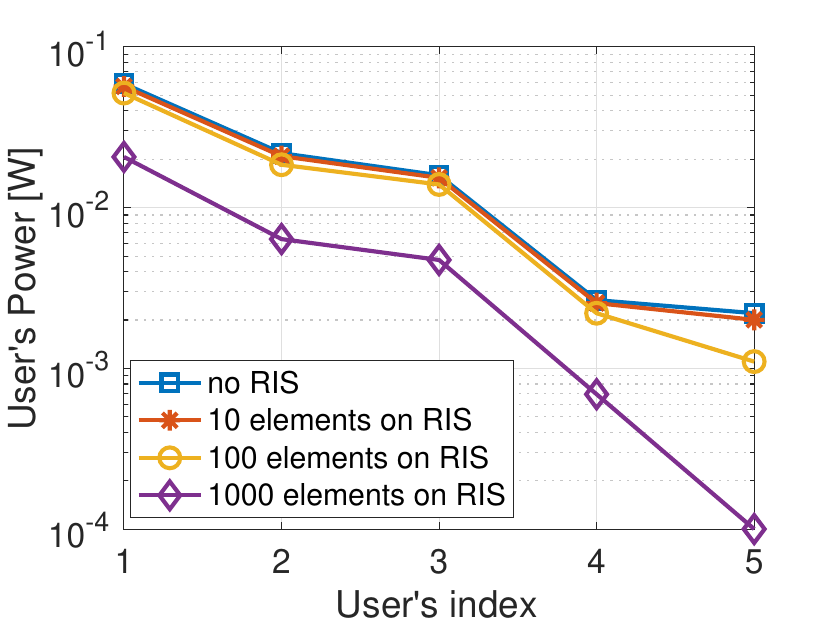} 
        \caption{mmWave}
        \label{fig:mmwave_p}
    \end{subfigure}
    \hspace{0.000001\textwidth}
    \begin{subfigure}[b]{0.4899\columnwidth}
        \centering
        \includegraphics[width=4.5cm, height=3cm]{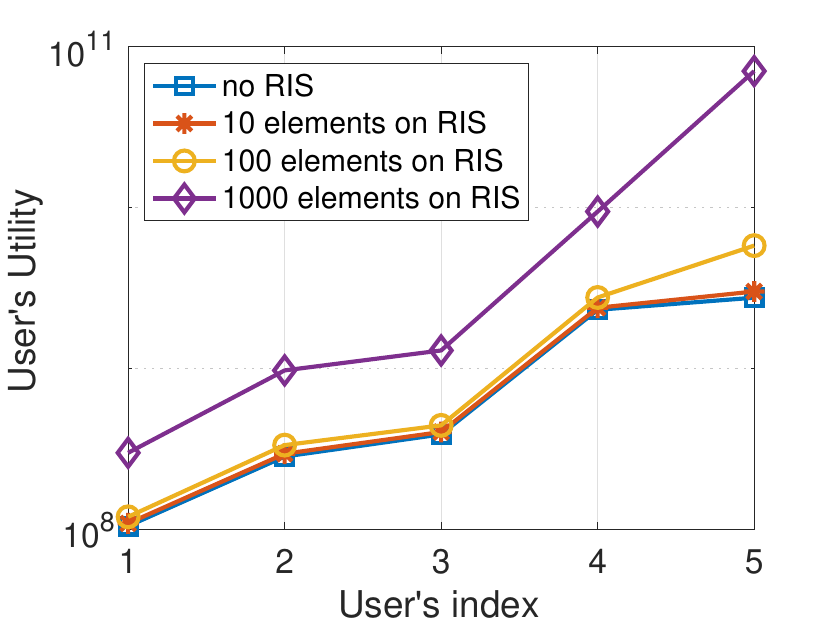} 
        \caption{mmWave}
        \label{fig:mmwave_utility}
    \end{subfigure}
     \caption{Evaluation of the hierarchical game-theoretic power control framework per \gls{ue}, under different number of \gls{ris} elements and different bands.}
     \label{fig:powercontrol}
     \vspace{-0.45cm}
\end{figure}

The outcomes of the hierarchical game-theoretic power allocation in the Sub-6 GHz and \gls{mmwave} bands are observed by considering the allocated \gls{ul} \gls{tx} power level and achieved \gls{ue} utility both independently and collectively. In Fig.~\ref{fig:powercontrol}, we observe each \gls{ue}'s allocated \gls{ul} \gls{tx} power (Figs.~\ref{fig:sub6ghz_p}, \ref{fig:mmwave_p}) and achieved utility (Figs. \ref{fig:sub6ghz_utility}, \ref{fig:mmwave_utility}), versus their respective index  under different number of \gls{ris}'s reflecting elements both  for the Sub-6 GHz and the \gls{mmwave} band. All \glspl{ue} are sorted in descending order with reference to their channel gains, where the lowest index represents the \gls{ue} with the highest channel gain, and the highest \gls{ue} index corresponds to the one with the lowest channel gain. Notably, the framework's power control \cite{diamanti2021energy} enables low channel gain \glspl{ue} to transmit with lower power levels and achieve higher utilities. This behavior can be explained by the utility function which is affected by the \gls{ue}'s channel-gain-to-interference ratio, i.e., $\frac{G_i}{I_i}$, at the exponent of the utility's numerator (see~\eqref{eq:eq3}). In detail, low channel gain \glspl{ue} usually exhibit a low channel-gain-to-interference ratio which  forces the maximum point of the utility function to take a low power value (i.e., $P_{i}$). This pattern is clearly illustrated in Fig. \ref{fig:powercontrol} where we also observe that as the number of \gls{ris}'s reflecting elements increases, their achieved utilities also increase (Figs. \ref{fig:sub6ghz_utility}, \ref{fig:mmwave_utility}), whereas the \glspl{ue}' allocated \gls{ul} \gls{tx} powers decrease (Figs. \ref{fig:sub6ghz_p}, \ref{fig:mmwave_p}), as expected. With regards to the two bands under study, in Fig. \ref{fig:sub6ghz_p} we observe that for the Sub-6 GHz band, $100$ reflecting elements on the \gls{ris} are enough for efficient power control, while for the \gls{mmwave} band, a minimum of $1000$ elements is needed as seen in Fig. \ref{fig:mmwave_p}. This is due to the high propagation losses of the latter band~\cite{tapio2021survey}, which stress the need for bigger \gls{ris} surfaces~\cite{da2023varactor} to enable effective beamforming and therefore more efficient power allocation. 

\begin{figure}[ht]
    \centering
    \begin{subfigure}[b]{0.4899\columnwidth}
        \centering
        \includegraphics[width=\columnwidth]{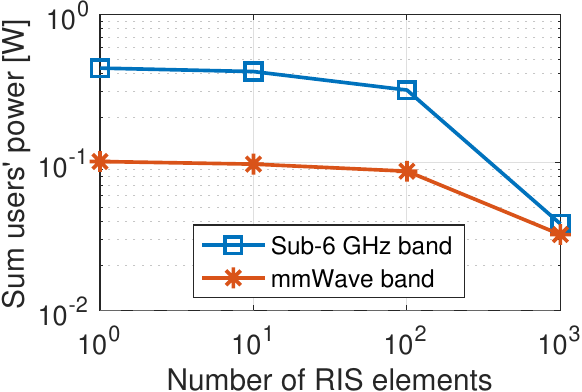}
        \label{fig:sub6ghz_p_sum}
    \end{subfigure}
    \hfill
    \begin{subfigure}[b]{0.4899\columnwidth}
        \centering
        \includegraphics[width=\columnwidth]{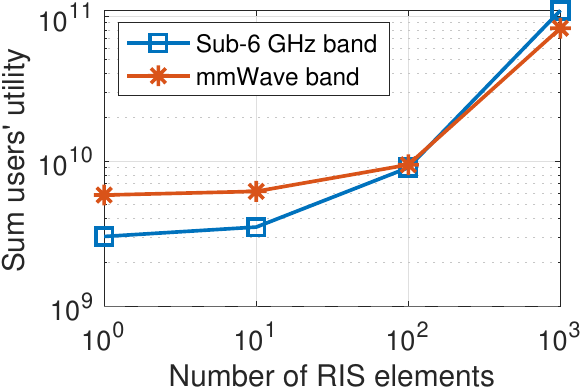}
        \label{fig:sub6ghz_utility_sum}
    \end{subfigure}
    \hfill
     \caption{Evaluation of the hierarchical game-theoretic power control framework on the entire system, under different number of RIS elements and different bands.}
     \label{fig:powercontrol_sum}
     \vspace{-0.45cm}
\end{figure}

The overall performance gain of the entire system (e.g., the \gls{noma} cluster) is illustrated in Fig. \ref{fig:powercontrol_sum}, where we observe both the sum \glspl{ue}' allocated power levels and utilities as a function of a different number of \gls{ris} elements both for the Sub-6 GHz and the \gls{mmwave} band. We can deduce that for a small number \gls{ris} elements (e.g., $10^1$ RIS elements), the system's performance is comparable to the case without a \gls{ris} (i.e., $10^0$ \gls{ris} elements). When the number of \gls{ris} elements increases, the sum \glspl{ue}' powers decreases while their sum utility/satisfaction increases, as expected.
Notably, the increased number of \gls{ris} elements results in the reduced sum of \glspl{ue}’ powers and thus in their increased sum utility/satisfaction.
We also observe that even if the system follows the same power allocation trend in the two bands, the sum of the allocated powers to the \gls{mmwave} \glspl{ue} is lower compared to the case of their Sub-6 GHz counterparts. \blue{It is reminded that the devised utility function for energy efficiency, captures the trade-off between increasing $\gamma_i$ (which improves the \gls{qos}) and minimizing $P_i$ (which reduces power consumption). Since the increase in $\gamma_i$ is limited by a low $\frac{G_i}{I_i}$, increasing $P_i$ will lead to diminishing returns in the numerator of the utility function. In detail, when the ratio $\frac{G_i}{I_i}$ is low, the \gls{sinr} i.e., $\gamma_i = \frac{P_i G_i}{I_i}$ becomes small. In this case, the exponential term $e^{-\alpha \gamma_i}$ approaches 1, meaning that even if the \gls{ue} increases $P_i$, the utility does not increase significantly because the \gls{sinr} is limited by the high interference. Thus, the utility function reaches its maximum at a lower power level $P_i$ because increasing power doesn't significantly improve the \gls{sinr}. Therefore, \glspl{ue} with low $\frac{G_i}{I_i}$ transmit at lower power levels to avoid unnecessary energy consumption, as increasing $P_i$ will not substantially improve their utility. Our power control framework favors \glspl{ue} with lower $\frac{G_i}{I_i}$  to transmit with lower power levels. In our work, \gls{mmwave} \glspl{ue} experience low $\frac{G_i}{I_i}$ levels, and hence have a lower \gls{cqi} compared to the \glspl{ue} of the Sub-6 GHz band they are compared to. As a result when ultimately compared, given the low $\frac{G_i}{I_i}$  levels of the former, they get to transmit with lower power levels compared to their Sub-6 GHz counterparts.} 

\begin{figure*}[ht]
    \centering
    \hspace{-0.09\textwidth}
    \begin{subfigure}[b]{1\columnwidth}
        \centering
        \includegraphics[width=10.1cm, height=3.5cm]{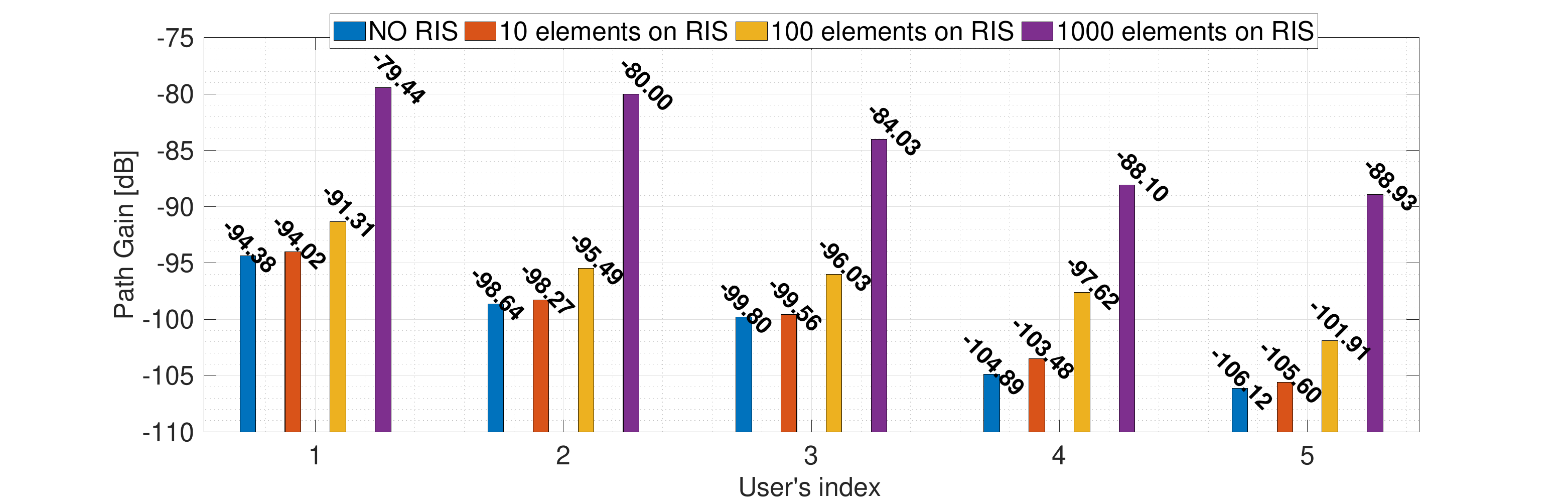} 
        \centering
        \caption{Sub-6 GHz band}
        \label{fig:sub6ghzpg}
    \end{subfigure}
    \hspace{0.01\textwidth}
    \begin{subfigure}[b]{1\columnwidth}
        \centering
        \includegraphics[width=10.1cm, height=3.5cm]{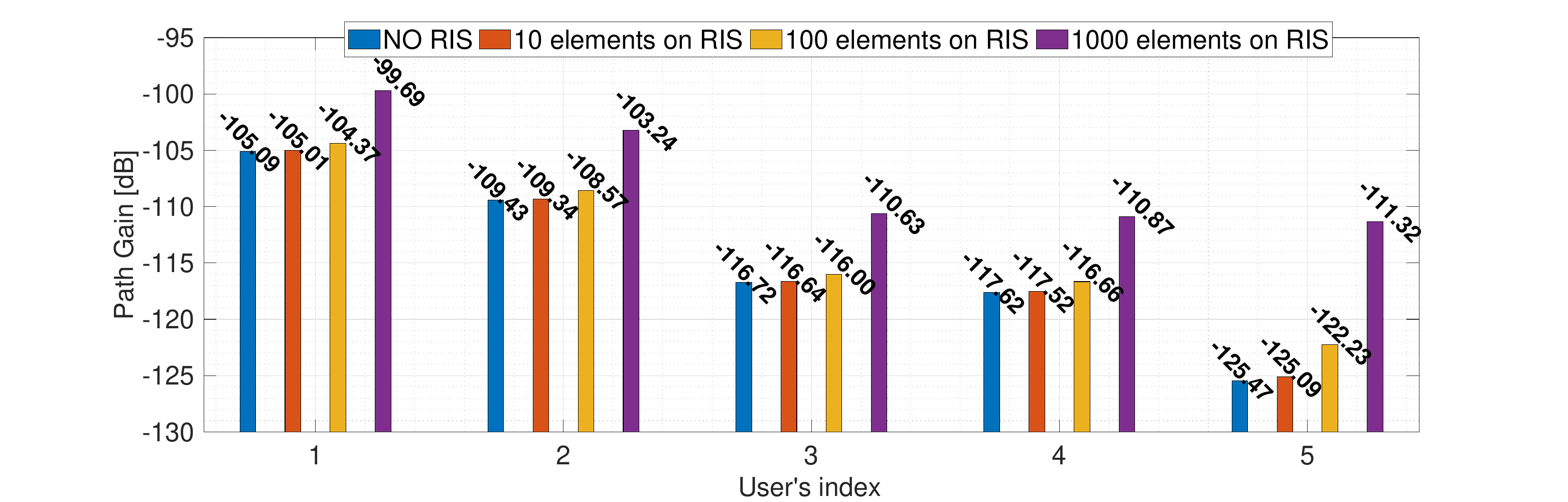} 
        \centering
        \caption{mmWave band}
        \label{fig:mmwavepg}
    \end{subfigure}
     \centering
     \caption{Path Gains for two different Spectrum bands.}
     \label{fig:pathgainsris}
     \vspace{-0.45cm}
\end{figure*}

Finally, in Fig.~\ref{fig:pathgainsris}, we present the path gains, $PG_i$, for every \gls{ue}, $i$, with and without the inclusion of the \gls{ris} both for the Sub-6 GHz and the \gls{mmwave} band. All the \glspl{ue} are sorted in descending order, and thus the \gls{ue} with the lowest index is the \gls{ue} with the best channel gain conditions, while the \gls{ue} with the biggest index corresponds to the \gls{ue} with the worst channel. 
Specifically, Fig.~\ref{fig:sub6ghzpg} depicts the path gains for the Sub-6 GHz band with the carrier frequency centered at $5.9$ GHz. Based on the reported findings, the performance achieved with a minimum of~$10$ \gls{ris} elements closely resembles the performance without the \gls{ris}. A noticeable difference at $\sim|1.41|$~dB between the aforementioned cases is only observed for index $4$, which, within the scope of this work, indicates a \gls{ue} with \emph{poor} \gls{cqi}.
Discernible differences can be observed for all the \glspl{ue} when focusing on the cases of $100$ or $1000$ \gls{ris} elements. In detail, an average difference of $\sim|4|$~dB is observed when collectively comparing all the \glspl{ue} without and with the existence of \gls{ris} in the topology under~$100$~\gls{ris} elements. Specifically, among all the \glspl{ue}, the \gls{ue} with index~$4$, achieves the most notable difference, by achieving a $|7.27|$~dB higher value at $-97.62$~dB in the presence of $100$~\gls{ris} elements. In the absence of \gls{ris}, the corresponding value is reported at $-104.89$~dB. Last, the benefits of including $1000$ elements on \gls{ris} are clearly illustrated in Fig.~\ref{fig:sub6ghzpg}, where the average path gain value for all the \glspl{ue} considered is marked at $-84.1$~dB, $\sim16.5\%$ higher compared to the average value of $-100.77$~dB achieved without the \gls{ris} inclusion.

Fig.~\ref{fig:mmwavepg} depicts the path gains obtained for the \gls{mmwave} band with the carrier frequency centered at~$28$~GHz. Contrary to the findings of the conducted evaluation in the Sub-6 GHz, in the \gls{mmwave} band, a minimum number of~$1000$ \gls{ris} elements is required to achieve distinguishable differences between the cases with and without a \gls{ris}. In this band, the \gls{ue} that significantly benefits from the \gls{ris} technology with $1000$ onboarded elements is the \gls{ue} that experiences the worst \gls{cqi} conditions, identified by \gls{ue} ID~$5$. Specifically, the latter \gls{ue} reports a $|14.15|$~dB higher value at~$-111.32$~dB compared to the case without the \gls{ris}, where the path gain is marked at $-125.47$~dB. On average, \glspl{ue} under the presence of $1000$~\gls{ris} elements in the topology report path gains $\sim|7.72|$~dB or $\sim7\%$ higher than in the absence of \gls{ris}. Precisely, the average path gain value of the \glspl{ue} in the case of~$1000$ \gls{ris} elements is reported at $-107.15$~dB, while in the case of \gls{ris} absence this value is indicated at~$-114.87$~dB.

Conclusively, in the Sub-6 GHz band, the average path gain value reported with $100$ \gls{ris} elements is marked at $-96.5$ dB, a value $\sim4\%$ higher than the average path gain of $-100.77$ dB reported in the absence of \gls{ris}. In the \gls{mmwave} band, with $1000$ \gls{ris} elements, the corresponding increase is found to be $\sim7\%$. Based on these findings, a minimum number of $100$ \gls{ris} elements is deemed necessary to support the deployment of \gls{ris} in the Sub-6 GHz band, while in the \gls{mmwave} this number is found to be $1000$.

\begin{figure}[ht]
    \centering
    \begin{subfigure}[b]{0.4899\columnwidth}
        \centering
        \includegraphics[width=4.5cm, height=3cm]{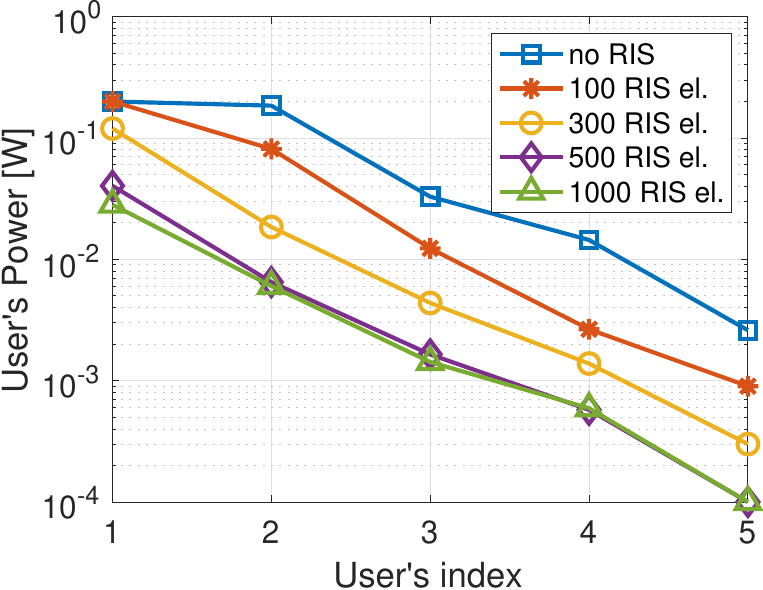} 
        \caption{Sub-6 GHz}
        \label{fig:sub6ghz_p_more_elems}
    \end{subfigure}
    \hspace{0.000001\textwidth}
    \begin{subfigure}[b]{0.4899\columnwidth}
        \centering
        \includegraphics[width=4.5cm, height=3cm]{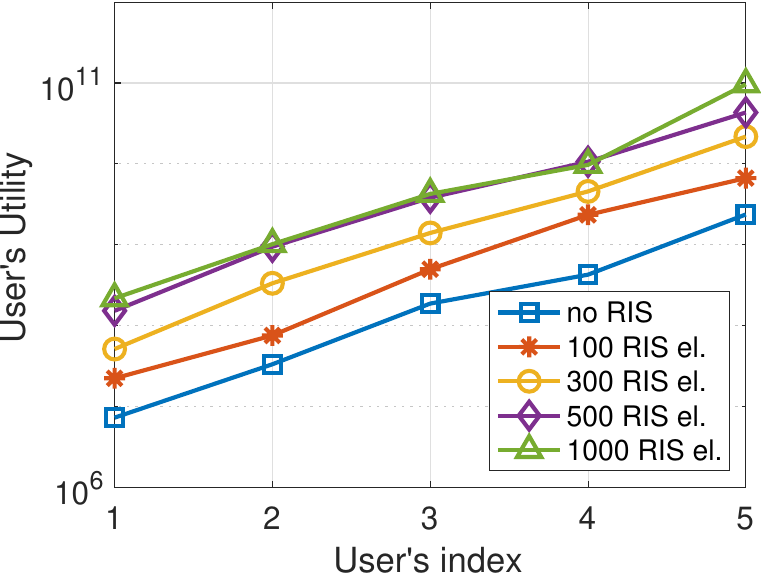} 
        \caption{Sub-6 GHz}
        \label{fig:sub6ghz_utility_more_elems}
    \end{subfigure}
    \hspace{0.000001\textwidth}
    \begin{subfigure}[b]{0.4899\columnwidth}
        \centering
        \includegraphics[width=4.5cm, height=3cm]{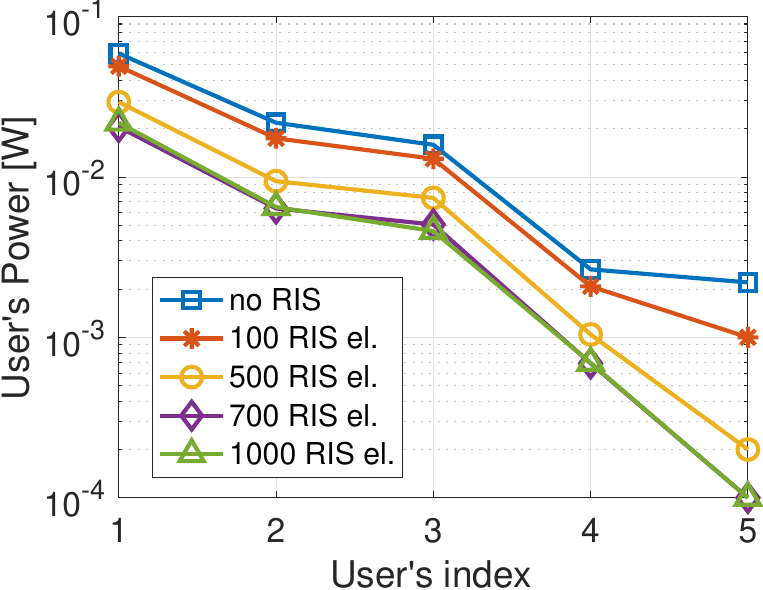} 
        \caption{mmWave}
        \label{fig:mmwave_p_more_elems}
    \end{subfigure}
    \hspace{0.000001\textwidth}
    \begin{subfigure}[b]{0.4899\columnwidth}
        \centering
        \includegraphics[width=4.5cm, height=3cm]{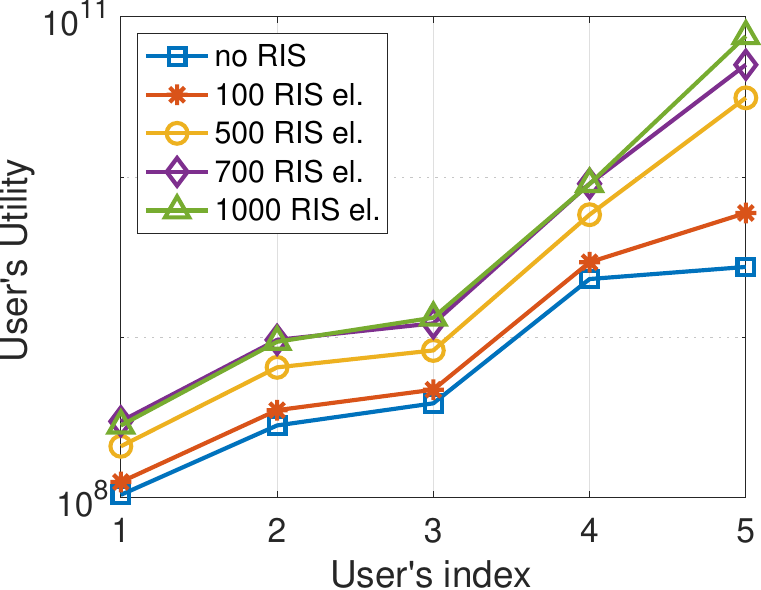} 
        \caption{mmWave}
        \label{fig:mmwave_utility_more_elems}
    \end{subfigure}
     \caption{\blue{Evaluation of the hierarchical game-theoretic power control framework per \gls{ue}, under different number of \gls{ris} elements per each band.}}
     \label{fig:powercontrol_more_elems}
     \vspace{-0.45cm}
\end{figure}

\blue{It is finally noted that the performance evaluation of the Stackelberg game-theoretic optimization process involved elements in increasing orders of magnitude, representing successive powers of $10$. The aforementioned numbers not only represent typical values encountered in the literature~\cite{rossanese2022designing,bjornson2022reconfigurable}, but also meet the requirements outlined in~\cite{ETSI_RIS001}, which state that existing \gls{ris} elements can range from $10$ elements and above.
In the following, we also provide a performance evaluation under different sets of \gls{ris} elements for the respective bands, aiming to determine the minimal number of \gls{ris} elements each band requires to achieve higher levels of energy efficiency. Based on the experimental results in Fig.~\ref{fig:powercontrol_more_elems}, it is noted that for the Sub-6 Ghz band, $500$ \gls{ris} elements result in similar performance to $1000$ (Figs.~\ref{fig:sub6ghz_p_more_elems} and~\ref{fig:sub6ghz_utility_more_elems}), while discernible performance improvements are observed from $100$ elements and above, with $300$ being and indicative value. For the \gls{mmwave} band, it is noted that the respective values are $700$ (resulting in similar performance to $1000$) and $500$ (indicative value for performance improvement), correspondingly (Figs.~\ref{fig:mmwave_p_more_elems} and~\ref{fig:mmwave_utility_more_elems}).}

\subsection{Comparison with the state-of-the-art}\label{soa}
\blue{In this section, we evaluate our optimization framework using two different utilities for energy efficiency, as shown in Table~\ref{table:utility-functions-comparison}. The utility function proposed in our work is labeled \textit{Proposed Utility} and is compared with the utility presented in~\cite{diamanti2021prospect}, referred to as \textit{Literature Utility}. To assess the performance of our proposed method in comparison with~\cite{diamanti2021prospect}, we also consider a \gls{noma} cluster of $4$ \glspl{ue}, a bandwidth of $5$~MHz, and the same noise levels as in~\cite{diamanti2021prospect}, and we evaluate the performance in the Sub-6~GHz band. Finally, we consider two cases: one without the inclusion of a \gls{ris} and the other with $100$ \gls{ris} elements onboard.}

\begin{table}[ht]
\vspace{-0.1cm}
\centering
\small
\setlength\abovecaptionskip{-.01cm}
\caption{\blue{Utility Functions Comparison}}
\begin{adjustbox}{width=0.75\linewidth}
\renewcommand{\arraystretch}{2} 
\begin{tabular}{@{}l l@{}}
\toprule
\textbf{\blue{Proposed Utility:}} & $U_{i}\left(P_{i}, \mathbf{P}_{-i}\right)=\frac{W \cdot\left(1-e^{-\alpha \gamma_{i}}\right)^{M}}{P_{i}}$  \\
\textbf{\blue{Literature Utility:}} & $EE_i({P}_i, \mathbf{P}_{-i}) = \frac{ W \cdot \log_2 \left( 1 + \gamma_i \right)}{P_i}$  \\
\bottomrule
\end{tabular}
\end{adjustbox}
\label{table:utility-functions-comparison}
\end{table}

\begin{figure}[ht]
    \centering
    \begin{subfigure}[b]{0.4899\columnwidth}
        \centering
        \includegraphics[width=\columnwidth]{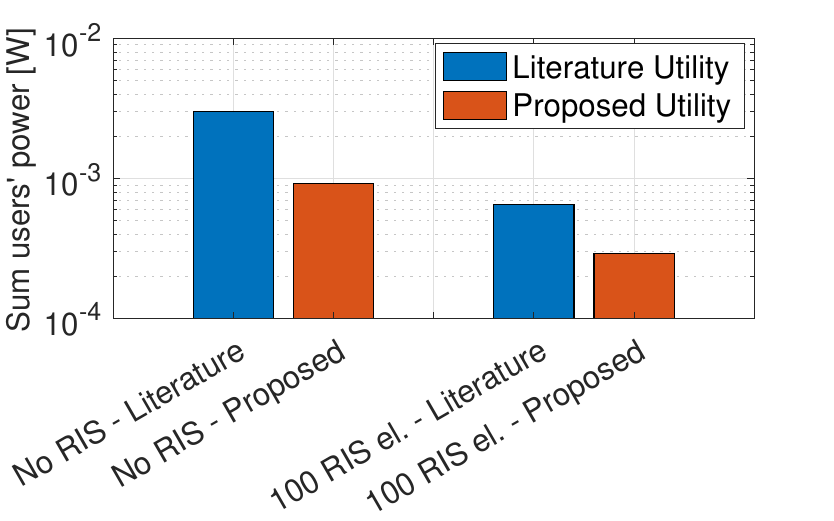}
        \caption{}
        \label{fig:sub6ghz_propp}
    \end{subfigure}
    \hfill
    \begin{subfigure}[b]{0.4899\columnwidth}
        \centering
        \includegraphics[width=\columnwidth]{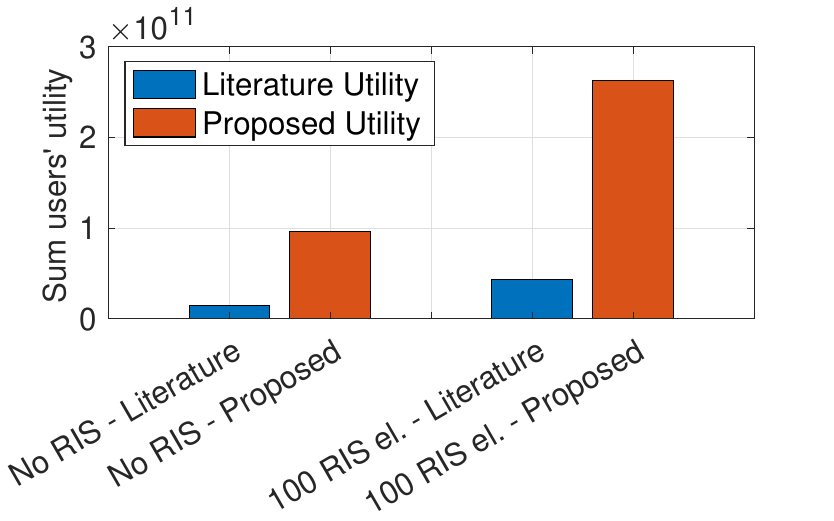}
        \caption{}
        \label{fig:propo_uti}
    \end{subfigure}
    \hfill
     \caption{Evaluation of the hierarchical game-theoretic power control framework under two different utilities given in Table~\ref{table:utility-functions-comparison}.}
     \label{fig:powercontrol_prop}
     \vspace{-0.45cm}
\end{figure}

\blue{The performance evaluation results reported in Fig.~\ref{fig:powercontrol_prop} clearly indicate the suitability of the proposed utility function in yielding lower transmission power levels (Fig.~\ref{fig:sub6ghz_propp}) and achieving higher energy efficiency (Fig.~\ref{fig:propo_uti}). This is due to the flexible design of our proposed utility through the control parameters, as well as its effective capture of the trade-off between achieving higher \gls{qos} and lower costs (e.g., reduced \gls{tx} power levels).}

\section{Performance Evaluation of RIS-Assisted \gls{cv2x} Communication Systems on Colosseum}\label{Section IV}

As mentioned in Section \ref{Section IIIB}, the optimization problem solved by the \gls{uav} and defined in Eqs. \ref{eq4a}-\ref{eq4b}, will result in the maximization of its total received signal strength through the individual maximization of each \gls{ue}'s incoming signal strength. Indeed, the latter is achieved through a linear combination of the \gls{ris} elements’ effective phase shifts as given in~\eqref{eq8}, which will enable the creation of the cascaded channels denoted as $\mathbf{{h}}_{RU}^H\boldsymbol{\Theta}\mathbf{{h}}_{iR}$. In consistency with Section \ref{Section IIIA}, all the incoming \gls{ue} signals at the \gls{uav} are the result of the coherent summation of the direct, i.e., $h_{iU}$, and reflected signals. It is reminded that the path gains of the corresponding generated channels, without and with the inclusion of a different number of \gls{ris} elements  
are provided and illustrated in Fig.~\ref{fig:pathgainsris}.

As a next step, we proceed by installing the generated channels in Colosseum. In summary, the reference input scenario to be installed includes two key components: the path gain, represented as a complex coefficient value in the form of \gls{fir}, and the \gls{toa} value. The path gain is derived from the coherent summation of generated \glspl{mpc} across the wireless links and has been averaged across multiple experiments, as detailed in Section~\ref{Section IIIA}. Meanwhile, the \gls{toa} values have been computed based on the Euclidean distance between the entities considered in the topology (i.e., \gls{uav}, \gls{ris}, and \glspl{ue}), given the speed of light (i.e., $3 \times 10^8$ [m/s]). These values collectively contribute to generating the time-variant \gls{cir} for each node pair within the topology. Finally, a tutorial on the installation of \gls{rf} scenarios in Colosseum can be found in~\cite{bonati2021colosseum,villa2022cast}, but as it is not the focus of this paper, the detailed explanation is omitted.

\begin{figure*}[ht]
    \centering
    \begin{subfigure}[b]{0.45\textwidth} 
        \centering
        \includegraphics[width=\columnwidth]{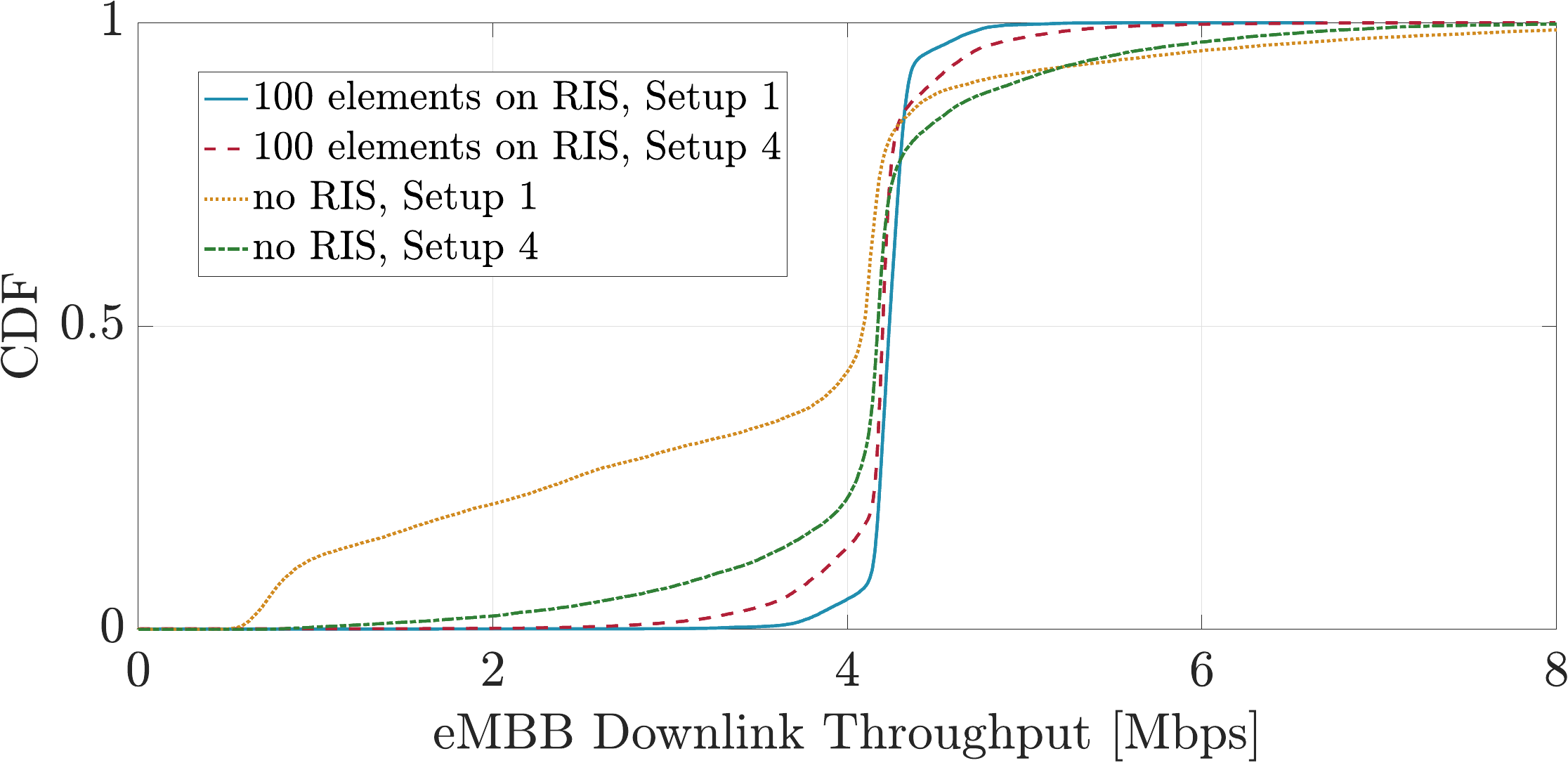}
        \caption{Sub-6 GHz}
        \label{fig:ris_on_col1}
    \end{subfigure}
    \hspace{0.05\textwidth} 
    \begin{subfigure}[b]{0.45\textwidth} 
        \centering
        \includegraphics[width=\columnwidth]{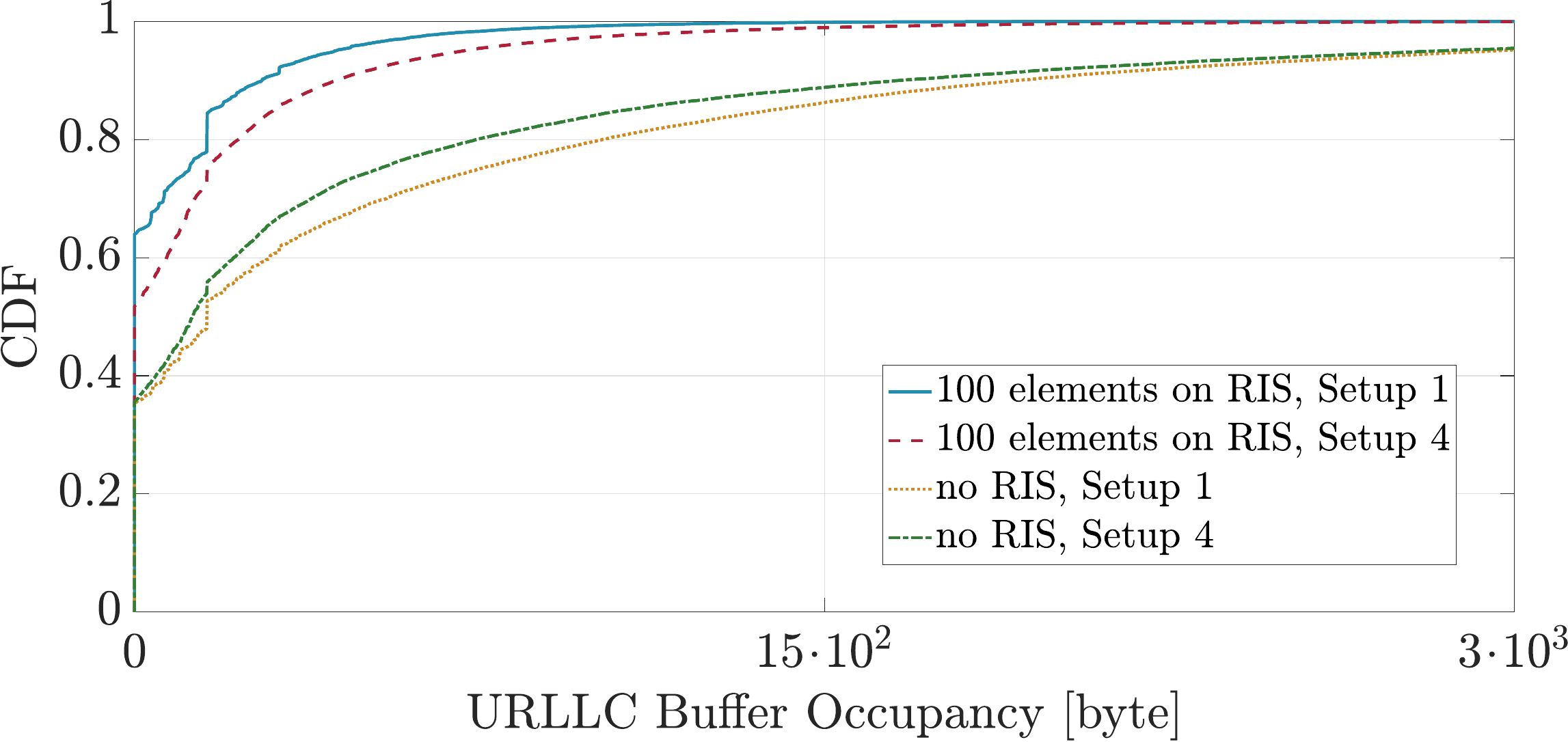}
        \caption{Sub-6 GHz}
        \label{fig:ris_on_col2}
    \end{subfigure}
    
    \begin{subfigure}[b]{0.45\textwidth} 
        \centering
        \includegraphics[width=\columnwidth]{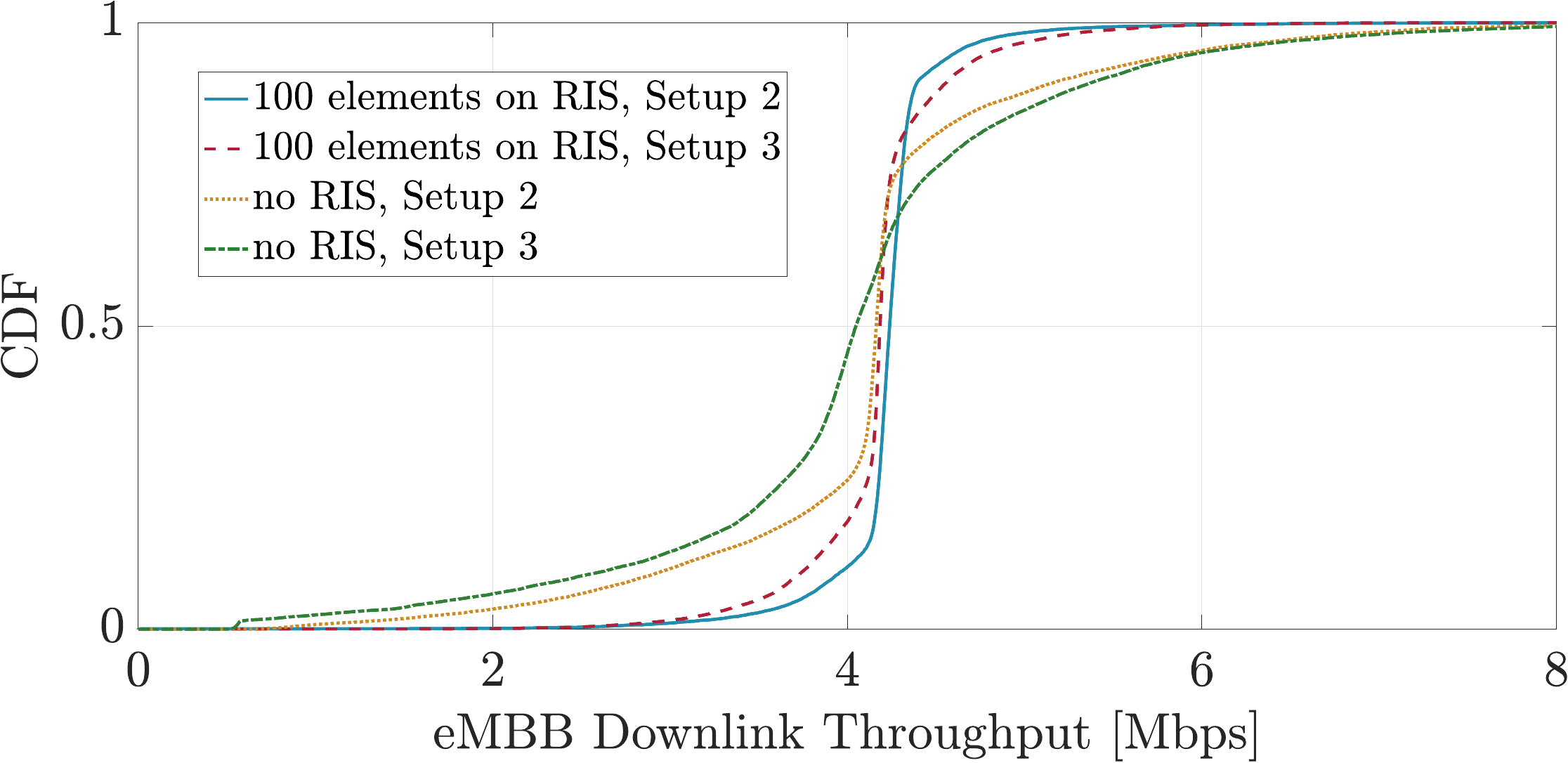}
        \caption{Sub-6 GHz}
        \label{fig:ris_on_col1_caseb}
    \end{subfigure}
    \hspace{0.05\textwidth} 
    \begin{subfigure}[b]{0.45\textwidth} 
        \centering
        \includegraphics[width=\columnwidth]{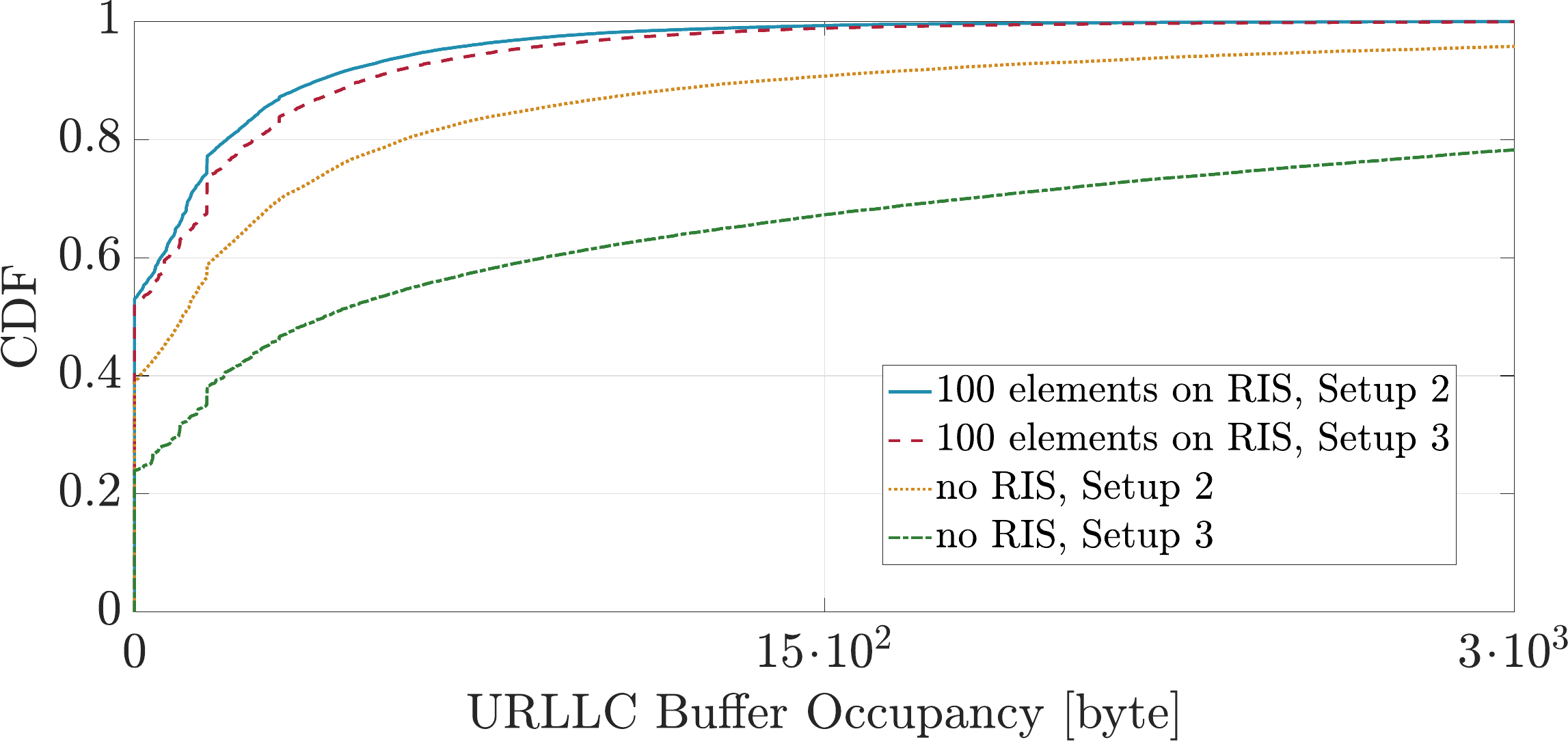}
        \caption{Sub-6 GHz}
        \label{fig:ris_on_col2_caseb}
    \end{subfigure}
    
    \caption{Evaluation of a slice-based Multi-\gls{ue} \gls{cv2x} communication system on Colosseum with and without the inclusion of \gls{ris} under different scheduling profiles.}
    \label{fig:ris_on_col_total}
    \vspace{-0.45cm}
\end{figure*}

We focus on the case of the Sub-6 GHz band, where all the wireless channels were simulated with \gls{quadriga} at the center frequency of~$5.9$ GHz. We only consider the scenario with $100$~\gls{ris} elements for all the \gls{ris}-assisted generated channels. This approach allows us to explore how a relatively small  number of \gls{ris} elements (typically large arrays are required \cite{bjornson2020reconfigurable,siddiqi2022reconfigurable}), and an average numerical increase of $\sim4\%$ in the path gains between cases with and without \gls{ris} in the topology, can significantly improve network performance.
Additionally, since previous results (see Section~\ref{Section III}) have shown that \gls{mmwave}~\glspl{ue} are mostly benefited in terms of energy efficiency (i.e., they get to transmit with lower power levels, especially for $1000$ \gls{ris} elements, while simultaneously satisfying their \gls{qos} demands), we aim at examining the advantages introduced by \gls{ris} technology in a realistic \gls{nextg} \gls{ran} scenario tested in the Sub-6 GHz band.
It is noted that although the channels were created at the frequency of~$5.9$ GHz using the \gls{quadriga} channel modeling simulator, the testing frequency for this scenario in Colosseum is set to $1$~GHz, which is the frequency that the \gls{mchem} is optimized to work \cite{villa2023twinning}. 

We leverage the capabilities of the SCOPE framework \cite{bonati2021scope} to deploy an 
end-to-end \gls{ran} and core network  though the srsRAN~\cite{gomez2016srslte} softwarized open-source protocol stack on Colosseum. Specifically, we deploy a 3GPP-compliant cellular network comprising a single \gls{bs} (i.e., the \gls{uav}) and $5$~\glspl{ue} distributed across $2$ different slices. These are: (i)~\gls{embb} that concerns high traffic modeling of high-quality multimedia content and streaming applications, and (ii)~\gls{urllc} for time-sensitive applications, such as autonomous driving in \gls{v2x} scenarios as detailed in~\cite{tsampazi2023comparative}. Slice-based traffic is generated
by the \gls{mgen} \gls{tcp}/\gls{udp} traffic generator \cite{mgen}, based on the following specifications: \gls{embb} \glspl{ue} request $4$\:Mbps constant bitrate, while \gls{urllc} \glspl{ue} generate $89.3$\:kbps Poisson traffic.
The bandwidth of the \gls{bs} is set equal to $10$~MHz (i.e., $50$ \glspl{prb}), and it is divided between the $2$ slices.
Specifically, we allocate 
$45$ \glspl{prb} on the \gls{embb}, while the remaining  $\sim5$ \glspl{prb} are assigned on the \gls{urllc}. In this way, we allocate $90$\% of the available resources to \gls{embb}, which has the greatest demand for higher throughput. Consequently, $10$\% of the available resources are allocated to support the requirements for low latency on the \gls{urllc}. Among the $5$ users considered, $3$ of them are \gls{urllc} \glspl{ue}, while the remaining $2$ are assigned on the \gls{embb} slice. It is also noted that the \glspl{ue} of the \gls{embb} are those who reported the highest path gain values, while the \glspl{ue} of the \gls{urllc} are the ones with the lowest reported \gls{cqi} values. With this allocation, we aim at exploring how the presence of the \gls{ris} in the topology will benefit the \gls{urllc} slice in a scenario where limited resources are available to serve $60\%$ of the available \glspl{ue}, especially when they experience poor transmission conditions (e.g., in the absence of \gls{ris} in the topology).
Finally, the \gls{kpm} observed for \gls{embb} is the \gls{dl} throughtput, while the respective value for \gls{urllc} is the buffer occupancy as a proxy for latency~\cite{tsampazi2023comparative}.

By using SCOPE's open \glspl{api}, we fine-tune the \gls{ran}'s functionalities, in terms of scheduling profile selection. Specifically, two \gls{mac}-layer scheduling algorithms are compared, namely the \gls{rr} and a fairer one, the \gls{wf}, in all possible combinations. \blue{In Table~\ref{table:sched-profiles}, we present all the configurations considered in terms of scheduling profiles (defined in Table~\ref{table:sched-profiles-catalog}) for the \gls{embb} and \gls{urllc} slices, which define how \glspl{prb} are internally allocated to \glspl{ue} belonging to
each slice~\cite{tsampazi2024pandora}.}

\blue{
\begin{table}[ht]
\vspace{-0.1cm}
\centering
\small
\setlength\abovecaptionskip{-.01cm}
\caption{\blue{Scheduling Profiles Catalog}}
\begin{adjustbox}{width=0.35\linewidth}
\begin{tabular}{@{}l l@{}}
\toprule
\textbf{\blue{RR:}} & \text{\blue{Round Robin}}  \\
\textbf{\blue{WF:}} & \text{\blue{Waterfilling}}  \\
\bottomrule
\end{tabular}
\end{adjustbox}
\label{table:sched-profiles-catalog}
\end{table}
}

\begin{table}[ht]
\vspace{-0.1cm}
\centering
\small
\setlength\abovecaptionskip{-.01cm}
\caption{Scheduling Profiles Configuration}
\begin{adjustbox}{width=0.45\linewidth}
\begin{tabular}{@{}c@{}c@{}c@{}}
\toprule
\multicolumn{1}{c}{\textbf{Setup ID}} & \multicolumn{1}{c}{\textbf{\texttt{eMBB}}} & \multicolumn{1}{c}{\textbf{\texttt{URLLC}}} \\ \midrule
\textbf{1} & \text{RR} & \text{RR}  \\
\textbf{2} & \text{RR} & \text{WF}  \\
\textbf{3} & \text{WF} & \text{RR} \\
\textbf{4} & \text{WF} & \text{WF} \\
\bottomrule
\end{tabular}
\end{adjustbox}
\label{table:sched-profiles}
\vspace{-0.25cm}
\end{table}
\noindent
The results of this comparative performance evaluation can be seen in Fig.~\ref{fig:ris_on_col_total}.
Finally, all subsequent results were produced by taking the median as the most representative statistical value of a dataset, and averaged over multiple repetitions of experiments in the \gls{dl} direction of the communication system\footnote{Note that all channels generated with \gls{quadriga} in the \gls{ris}-assisted and \gls{uav}-enabled scenarios are reciprocal. The focus of this Section is  the evaluation of a \gls{ris}-assisted and \gls{uav}-enabled communication system on the \gls{dl} direction.}.

In Figs.~\ref{fig:ris_on_col1} and~\ref{fig:ris_on_col2}, we focus on the case where both slices are served under the same scheduling policy. Based on the reported findings, when focusing on the \gls{embb} slice, all configurations, both with and without the \gls{ris} inclusion, deliver a median throughput value of $\sim4$\:Mbps. However, Setup $1$ achieves one of the highest reported throughput values of $4.236$ Mbps with $100$ \gls{ris} elements. The corresponding value in the absence of \gls{ris} is found to be $4.091$\:Mbps. With Setup~$4$ and $100$ onboarded \gls{ris} elements, the median throughput value reaches $4.2$\:Mbps. The respective value in the absence of \gls{ris} is reported at $4.172$\:Mbps. On the \gls{urllc} slice, however, in the absence of \gls{ris}, both Setups~$1$ and $4$ underperform compared to the case where \gls{ris} is included in the topology. Specifically, with Setup~$1$, the median value of the buffer size is reported at $158$~byte, while for Setup~$4$, the corresponding value is marked at $127$~byte. In both setups with \gls{ris}, a median value of $0$~byte buffer occupancy is achieved, indicating zero latency. 

In Figs.~\ref{fig:ris_on_col1_caseb} and~\ref{fig:ris_on_col2_caseb} we include the results collected when both slices are served under different scheduling policies. On the \gls{embb}, Setup~$2$ achieves the highest throughput value reported in the presence of $100$ elements on \gls{ris} at $4.238$\:Mbps, while Setup~$3$ achieves one of the lowest at $4.184$\:Mbps. The respective values in the absence of \gls{ris} are reported at $4.163$\:Mbps and $4.048$\:Mbps, correspondingly. On the \gls{urllc}, in the presence of \gls{ris}, all setups resulted in empty buffers, reporting a median value of~$0$~buffer occupancy. In the absence of \gls{ris}, with Setup~$2$ the median value for buffer occupancy is marked at~$105$~byte, while with Setup~$3$ the corresponding value is found to be~$412$ byte.

In Tables~\ref{table:ris-setups} and~\ref{table:no-ris-setups}, all the results obtained without and with the inclusion of $100$ \gls{ris} elements under different combinations of scheduling policies are collectively presented. Based on the reported findings in Table~\ref{table:ris-setups} we observe that in the presence of \gls{ris} in the topology, Setups~$1$ and~$2$ deliver almost identical performance on the \gls{embb}, reporting median throughput values of $4.236$\:Mbps and $4.238$\:Mbps, respectively. Setups~$3$~and~$4$ deliver slightly worse performance, at $4.184$\:Mbps and $4.2$\:Mbps, correspondingly. On the \gls{urllc}, all setups resulted in optimal performance by achieving a median value of $0$ byte buffer occupancy. The findings of Table~\ref{table:no-ris-setups} indicate that in the absence of \gls{ris} in the topology, Setups~$2$~and~$4$ yield similar performance, achieving median \gls{embb} throughput values of $4.163$\:Mbps and $4.172$\:Mbps. The aforementioned setups also lead to improved \gls{urllc} performance among those compared, achieving buffer occupancies of $105$~and~$127$ byte, respectively. Setups~$1$~and~$3$ achieve \gls{embb} throughputs of $4.091$\:Mbps and $4.048$\:Mbps \gls{embb}, respectively. The median values for buffer occupancy on the \gls{urllc} are $158$~and~$412$~byte, with the latter being the highest recorded in the absence of \gls{ris} in the topology. It is noted that all setups resulted in empty buffers on the \gls{urllc} when \gls{ris} was included in the topology. In the absence of \gls{ris}, the reported median values for buffer occupancy vary around the average value of $200$ byte and are included in Table~\ref{table:no-ris-setups}. When \gls{ris} is not included in the topology, all configurations result in increased buffer sizes, indicating higher levels of latency in the communication links.

Conclusively, in resource-constrained environments (e.g., limited availability of \glspl{prb}), where the \glspl{ue} may experience relatively poorer \gls{cqi} conditions, the presence of \gls{ris} enhances the \gls{qos} experience of the \gls{urllc} by resulting in a buffer occupancy of $0$ byte.

\begin{table}[ht]
\vspace{-0.1cm}
\centering
\small
\setlength\abovecaptionskip{-.01cm}
\caption{\glspl{kpm} for \gls{ris}-assisted and \gls{uav}-enabled \gls{cv2x} topologies.}
\begin{adjustbox}{width=0.65\linewidth}
\begin{tabular}{@{}c@{}c@{}c@{}}
\toprule
\multicolumn{1}{c}{\textbf{Setup ID}} & \multicolumn{1}{c}{\textbf{\texttt{\gls{embb}}} [Mbps]} & \multicolumn{1}{c}{\textbf{\texttt{\gls{urllc}}} [byte]} \\ \midrule
\textbf{1} & $4.236$ & $0$  \\
\textbf{2} & $4.238$ & $0$  \\
\textbf{3} & $4.184$ & $0$ \\
\textbf{4} & $4.2$ & $0$ \\
\bottomrule
\end{tabular}
\end{adjustbox}
\label{table:ris-setups}
\end{table}

\begin{table}[ht]
\vspace{-0.1cm}
\centering
\small
\setlength\abovecaptionskip{-.01cm}
\caption{\glspl{kpm} for \gls{uav}-enabled \gls{cv2x} topologies without \gls{ris}.}
\begin{adjustbox}{width=0.65\linewidth}
\begin{tabular}{@{}c@{}c@{}c@{}}
\toprule
\multicolumn{1}{c}{\textbf{Setup ID}} & \multicolumn{1}{c}{\textbf{\texttt{\gls{embb}}} [Mbps]} & \multicolumn{1}{c}{\textbf{\texttt{\gls{urllc}}} [byte]} \\ \midrule
\textbf{1} & $4.091$ & $158$  \\
\textbf{2} & $4.163$ & $105$  \\
\textbf{3} & $4.048$ & $412$ \\
\textbf{4} & $4.172$ & $127$ \\
\bottomrule
\end{tabular}
\end{adjustbox}
\label{table:no-ris-setups}
\vspace{-0.15cm}
\end{table}

\section{Conclusion}\label{Section V}

In this work, we have
leveraged \quadriga's open-source code to simulate a \gls{ris}-assisted and \gls{uav}-enabled \gls{cv2x} topology which we evaluated under a different number of \gls{ris} elements both in the Sub-6 GHz and the \gls{mmwave} portion of the spectrum. Through the formulation of a hierarchical - one leader - multiple followers - Stackelberg-based game-theoretic optimization framework, we were able to study the achieved energy efficiency of the topology on the aforementioned bands. Our results on power control indicate significant power savings and \gls{ue} satisfaction, specifically for the \gls{mmwave} band. Finally, while focusing on the case of $100$ elements mounted on the \gls{ris} and the Sub-6 GHz band, we installed the generated channels on the Colosseum wireless network emulator and conducted an evaluation campaign under different setups of scheduling profiles, considering a fixed \gls{prb} allocation for both the \gls{embb} and \gls{urllc} network slices. The performance evaluation results from network emulation on Colosseum indicate that in resource-limited conditions (i.e., a low number of allocated \glspl{prb}), the presence of \gls{ris} will counteract potential network degradation by resulting in empty buffers, thus ensuring low latency levels on the \gls{urllc} slice. Some interesting extensions of the current work include focusing on the study of stochastic \gls{ue} mobility on the Colosseum Network Emulator and exploring the expected improvements introduced by the coexistence of \glspl{ris} and \glspl{uav}, particularly in eliminating potential network coverage holes. \blue{Additionally, comparing \gls{quadriga} \gls{ris}-assisted channels with other \gls{ris}-aided schemes and traditional beamforming represents an interesting direction for future research.}

\bibliography{References}

\begin{thebibliography}{100}
\providecommand{\url}[1]{#1}
\csname url@samestyle\endcsname
\providecommand{\newblock}{\relax}
\providecommand{\bibinfo}[2]{#2}
\providecommand{\BIBentrySTDinterwordspacing}{\spaceskip=0pt\relax}
\providecommand{\BIBentryALTinterwordstretchfactor}{4}
\providecommand{\BIBentryALTinterwordspacing}{\spaceskip=\fontdimen2\font plus
\BIBentryALTinterwordstretchfactor\fontdimen3\font minus \fontdimen4\font\relax}
\providecommand{\BIBforeignlanguage}[2]{{%
\expandafter\ifx\csname l@#1\endcsname\relax
\typeout{** WARNING: IEEEtran.bst: No hyphenation pattern has been}%
\typeout{** loaded for the language `#1'. Using the pattern for}%
\typeout{** the default language instead.}%
\else
\language=\csname l@#1\endcsname
\fi
#2}}
\providecommand{\BIBdecl}{\relax}
\BIBdecl

\bibitem{attar20225g}
H.~Attar, H.~Issa, J.~Ababneh, M.~Abbasi, A.~A. Solyman, M.~Khosravi, R.~Said~Agieb \emph{et~al.}, ``{5G System Overview for Ongoing Smart Applications: Structure, Requirements, and Specifications},'' \emph{Computational Intelligence and Neuroscience}, vol. 2022, 2022.

\bibitem{shehab20215g}
M.~J. Shehab, I.~Kassem, A.~A. Kutty, M.~Kucukvar, N.~Onat, and T.~Khattab, ``{5G networks towards smart and sustainable cities: A review of recent developments, applications and future perspectives},'' \emph{IEEE Access}, vol.~10, pp. 2987--3006, 2021.

\bibitem{sobhi2020energy}
S.~Sobhi-Givi, M.~G. Shayesteh, and H.~Kalbkhani, ``{Energy-efficient power allocation and user selection for mmWave-NOMA transmission in M2M communications underlaying cellular heterogeneous networks},'' \emph{IEEE Transactions on Vehicular Technology}, vol.~69, no.~9, pp. 9866--9881, 2020.

\bibitem{zhang2017network}
H.~Zhang, N.~Liu, X.~Chu, K.~Long, A.-H. Aghvami, and V.~C. Leung, ``{Network slicing based 5G and future mobile networks: Mobility, resource management, and challenges},'' \emph{IEEE communications magazine}, vol.~55, no.~8, pp. 138--145, 2017.

\bibitem{wang2018survey}
C.-X. Wang, J.~Bian, J.~Sun, W.~Zhang, and M.~Zhang, ``{A survey of 5G channel measurements and models},'' \emph{IEEE Communications Surveys \& Tutorials}, vol.~20, no.~4, pp. 3142--3168, 2018.

\bibitem{lopes20235g}
I.~Lopes, T.~Guarda, A.~Fernandes, and M.~I. Ribeiro, ``{How 5G Will Transform Smart Cities: A Literature Review},'' in \emph{International Conference on Computational Science and Its Applications}.\hskip 1em plus 0.5em minus 0.4em\relax Springer, 2023, pp. 70--81.

\bibitem{liu2021promoting}
L.~Liu, X.~Guo, and C.~Lee, ``{Promoting smart cities into the 5G era with multi-field Internet of Things (IoT) applications powered with advanced mechanical energy harvesters},'' \emph{Nano Energy}, vol.~88, p. 106304, 2021.

\bibitem{sharif2019compact}
A.~Sharif, J.~Guo, J.~Ouyang, S.~Sun, K.~Arshad, M.~A. Imran, and Q.~H. Abbasi, ``{Compact base station antenna based on image theory for UWB/5G RTLS embraced smart parking of driverless cars},'' \emph{IEEE Access}, vol.~7, pp. 180\,898--180\,909, 2019.

\bibitem{marabissi2018real}
D.~Marabissi, L.~Mucchi, R.~Fantacci, M.~R. Spada, F.~Massimiani, A.~Fratini, G.~Cau, J.~Yunpeng, and L.~Fedele, ``{A real case of implementation of the future 5G city},'' \emph{Future Internet}, vol.~11, no.~1, p.~4, 2018.

\bibitem{jiang2019smart}
Y.~Jiang, W.~Xiao, R.~Wang, and A.~Barnawi, ``{Smart urban living: Enabling emotion-guided interaction with next generation sensing fabric},'' \emph{IEEE Access}, vol.~8, pp. 28\,395--28\,402, 2019.

\bibitem{zhao2021nanogenerators}
X.~Zhao, H.~Askari, and J.~Chen, ``{Nanogenerators for smart cities in the era of 5G and Internet of Things},'' \emph{Joule}, vol.~5, no.~6, pp. 1391--1431, 2021.

\bibitem{sanchez2021review}
J.~E. S{\'a}nchez-Cano, W.~X. Garc{\'\i}a-Quilachamin, J.~P{\'e}rez-V{\'e}liz, J.~Herrera-Tapia, and K.~A. Fuentes, ``{Review of methods to reduce energy consumption in a smart city based on IoT and 5G technology},'' \emph{iJOE}, vol.~17, no.~08, p.~5, 2021.

\bibitem{rana2023review}
B.~Rana, S.-S. Cho, and I.-P. Hong, ``{Review paper on hardware of reconfigurable intelligent surfaces},'' \emph{IEEE Access}, 2023.

\bibitem{zivuku2022maximizing}
P.~Zivuku, S.~Kisseleff, V.-D. Nguyen, K.~Ntontin, W.~A. Martins, S.~Chatzinotas, and B.~Ottersten, ``{Maximizing the Number of Served Users in a Smart City using Reconfigurable Intelligent Surfaces},'' in \emph{IEEE Wireless Communications and Networking Conference (WCNC)}, 2022, pp. 494--499.

\bibitem{al2023emerging}
A.~Al~Amin, J.~Hong, V.-H. Bui, and W.~Su, ``{Emerging 6G/B6G wireless communication for the power infrastructure in smart cities: Innovations, challenges, and future perspectives},'' \emph{Algorithms}, vol.~16, no.~10, p. 474, 2023.

\bibitem{kisseleff2020reconfigurable}
S.~Kisseleff, W.~A. Martins, H.~Al-Hraishawi, S.~Chatzinotas, and B.~Ottersten, ``{Reconfigurable intelligent surfaces for smart cities: Research challenges and opportunities},'' \emph{IEEE Open Journal of the Communications Society}, vol.~1, pp. 1781--1797, 2020.

\bibitem{liu2023integrated}
R.~Liu, M.~Li, H.~Luo, Q.~Liu, and A.~L. Swindlehurst, ``{Integrated sensing and communication with reconfigurable intelligent surfaces: Opportunities, applications, and future directions},'' \emph{IEEE Wireless Communications}, vol.~30, no.~1, pp. 50--57, 2023.

\bibitem{salah2022paving}
M.~Salah, A.~Pitsillides, and A.~S. Mubarak, ``{Paving the way for economically accepted and technically pronounced smart radio environment},'' \emph{China Communications}, vol.~19, no.~8, pp. 247--266, 2022.

\bibitem{makarfi2020reconfigurable}
A.~U. Makarfi, K.~M. Rabie, O.~Kaiwartya, K.~Adhikari, X.~Li, M.~Quiroz-Castellanos, and R.~Kharel, ``{Reconfigurable intelligent surfaces-enabled vehicular networks: A physical layer security perspective},'' \emph{arXiv preprint arXiv:2004.11288}, 2020.

\bibitem{kamruzzaman2022key}
M.~Kamruzzaman, ``{Key technologies, applications and trends of internet of things for energy-efficient 6G wireless communication in smart cities},'' \emph{Energies}, vol.~15, no.~15, p. 5608, 2022.

\bibitem{bariah2023digital}
L.~Bariah, H.~Sari, and M.~Debbah, ``{Digital twin-empowered smart cities: A new frontier of wireless networks},'' \emph{Authorea Preprints}, 2023.

\bibitem{dagiuklas2023journey}
T.~Dagiuklas, ``{The journey from 5G towards 6G},'' in \emph{8th International Symposium on Electrical and Electronics Engineering (ISEEE)}.\hskip 1em plus 0.5em minus 0.4em\relax IEEE, 2023, pp. 14--18.

\bibitem{li2023liquid}
J.~Li, ``{From Liquid Crystal on Silicon and Liquid Crystal Reflectarray to Reconfigurable Intelligent Surfaces for Post-5G Networks},'' \emph{Applied Sciences}, vol.~13, no.~13, p. 7407, 2023.

\bibitem{mishra20236g}
P.~Mishra and G.~Singh, ``{6G-IoT Framework for Sustainable Smart City: Vision and Challenges},'' in \emph{Sustainable Smart Cities: Enabling Technologies, Energy Trends and Potential Applications}.\hskip 1em plus 0.5em minus 0.4em\relax Springer, 2023, pp. 97--117.

\bibitem{masouros2023guest}
C.~Masouros, J.~A. Zhang, F.~Liu, L.~Zheng, H.~Wymeersch, and M.~Di~Renzo, ``{Guest Editorial: Integrated Sensing and Communications for 6G},'' \emph{IEEE Wireless Communications}, vol.~30, no.~1, pp. 14--15, 2023.

\bibitem{renzo2019smart}
M.~D. Renzo, M.~Debbah, D.-T. Phan-Huy, A.~Zappone, M.-S. Alouini, C.~Yuen, V.~Sciancalepore, G.~C. Alexandropoulos, J.~Hoydis, H.~Gacanin \emph{et~al.}, ``{Smart radio environments empowered by reconfigurable AI meta-surfaces: An idea whose time has come},'' \emph{EURASIP Journal on Wireless Communications and Networking}, vol. 2019, no.~1, pp. 1--20, 2019.

\bibitem{hou2020reconfigurable}
T.~Hou, Y.~Liu, Z.~Song, X.~Sun, Y.~Chen, and L.~Hanzo, ``{Reconfigurable intelligent surface aided NOMA networks},'' \emph{IEEE Journal on Selected Areas in Communications}, vol.~38, no.~11, pp. 2575--2588, 2020.

\bibitem{9079457}
{Zhou, Shaoqing and Xu, Wei and Wang, Kezhi and Di Renzo, Marco and Alouini, Mohamed-Slim}, ``Spectral and energy efficiency of irs-assisted miso communication with hardware impairments,'' \emph{IEEE Wireless Communications Letters}, vol.~9, no.~9, pp. 1366--1369, 2020.

\bibitem{guan2022irs}
X.~GUAN and Q.~WU, ``{IRS-Enabled Spectrum Sharing: Interference Modeling, Channel Estimation and Robust Passive Beamforming},'' \emph{ZTE Communications}, vol.~20, no.~1, pp. 28--35, 2022.

\bibitem{ji2022reconfigurable}
Z.~Ji, Z.~Qin, and C.~G. Parini, ``{Reconfigurable intelligent surface aided cellular networks with device-to-device users},'' \emph{IEEE Transactions on Communications}, vol.~70, no.~3, pp. 1808--1819, 2022.

\bibitem{liaskos2022software}
C.~Liaskos, L.~Mamatas, A.~Pourdamghani, A.~Tsioliaridou, S.~Ioannidis, A.~Pitsillides, S.~Schmid, and I.~F. Akyildiz, ``Software-defined reconfigurable intelligent surfaces: From theory to end-to-end implementation,'' \emph{Proceedings of the IEEE}, vol. 110, no.~9, pp. 1466--1493, 2022.

\bibitem{banchs2021network}
A.~Banchs, M.~Fiore, A.~Garcia-Saavedra, and M.~Gramaglia, ``{Network intelligence in 6G: Challenges and opportunities},'' in \emph{Proceedings of the 16th ACM Workshop on Mobility in the Evolving Internet Architecture}, 2021, pp. 7--12.

\bibitem{basharat2022exploring}
S.~Basharat, M.~Khan, M.~Iqbal, U.~S. Hashmi, S.~A.~R. Zaidi, and I.~Robertson, ``{Exploring reconfigurable intelligent surfaces for 6G: State-of-the-art and the road ahead},'' \emph{IET Communications}, vol.~16, no.~13, pp. 1458--1474, 2022.

\bibitem{liu2021reconfigurable}
Y.~Liu, X.~Liu, X.~Mu, T.~Hou, J.~Xu, M.~Di~Renzo, and N.~Al-Dhahir, ``{Reconfigurable intelligent surfaces: Principles and opportunities},'' \emph{IEEE communications surveys \& tutorials}, vol.~23, no.~3, pp. 1546--1577, 2021.

\bibitem{mu2021capacity}
X.~Mu, Y.~Liu, L.~Guo, J.~Lin, and N.~Al-Dhahir, ``{Capacity and optimal resource allocation for IRS-assisted multi-user communication systems},'' \emph{IEEE Transactions on Communications}, vol.~69, no.~6, pp. 3771--3786, 2021.

\bibitem{elzanaty2021reconfigurable}
A.~Elzanaty, A.~Guerra, F.~Guidi, and M.-S. Alouini, ``{Reconfigurable intelligent surfaces for localization: Position and orientation error bounds},'' \emph{IEEE Transactions on Signal Processing}, vol.~69, pp. 5386--5402, 2021.

\bibitem{yang2020intelligent}
H.~Yang, Z.~Xiong, J.~Zhao, D.~Niyato, Q.~Wu, H.~V. Poor, and M.~Tornatore, ``{Intelligent reflecting surface assisted anti-jamming communications: A fast reinforcement learning approach},'' \emph{IEEE transactions on wireless communications}, vol.~20, no.~3, pp. 1963--1974, 2020.

\bibitem{salih2023enhancing}
N.~M. Salih, M.~Aldababsa, and K.~Yahya, ``{Enhancing UAV communication links with Reconfigurable intelligent surfaces},'' \emph{AEU-International Journal of Electronics and Communications}, vol. 171, p. 154933, 2023.

\bibitem{diamanti2021energy}
M.~Diamanti, M.~Tsampazi, E.~E. Tsiropoulou, and S.~Papavassiliou, ``{Energy efficient multi-user communications aided by reconfigurable intelligent surfaces and UAVs},'' in \emph{IEEE International Conference on Smart Computing (SMARTCOMP)}, 2021, pp. 371--376.

\bibitem{li2020reconfigurable}
S.~Li, B.~Duo, X.~Yuan, Y.-C. Liang, and M.~Di~Renzo, ``{Reconfigurable intelligent surface assisted UAV communication: Joint trajectory design and passive beamforming},'' \emph{Wireless Communications Letters}, vol.~9, no.~5, pp. 716--720, 2020.

\bibitem{basharat2021reconfigurable}
S.~Basharat, S.~A. Hassan, H.~Pervaiz, A.~Mahmood, Z.~Ding, and M.~Gidlund, ``{Reconfigurable intelligent surfaces: Potentials, applications, and challenges for 6G wireless networks},'' \emph{Wireless Communications}, vol.~28, no.~6, pp. 184--191, 2021.

\bibitem{ning2023intelligent}
Z.~Ning, T.~Li, Y.~Wu, X.~Wang, Q.~Wu, F.~R. Yu, and S.~Guo, ``{Intelligent-Reflecting-Surface-Assisted UAV Communications for 6G Networks},'' \emph{arXiv preprint arXiv:2310.20242}, 2023.

\bibitem{yu2022fair}
Y.~Yu, X.~Liu, and V.~C. Leung, ``{Fair downlink communications for RIS-UAV enabled mobile vehicles},'' \emph{Wireless Communications Letters}, vol.~11, no.~5, pp. 1042--1046, 2022.

\bibitem{MATLAB:2021a}
\emph{{MATLAB R2021a}}, The Mathworks, Inc., Natick, Massachusetts, 2021.

\bibitem{quadrigacode}
{QuaDRiGa Source Code GitHub repository}. \url{https://github.com/fraunhoferhhi/QuaDRiGa}.

\bibitem{bonati2021colosseum}
L.~Bonati, P.~Johari, M.~Polese, S.~D'Oro, S.~Mohanti, M.~Tehrani-Moayyed, D.~Villa, S.~Shrivastava, C.~Tassie, K.~Yoder, A.~Bagga, P.~Patel, V.~Petkov, M.~Seltser, F.~Restuccia, A.~Gosain, K.~R. Chowdhury, S.~Basagni, and T.~Melodia, ``{{Colosseum: Large-Scale Wireless Experimentation Through Hardware-in-the-Loop Network Emulation}},'' in \emph{Proc. of IEEE Intl. Symp. on Dynamic Spectrum Access Networks (DySPAN)}, Virtual Conference, December 2021.

\bibitem{basar2021reconfigurable}
E.~Basar and I.~Yildirim, ``{Reconfigurable intelligent surfaces for future wireless networks: A channel modeling perspective},'' \emph{IEEE Wireless Communications}, vol.~28, no.~3, pp. 108--114, 2021.

\bibitem{basar2020simris}
{E. Basar, and I. Yildirim}, ``{SimRIS channel simulator for reconfigurable intelligent surface-empowered communication systems},'' in \emph{Latin-American Conference on Communications (LATINCOM)}.\hskip 1em plus 0.5em minus 0.4em\relax IEEE, 2020, pp. 1--6.

\bibitem{sihlbom2022reconfigurable}
B.~Sihlbom, M.~I. Poulakis, and M.~Di~Renzo, ``{Reconfigurable intelligent surfaces: Performance assessment through a system-level simulator},'' \emph{Wireless Communications}, 2022.

\bibitem{sun20213d}
Y.~Sun, C.-X. Wang, J.~Huang, and J.~Wang, ``{A 3D non-stationary channel model for 6G wireless systems employing intelligent reflecting surfaces with practical phase shifts},'' \emph{IEEE Transactions on Cognitive Communications and Networking}, vol.~7, no.~2, pp. 496--510, 2021.

\bibitem{dang2021geometry}
J.~Dang, S.~Gao, Y.~Zhu, R.~Guo, H.~Jiang, Z.~Zhang, L.~Wu, B.~Zhu, and L.~Wang, ``{A geometry-based stochastic channel model and its application for intelligent reflecting surface assisted wireless communication},'' \emph{IET Communications}, vol.~15, no.~3, pp. 421--434, 2021.

\bibitem{dajer2022reconfigurable}
M.~Dajer, Z.~Ma, L.~Piazzi, N.~Prasad, X.-F. Qi, B.~Sheen, J.~Yang, and G.~Yue, ``{Reconfigurable intelligent surface: Design the channel--A new opportunity for future wireless networks},'' \emph{Digital Communications and Networks}, vol.~8, no.~2, pp. 87--104, 2022.

\bibitem{zhou2023survey}
H.~Zhou, M.~Erol-Kantarci, Y.~Liu, and H.~V. Poor, ``A survey on model-based, heuristic, and machine learning optimization approaches in ris-aided wireless networks,'' \emph{IEEE Communications Surveys \& Tutorials}, 2023.

\bibitem{kong2021channel}
L.~Kong, J.~He, Y.~Ai, S.~Chatzinotas, and B.~Ottersten, ``{Channel modeling and analysis of reconfigurable intelligent surfaces assisted vehicular networks},'' in \emph{International Conference on Communications Workshops (ICC Workshops)}.\hskip 1em plus 0.5em minus 0.4em\relax IEEE, 2021, pp. 1--6.

\bibitem{tian2023reconfigurable}
Z.~Tian, Z.~Chen, M.~Wang, Y.~Jia, and W.~Wen, ``{Reconfigurable intelligent surface-aided spectrum sharing coexisting with multiple primary networks},'' in \emph{IEEE Wireless Communications and Networking Conference (WCNC)}, 2023, pp. 1--6.

\bibitem{chen2021qos}
Y.~Chen, Y.~Wang, J.~Zhang, and M.~Di~Renzo, ``{QoS-driven spectrum sharing for reconfigurable intelligent surfaces (RISs) aided vehicular networks},'' \emph{IEEE Transactions on Wireless Communications}, vol.~20, no.~9, pp. 5969--5985, 2021.

\bibitem{boulogeorgos2021coverage}
A.-A.~A. Boulogeorgos and A.~Alexiou, ``{Coverage analysis of reconfigurable intelligent surface assisted THz wireless systems},'' \emph{IEEE Open Journal of Vehicular Technology}, vol.~2, pp. 94--110, 2021.

\bibitem{bjornson2022reconfigurable}
E.~Bj{\"o}rnson, H.~Wymeersch, B.~Matthiesen, P.~Popovski, L.~Sanguinetti, and E.~de~Carvalho, ``{Reconfigurable intelligent surfaces: A signal processing perspective with wireless applications},'' \emph{IEEE Signal Processing Magazine}, vol.~39, no.~2, pp. 135--158, 2022.

\bibitem{yu2021smart}
X.~Yu, V.~Jamali, D.~Xu, D.~W.~K. Ng, and R.~Schober, ``Smart and reconfigurable wireless communications: From irs modeling to algorithm design,'' \emph{Wireless Communications}, vol.~28, no.~6, pp. 118--125, 2021.

\bibitem{xu2023exploiting}
J.~Xu, X.~Mu, and Y.~Liu, ``{Exploiting STAR-RISs in near-field communications},'' \emph{IEEE Transactions on Wireless Communications}, 2023.

\bibitem{tang2021channel}
W.~Tang, X.~Chen, M.~Z. Chen, J.~Y. Dai, Y.~Han, S.~Jin, Q.~Cheng, G.~Y. Li, and T.~J. Cui, ``{On channel reciprocity in reconfigurable intelligent surface assisted wireless networks},'' \emph{Wireless Communications}, vol.~28, no.~6, pp. 94--101, 2021.

\bibitem{da2023varactor}
L.~G. da~Silva, Z.~Chu, P.~Xiao, and A.~Cerqueira S~Jr, ``{A varactor-based 1024-element RIS design for mm-waves},'' \emph{Frontiers in Communications and Networks}, vol.~4, p. 1086011, 2023.

\bibitem{liu2022reconfigurable}
Y.~Liu, J.~Dou, Y.~Cui, Y.~Chen, J.~Yang, F.~Qin, and Y.~Wang, ``{Reconfigurable Intelligent Surface Physical Model in Channel Modeling},'' \emph{Electronics}, vol.~11, no.~17, p. 2798, 2022.

\bibitem{liu2022simulation}
R.~Liu, J.~Dou, P.~Li, J.~Wu, and Y.~Cui, ``Simulation and field trial results of reconfigurable intelligent surfaces in 5g networks,'' \emph{IEEE Access}, vol.~10, pp. 122\,786--122\,795, 2022.

\bibitem{rossanese2022designing}
M.~Rossanese, P.~Mursia, A.~Garcia-Saavedra, V.~Sciancalepore, A.~Asadi, and X.~Costa-Perez, ``{Designing, building, and characterizing RF switch-based reconfigurable intelligent surfaces},'' in \emph{Proceedings of the 16th ACM Workshop on Wireless Network Testbeds, Experimental evaluation \& CHaracterization}, 2022, pp. 69--76.

\bibitem{8741198}
C.~Huang, A.~Zappone, G.~C. Alexandropoulos, M.~Debbah, and C.~Yuen, ``{Reconfigurable Intelligent Surfaces for Energy Efficiency in Wireless Communication},'' \emph{IEEE Transactions on Wireless Communications}, vol.~18, no.~8, pp. 4157--4170, 2019.

\bibitem{9950621}
X.~Yu, G.~Wang, X.~Huang, K.~Wang, W.~Xu, and Y.~Rui, ``{Energy Efficient Resource Allocation for Uplink RIS-Aided Millimeter-Wave Networks With NOMA},'' \emph{IEEE Transactions on Mobile Computing}, vol.~23, no.~1, pp. 423--436, 2024.

\bibitem{9810453}
A.~A. Deshpande, C.~J. Vaca-Rubio, S.~Mohebi, D.~Salami, E.~de~Carvalho, P.~Popovski, S.~Sigg, M.~Zorzi, and A.~Zanella, ``Energy-efficient design for ris-assisted uav communications in beyond-5g networks,'' in \emph{2022 20th Mediterranean Communication and Computer Networking Conference (MedComNet)}, 2022, pp. 158--165.

\bibitem{9628234}
M.~Ahsan, S.~Jamil, M.~T. Ejaz, and M.~S. Abbas, ``Energy efficiency maximization in ris-assisted wireless networks,'' in \emph{2021 International Conference on Computing, Electronic and Electrical Engineering (ICE Cube)}, 2021, pp. 1--6.

\bibitem{10096617}
R.~K. Fotock, A.~Zappone, and M.~D. Renzo, ``{Energy Efficiency Maximization in RIS-aided Networks with Global Reflection Constraints},'' in \emph{ICASSP 2023 - 2023 IEEE International Conference on Acoustics, Speech and Signal Processing (ICASSP)}, 2023, pp. 1--5.

\bibitem{10107766}
Y.~Zhang, Y.~Lu, R.~Zhang, B.~Ai, and D.~Niyato, ``Deep reinforcement learning for secrecy energy efficiency maximization in ris-assisted networks,'' \emph{IEEE Transactions on Vehicular Technology}, vol.~72, no.~9, pp. 12\,413--12\,418, 2023.

\bibitem{9964251}
Y.~Guo, F.~Fang, D.~Cai, and Z.~Ding, ``{Energy-Efficient Design for a NOMA Assisted STAR-RIS Network With Deep Reinforcement Learning},'' \emph{IEEE Transactions on Vehicular Technology}, vol.~72, no.~4, pp. 5424--5428, 2023.

\bibitem{diamanti2021prospect}
M.~Diamanti, P.~Charatsaris, E.~E. Tsiropoulou, and S.~Papavassiliou, ``{The prospect of reconfigurable intelligent surfaces in integrated access and backhaul networks},'' \emph{IEEE Transactions on Green Communications and Networking}, vol.~6, no.~2, pp. 859--872, 2021.

\bibitem{tehrani2021creating}
M.~Tehrani-Moayyed, L.~Bonati, P.~Johari, T.~Melodia, and S.~Basagni, ``{Creating {RF} Scenarios for Large-Scale, Real-Time Wireless Channel Emulators},'' in \emph{Proceedings of IEEE MedComNet}, 2021.

\bibitem{villa2022cast}
D.~Villa, M.~Tehrani-Moayyed, P.~Johari, S.~Basagni, and T.~Melodia, ``{CaST: a toolchain for creating and characterizing realistic wireless network emulation scenarios},'' in \emph{Proceedings of the 16th ACM Workshop on Wireless Network Testbeds, Experimental evaluation \& CHaracterization}, 2022, pp. 45--52.

\bibitem{rusca2023mobile}
R.~Rusca, F.~Raviglione, C.~Casetti, P.~Giaccone, and F.~Restuccia, ``{Mobile RF Scenario Design for Massive-Scale Wireless Channel Emulators},'' in \emph{Joint European Conference on Networks and Communications \& 6G Summit (EuCNC/6G Summit)}.\hskip 1em plus 0.5em minus 0.4em\relax IEEE, 2023, pp. 675--680.

\bibitem{lasaulce2009introducing}
S.~Lasaulce, Y.~Hayel, R.~El~Azouzi, and M.~Debbah, ``{Introducing hierarchy in energy games},'' \emph{IEEE Transactions on Wireless Communications}, vol.~8, no.~7, pp. 3833--3843, 2009.

\bibitem{he2011stackelberg}
G.~He, S.~Lasaulce, and Y.~Hayel, ``{Stackelberg games for energy-efficient power control in wireless networks},'' in \emph{2011 Proceedings IEEE INFOCOM}.\hskip 1em plus 0.5em minus 0.4em\relax IEEE, 2011, pp. 591--595.

\bibitem{srsran}
{srsRAN website}. \url{https://www.srslte.com/}. Accessed September 2022.

\bibitem{googlemaps}
``{Google Maps},'' [Online]. Available: \url{https://www.google.com/maps}.

\bibitem{ju2019millimeter}
S.~Ju, O.~Kanhere, Y.~Xing, and T.~S. Rappaport, ``{A millimeter-wave channel simulator NYUSIM with spatial consistency and human blockage},'' in \emph{IEEE global communications conference (GLOBECOM)}, 2019, pp. 1--6.

\bibitem{kyosti20074}
P.~Kyosti, J.~Meinila, L.~Hentila, X.~Zhao, T.~Jamsa, C.~Schneider, M.~Narandzic, M.~Milojevic, A.~Hong, J.~Ylitalo \emph{et~al.}, ``{IST-4-027756 WINNER II D1. 1.2 V1. 2: WINNER II channel models},'' \emph{Available: www. ist-winner. org}, 2007.

\bibitem{ademaj20163gpp}
F.~Ademaj, M.~Taranetz, and M.~Rupp, ``{3GPP 3D MIMO channel model: A holistic implementation guideline for open source simulation tools},'' \emph{EURASIP Journal on Wireless Communications and Networking}, no.~1, pp. 1--14, 2016.

\bibitem{pang2022investigation}
L.~Pang, J.~Zhang, Y.~Zhang, X.~Huang, Y.~Chen, and J.~LI, ``{Investigation and Comparison of 5G Channel Models: From QuaDRiGa, NYUSIM, and MG5G Perspectives},'' \emph{Chinese Journal of Electronics}, vol.~31, no.~1, pp. 1--17, 2022.

\bibitem{mondal20153d}
B.~Mondal, T.~A. Thomas, E.~Visotsky, F.~W. Vook, A.~Ghosh, Y.-H. Nam, Y.~Li, J.~Zhang, M.~Zhang, Q.~Luo \emph{et~al.}, ``{3D channel model in 3GPP},'' \emph{IEEE Communications Magazine}, vol.~53, no.~3, pp. 16--23, 2015.

\bibitem{jaeckel2019f}
S.~Jaeckel, L.~Raschkowski, K.~Boerner, and L.~Thiele, ``{F. Burkhardt, E. Eberlein,“QuaDRiGa-Quasi Deterministic Radio Channel Generator,” User Manual and Documentation},'' Tech. Rep. v2. 2.0, Fraunhofer Heinrich Hertz Institute, Tech. Rep., 2019, [Online]. Available: \url{https://github.com/fraunhoferhhi/QuaDRiGa/blob/main/quadriga_documentation_v2.6.1-0.pdf}.

\bibitem{jaeckel2014quadriga}
S.~Jaeckel, L.~Raschkowski, K.~B{\"o}rner, and L.~Thiele, ``{QuaDRiGa: A 3-D multi-cell channel model with time evolution for enabling virtual field trials},'' \emph{IEEE transactions on antennas and propagation}, vol.~62, no.~6, pp. 3242--3256, 2014.

\bibitem{ranjkesh2015optimized}
M.~E. Ranjkesh and A.~A. Khazaei, ``{The Optimized QuaDRiGa Wi-Fi Channel Model},'' \emph{International Journal of Science and Engineering Investigations}, vol.~4, no.~9, pp. 51--61, 2015.

\bibitem{etsi19deliv}
{``Intelligent Transport Systems (ITS); Access Layer; Part 1: Channel Models for the 5,9 GHz frequency band''}. Sophia Antipolis Cedex, FRANCE, ETSI TR 103 257-1 V1.1.1, Tech. Rep., May, 2019. [Online]. Available: \url{https://www.etsi.org/deliver/etsi_tr/103200_103299/10325701/01.01.01_60/tr_10325701v010101p.pdf}.

\bibitem{he2020investigation}
Y.~He, Y.~Zhang, J.~Zhang, L.~Pang, Y.~Chen, and G.~Ren, ``{Investigation and Comparison of QuaDRiGa, NYUSIM and MG5G Channel Models for 5G Wireless Communications},'' in \emph{IEEE 92nd Vehicular Technology Conference (VTC2020-Fall)}, 2020, pp. 1--5.

\bibitem{etsi5g}
{ETSI TR 138 901 V14.3.0 (2018-01)}. [Online]. Available: \url{https://www.etsi.org/deliver/etsi_tr/138900_138999/138901/14.03.00_60/tr_138901v140300p.pdf}.

\bibitem{arunachalaperumal2018enhanced}
C.~Arunachalaperumal, S.~Dhilipkumar, and G.~Abija, ``{Enhanced 3D MIMO channel for urban macro environment},'' \emph{International Journal of Pure and Applied Mathematics, Special Issue}, vol. 118, no.~10, pp. 259--269, 2018.

\bibitem{giordani2019path}
M.~Giordani, T.~Shimizu, A.~Zanella, T.~Higuchi, O.~Altintas, and M.~Zorzi, ``{Path loss models for V2V mmWave communication: performance evaluation and open challenges},'' in \emph{IEEE Connected and Automated Vehicles Symposium (CAVS)}.\hskip 1em plus 0.5em minus 0.4em\relax IEEE, 2019, pp. 1--5.

\bibitem{mecklenbrauker2011vehicular}
C.~F. Mecklenbrauker, A.~F. Molisch, J.~Karedal, F.~Tufvesson, A.~Paier, L.~Bernad{\'o}, T.~Zemen, O.~Klemp, and N.~Czink, ``{Vehicular channel characterization and its implications for wireless system design and performance},'' \emph{Proceedings of the IEEE}, vol.~99, no.~7, pp. 1189--1212, 2011.

\bibitem{mgen}
{U.S. Naval Research Laboratory, "Multi-Generator ({MGEN}) Network Test Tool"}. \url{https://www.nrl.navy.mil/itd/ncs/products/mgen}. 2019.

\bibitem{bloessl2017performance}
B.~Bloessl, M.~Segata, C.~Sommer, and F.~Dressler, ``{Performance assessment of IEEE 802.11 p with an open source SDR-based prototype},'' \emph{IEEE transactions on mobile computing}, vol.~17, no.~5, pp. 1162--1175, 2017.

\bibitem{khan2022vehicle}
A.~A. Khan, A.~A. Laghari, M.~Shafiq, S.~A. Awan, and Z.~Gu, ``{Vehicle to everything (V2X) and edge computing: A secure lifecycle for UAV-assisted vehicle network and offloading with blockchain},'' \emph{Drones}, vol.~6, no.~12, p. 377, 2022.

\bibitem{ETSI_RIS001}
\BIBentryALTinterwordspacing
ETSI, ``{Reconfigurable Intelligent Surfaces (RIS); Use Cases, Deployment Scenarios and Requirements},'' ETSI, ETSI Group Report RIS 001 V1.1.1, April 2023. [Online]. Available: \url{https://www.etsi.org/deliver/etsi_gr/RIS/001_099/001/01.01.01_60/gr_RIS001v010101p.pdf}
\BIBentrySTDinterwordspacing

\bibitem{paul2022omni}
L.~C. Paul, H.~K. Saha, T.~Rani, M.~Z. Mahmud, T.~K. Roy, W.-S. Lee \emph{et~al.}, ``{An Omni-Directional Wideband Patch Antenna with Parasitic Elements for Sub-6 GHz Band Applications},'' \emph{International Journal of Antennas and Propagation}, 2022.

\bibitem{azim2021multi}
R.~Azim, A.~M.~H. Meaze, A.~Affandi, M.~M. Alam, R.~Aktar, M.~S. Mia, T.~Alam, M.~Samsuzzaman, and M.~T. Islam, ``{A multi-slotted antenna for LTE/5G Sub-6 GHz wireless communication applications},'' \emph{International Journal of Microwave and Wireless Technologies}, vol.~13, no.~5, pp. 486--496, 2021.

\bibitem{rani2022development}
T.~Rani, S.~C. Das, M.~S. Hossen, L.~C. Paul, and T.~K. Roy, ``{Development of a Broadband Antenna for 5G Sub-6 GHz Cellular and IIoT Smart Automation Applications},'' in \emph{12th International Conference on Electrical and Computer Engineering (ICECE)}.\hskip 1em plus 0.5em minus 0.4em\relax IEEE, 2022, pp. 465--468.

\bibitem{sharawi2017two}
M.~S. Sharawi, M.~Ikram, and A.~Shamim, ``{A two concentric slot loop based connected array MIMO antenna system for 4G/5G terminals},'' \emph{IEEE Transactions on antennas and propagation}, vol.~65, no.~12, pp. 6679--6686, 2017.

\bibitem{de2021radio}
B.~De~Beelde, E.~Tanghe, M.~Yusuf, D.~Plets, and W.~Joseph, ``{Radio channel modeling in a ship hull: Path loss at 868 MHz and 2.4, 5.25, and 60 GHz},'' \emph{IEEE Antennas and Wireless Propagation Letters}, vol.~20, no.~4, pp. 597--601, 2021.

\bibitem{zaidi2020wide}
A.~Zaidi, W.~A. Awan, N.~Hussain, and A.~Baghdad, ``{A Wide and Tri-band Flexible Antennas with Independently Controllable Notch Bands for Sub-6-GHz Communication System.}'' \emph{Radioengineering}, vol.~29, no.~1, 2020.

\bibitem{mao2018planar}
C.-X. Mao, M.~Khalily, P.~Xiao, T.~W. Brown, and S.~Gao, ``{Planar sub-millimeter-wave array antenna with enhanced gain and reduced sidelobes for 5G broadcast applications},'' \emph{IEEE Transactions on Antennas and Propagation}, vol.~67, no.~1, pp. 160--168, 2018.

\bibitem{ranvier2008low}
S.~Ranvier, S.~Dudorov, M.~Kyro, C.~Luxey, C.~Icheln, R.~Staraj, and P.~Vainikainen, ``{Low-cost planar omnidirectional antenna for mm-wave applications},'' \emph{IEEE Antennas and Wireless Propagation Letters}, vol.~7, pp. 521--523, 2008.

\bibitem{fan2018wideband}
K.~Fan, Z.-C. Hao, Q.~Yuan, J.~Hu, G.~Q. Luo, and W.~Hong, ``{Wideband horizontally polarized omnidirectional antenna with a conical beam for millimeter-wave applications},'' \emph{IEEE Transactions on Antennas and Propagation}, vol.~66, no.~9, pp. 4437--4448, 2018.

\bibitem{maccartney2015millimeter}
G.~R. Maccartney, T.~S. Rappaport, M.~K. Samimi, and S.~Sun, ``{Millimeter-wave omnidirectional path loss data for small cell 5G channel modeling},'' \emph{IEEE access}, vol.~3, pp. 1573--1580, 2015.

\bibitem{hasan2019dual}
M.~N. Hasan, S.~Bashir, and S.~Chu, ``{Dual band omnidirectional millimeter wave antenna for 5G communications},'' \emph{Journal of Electromagnetic Waves and Applications}, vol.~33, no.~12, pp. 1581--1590, 2019.

\bibitem{sun2015synthesizing}
S.~Sun, G.~R. MacCartney, M.~K. Samimi, and T.~S. Rappaport, ``{Synthesizing omnidirectional antenna patterns, received power and path loss from directional antennas for 5G millimeter-wave communications},'' in \emph{IEEE Global Communications Conference (GLOBECOM)}.\hskip 1em plus 0.5em minus 0.4em\relax IEEE, 2015, pp. 1--7.

\bibitem{abirami2017review}
M.~Abirami, ``{A review of patch antenna design for 5G},'' in \emph{IEEE International Conference on Electrical, Instrumentation and Communication Engineering (ICEICE)}.\hskip 1em plus 0.5em minus 0.4em\relax IEEE, 2017, pp. 1--3.

\bibitem{niu2015survey}
Y.~Niu, Y.~Li, D.~Jin, L.~Su, and A.~V. Vasilakos, ``{A survey of millimeter wave communications (mmWave) for 5G: opportunities and challenges},'' \emph{Wireless networks}, vol.~21, pp. 2657--2676, 2015.

\bibitem{pan2011dual}
H.~K. Pan, ``{Dual-polarized Mm-wave phased array antenna for multi-Gb/s 60GHz communication},'' in \emph{IEEE International Symposium on Antennas and Propagation (APSURSI)}.\hskip 1em plus 0.5em minus 0.4em\relax IEEE, 2011, pp. 3279--3282.

\bibitem{mao2020compact}
C.~X. Mao, M.~Khalily, L.~Zhang, P.~Xiao, Y.~Sun, and D.~H. Werner, ``{Compact patch antenna with vertical polarization and omnidirectional radiation characteristics},'' \emph{IEEE Transactions on Antennas and Propagation}, vol.~69, no.~2, pp. 1158--1161, 2020.

\bibitem{tsiropoulou2015combined}
E.~E. Tsiropoulou, P.~Vamvakas, G.~K. Katsinis, and S.~Papavassiliou, ``{Combined power and rate allocation in self-optimized multi-service two-tier femtocell networks},'' \emph{Computer Communications}, vol.~72, pp. 38--48, 2015.

\bibitem{etsi2022}
{ETSI}, ``{{5G; NR; User Equipment (UE) radio transmission and reception; Part 1: Range 1 Standalone (3GPP TS 38.101-1 version 17.5.0 Release 17)}},'' \url{https://www.etsi.org/deliver/etsi_ts/138100_138199/13810101/17.05.00_60/ts_13810101v170500p.pdf}, May 2022, {ETSI TS 138 101-1 V17.5.0 (2022-05)}.

\bibitem{etsi2022b}
\BIBentryALTinterwordspacing
ETSI, ``{5G; NR; User Equipment (UE) radio transmission and reception; Part 2: Range 2 Standalone (3GPP TS 38.101-2 version 17.5.0 Release 17)},'' ETSI, Tech. Rep. ETSI TS 138 101-2 V17.5.0, April 2022, available online. [Online]. Available: \url{https://www.etsi.org/deliver/etsi_ts/138100_138199/13810102/17.05.00_60/ts_13810102v170500p.pdf}
\BIBentrySTDinterwordspacing

\bibitem{etsi2017}
{ETSI}, ``{{LTE; Evolved Universal Terrestrial Radio Access (E-UTRA); Physical layer procedures (3GPP TS 36.213 version 14.2.0 Release 14)}},'' \url{https://www.etsi.org/deliver/etsi_ts/136200_136299/136213/14.02.00_60/ts_136213v140200p.pdf}, April 2017, {ETSI TS 136 213 V14.2.0 (2017-04)}.

\bibitem{tapio2021survey}
V.~Tapio, I.~Hemadeh, A.~Mourad, A.~Shojaeifard, and M.~Juntti, ``{Survey on reconfigurable intelligent surfaces below 10 GHz},'' \emph{EURASIP Journal on Wireless Communications and Networking}, pp. 1--18, 2021.

\bibitem{bjornson2020reconfigurable}
E.~Bj{\"o}rnson, {\"O}.~{\"O}zdogan, and E.~G. Larsson, ``{Reconfigurable intelligent surfaces: Three myths and two critical questions},'' \emph{IEEE Communications Magazine}, vol.~58, no.~12, pp. 90--96, 2020.

\bibitem{siddiqi2022reconfigurable}
M.~Z. Siddiqi and T.~Mir, ``{Reconfigurable intelligent surface-aided wireless communications: An overview},'' \emph{Intelligent and Converged Networks}, vol.~3, no.~1, pp. 33--63, 2022.

\bibitem{villa2023twinning}
D.~Villa, D.~Uvaydov, L.~Bonati, P.~Johari, J.~M. Jornet, and T.~Melodia, ``{Twinning Commercial Radio Waveforms in the Colosseum Wireless Network Emulator},'' in \emph{Proceedings of the 17th ACM Workshop on Wireless Network Testbeds, Experimental evaluation \& Characterization}, 2023, pp. 33--40.

\bibitem{bonati2021scope}
L.~Bonati, S.~D'Oro, S.~Basagni, and T.~Melodia, ``{{SCOPE}: An open and softwarized prototyping platform for NextG systems},'' in \emph{Proceedings of the 19th Annual International Conference on Mobile Systems, Applications, and Services}, 2021, pp. 415--426.

\bibitem{gomez2016srslte}
I.~Gomez-Miguelez, A.~Garcia-Saavedra, P.~D. Sutton, P.~Serrano, C.~Cano, and D.~J. Leith, ``{{srsLTE}: An Open-Source Platform for {LTE} Evolution and Experimentation},'' in \emph{Proceedings of the Tenth ACM International Workshop on Wireless Network Testbeds, Experimental Evaluation, and Characterization}, 2016, pp. 25--32.

\bibitem{tsampazi2023comparative}
M.~Tsampazi, S.~D'Oro, M.~Polese, L.~Bonati, G.~Poitau, M.~Healy, and T.~Melodia, ``{A Comparative Analysis of Deep Reinforcement Learning-based xApps in O-RAN},'' \emph{IEEE Global Communications Conference (GLOBECOM)}, 2023.

\bibitem{tsampazi2024pandora}
M.~Tsampazi, S.~D'Oro, M.~Polese, L.~Bonati, G.~Poitau, M.~Healy, M.~Alavirad, and T.~Melodia, ``{PandORA: Automated Design and Comprehensive Evaluation of Deep Reinforcement Learning Agents for Open RAN},'' \emph{arXiv preprint arXiv:2407.11747}, 2024.

\end{thebibliography}
\bibliographystyle{IEEEtran}

\begin{IEEEbiography}[{\includegraphics[width=1in,height=1.25in,clip,keepaspectratio]{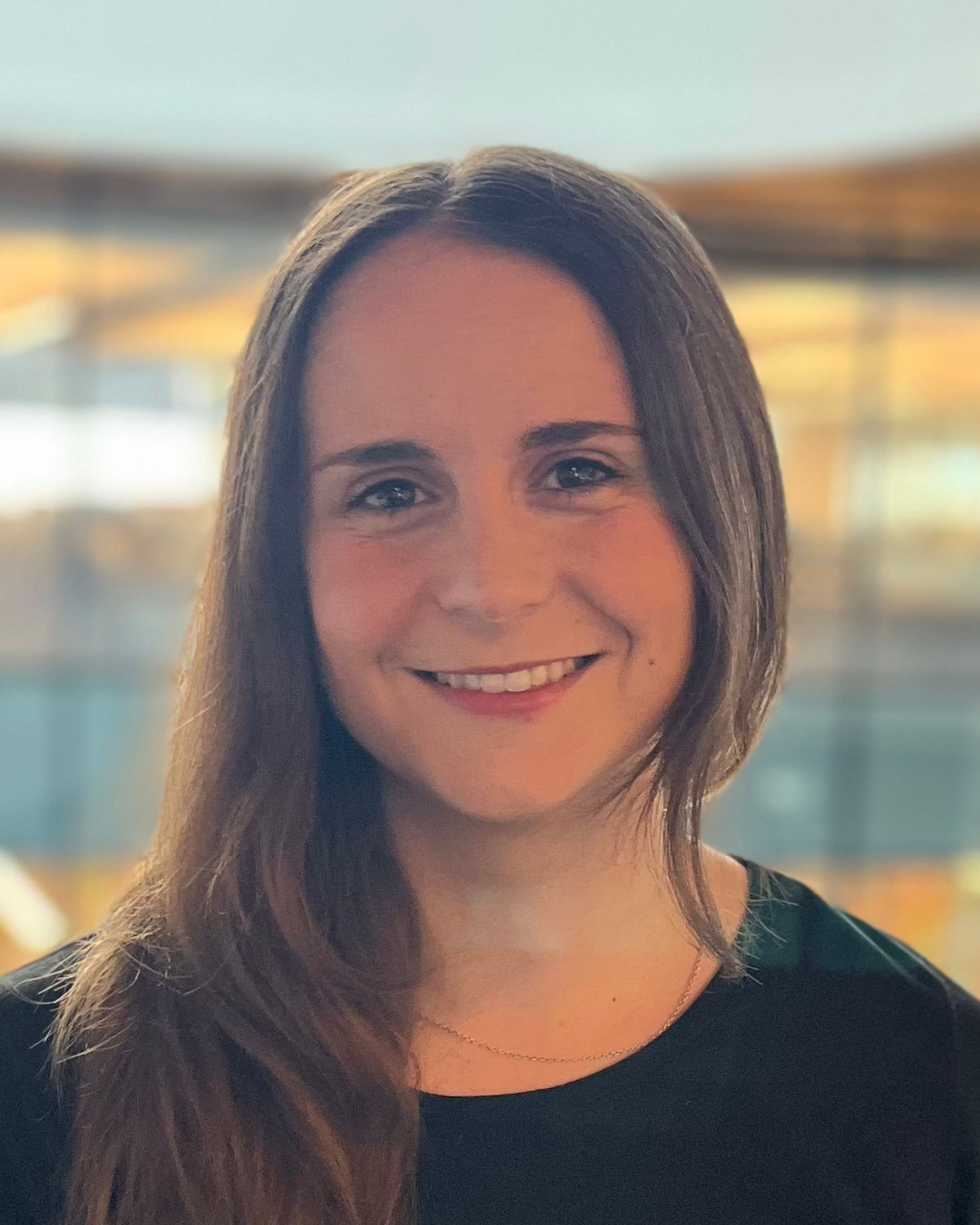}}]{Maria Tsampazi}
received her MEng Degree in Electrical and Computer Engineering from National Technical University of Athens, Greece in 2021. She is a Ph.D. Candidate in Electrical Engineering at the Institute for the Wireless Internet of Things at Northeastern University, Boston, MA, USA. Her research interests focus on NextG wireless networks and intelligent resource allocation in Open RAN. She has received academic awards sponsored by the U.S. National Science Foundation, the IEEE Communications Society, and Northeastern University, and is a 2024 recipient of the National Spectrum Consortium Women in Spectrum Scholarship. She has previously collaborated with both government and industry, including organizations such as the U.S. Department of Transportation and Dell Technologies.
\end{IEEEbiography}

\vspace{-10cm}

\begin{IEEEbiography}[{\includegraphics[width=1in,height=1.25in,clip,keepaspectratio]{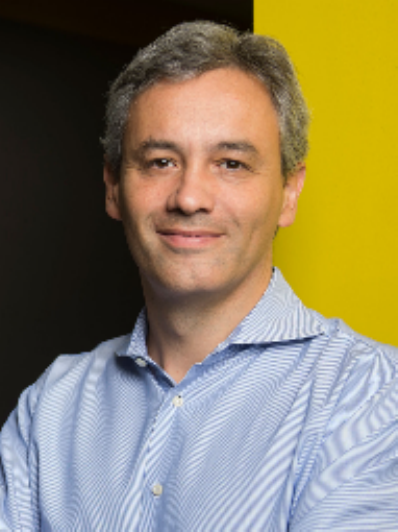}}]{Tommaso Melodia}
is the William Lincoln Smith
Chair Professor with the Department of Electrical and Computer Engineering at Northeastern
University in Boston. He is also the Founding
Director of the Institute for the Wireless Internet
of Things and the Director of Research for the
PAWR Project Office. He received his Ph.D. in
Electrical and Computer Engineering from the
Georgia Institute of Technology in 2007. He is
a recipient of the National Science Foundation
CAREER award. Prof. Melodia has served as
Associate Editor of IEEE Transactions on Wireless Communications,
IEEE Transactions on Mobile Computing, Elsevier Computer Networks,
among others. He has served as Technical Program Committee Chair
for IEEE INFOCOM 2018, General Chair for IEEE SECON 2019, ACM
Nanocom 2019, and ACM WUWnet 2014. Prof. Melodia is the Director of Research for the Platforms for Advanced Wireless Research
(PAWR) Project Office, a $100M$ public-private partnership to establish
4 city-scale platforms for wireless research to advance the US wireless
ecosystem in years to come. Prof. Melodia’s research on modeling, optimization, and experimental evaluation of Internet-of-Things and wireless
networked systems has been funded by the National Science Foundation, the Air Force Research Laboratory the Office of Naval Research,
DARPA, and the Army Research Laboratory. Prof. Melodia is a Fellow of
the IEEE and a Senior Member of the ACM.
\end{IEEEbiography}

\end{document}